\documentclass[a4paper,11pt]{article}
\pdfoutput=1 

\usepackage{lib/jinstpub} 

\usepackage{lineno, hyperref}
\usepackage{xspace}
\usepackage{booktabs}
\usepackage{siunitx}
\usepackage[table]{xcolor}
\newif\ifDraft \Draftfalse
\ifDraft
\usepackage[local,pcount,mark,markifdirty]{gitinfo2}
\hypersetup{
    pdftitle={Lycoris - a large-area, high resolution beam telescope - \gitAbbrevHash},
    pdfsubject={Commited on \gitAuthorDate},
    pdfauthor={\gitAuthorName},
    pdfkeywords={Telescope, SiD, KPiX, Silicon Strip Sensor, Charged particle tracking}
}
\else
\hypersetup{
    pdftitle={Lycoris - a large-area, high resolution beam telescope},
    pdfkeywords={Telescope, SiD, KPiX, Silicon Strip Sensor, Charged particle tracking}
}
\fi

\newcommand{\LYCORIS}{{\textsc{Lycoris}}\xspace}
\newcommand{\DESYII}{{\mbox{DESY II}}\xspace}
\newcommand{\DIITBF}{{\DESYII Test Beam Facility}\xspace}
\newcommand{\PIA}{{PIA}\xspace}
\newcommand{\LINACII}{{\mbox{LINAC II}}\xspace}
\newcommand{\AZALEA}{{\textsc{AZALEA}}\xspace}

\newcommand{\PCMAG}{{PCMAG}\xspace}
\newcommand{\SID}{{SiD}\xspace}
\newcommand{\KPIX}{{KPiX}\xspace}

\newcommand{\GEANT}{\textsc{Geant4}\xspace}
\newcommand{\EUDAQ}{\textsc{EUDAQ}\xspace}
\newcommand{\EUDAQII}{\textsc{EUDAQ2}\xspace}
\newcommand{\EUTELESCOPE}{\textsc{EUTelescope}\xspace}
\newcommand{\PYTHON}{\textsc{Python}\xspace}
\newcommand{\CPLUSPLUS}{\textsc{C++}\xspace}
\newcommand{\ROOT}{\textsc{ROOT}\xspace}

\newcommand{\CMAKE}{\textsc{CMake}\xspace}
\newcommand{\Rogue}{\textsc{Rogue}\xspace}
\newcommand{\BOOSTPYTHON}{\textsc{Boost.Python}\xspace}
\newcommand{\striplet}{\textit{Striplet}\xspace}
\newcommand{\striplets}{\textit{Striplets}\xspace}

\newcommand{\DATACONVERTER}{\texttt{DataConverter}\xspace}

\newcommand{\EUDET}{\textsc{EUDET}\xspace}
\newcommand{\AIDA}{\textsc{AIDA}\xspace}
\newcommand{\AIDAII}{\textsc{AIDA-2020}\xspace}

\newcommand{\RunControl}{\texttt{Run Control}\xspace}
\newcommand{\Producer}{\texttt{Producer}\xspace}
\newcommand{\Producers}{\texttt{Producers}\xspace}
\newcommand{\DataCollector}{\texttt{Data Collector}\xspace}

\newcommand{\StdEventMonitor}{\texttt{StdEventMonitor}\xspace}

\newcommand{\MIMOSA}{{Mimosa26}\xspace}
\newcommand{\ITWOC}{{I$^2$C}\xspace}

\newcommand{\KPIXAcqstart}{\textit{Acquisition Start Signal}\xspace}

\newcommand{\KPIXTIdle}{$T_{\mathrm{Idle}}$\xspace}
\newcommand{\KPIXTAcq}{$T_{\mathrm{Acquisition}}$\xspace}
\newcommand{\KPIXTDigi}{$T_{\mathrm{Digitization}}$\xspace}
\newcommand{\KPIXTPreC}{$T_{\mathrm{Pre-Charge}}$\xspace}
\newcommand{\KPIXTReadout}{$T_{\mathrm{Readout}}$\xspace}

\def\GeV{\ifmode {\mathrm{\ Ge\kern -0.1em V }}\else
								 {\textrm{Ge\kern -0.1em V }}\fi}%
\DeclareSIUnit[per-mode=symbol,per-symbol=p]{\fC}{\femto\coulomb}

\title{\textsc{Lycoris} - a large-area, high resolution beam telescope\note{(c)
All figures and pictures by the author(s) under a \href{https://creativecommons.org/licenses/by/4.0/}{CC BY 4.0} license}}

\author[e]{James Brau}
\author[b]{Martin Breidenbach}
\author[b]{Dietrich R. Freytag}
\author[a]{Claus Kleinwort}
\author[a]{Uwe Kraemer}
\author[b]{Benjamin A. Reese}
\author[a,c]{Sebastiaan Roelofs}
\author[a]{Marcel Stanitzki}
\author[e]{Amanda Steinhebel}
\author[a,d]{Dimitra Tsionou}
\author[a,f,2]{Mengqing Wu\note{Corresponding author.}}

\affiliation[a]{DESY, Notkestrasse 85, 22607 Hamburg, Germany}
\affiliation[b]{SLAC National Accelerator Laboratory, 2575 Sand Hill Rd, Menlo Park, CA, USA }
\affiliation[c]{now at: Delft University of Technology, 2600 AA Delft, The Netherlands}
\affiliation[d]{now at: Aristotle University of Thessaloniki, University Campus, 54124 Thessaloniki, Greece} 
\affiliation[e]{University of Oregon, Center for High Energy Physics, 1274 University of Oregon Eugene, Oregon 97403-1274 USA}
\affiliation[f]{now at: Institute for Mathematics, Astrophysics and Particle Physics, Radboud University/Nikhef, 6500 GL Nijmegen, The Netherlands}
\emailAdd{mengqing.wu@desy.de}

\ifDraft
\linenumbers
\else
\fi

\abstract{
A high-resolution beam telescope is one of the most important and demanding infrastructure
components at any test beam facility. Its main purpose is to
provide reference particle tracks from the incoming test beam particles to the
test beam users, which allows measurement of the performance of the device-under-test (DUT).
\LYCORIS, a six-plane compact beam telescope with an
active area of $\sim$10$\times$\SI{10}{\square\centi\metre} (extensible to
10$\times$\SI{20}{\square\centi\metre}) was installed at the \DIITBF in
2019, to provide a precise momentum measurement in a \SI{1}{\tesla} solenoid
magnet or to provide tracking over a large area.

The overall design of \LYCORIS will be described as well as the performance
of the chosen silicon sensor. The \SI{25}{\micro\metre} pitch micro-strip
sensor used for \LYCORIS was originally designed for the \SID detector concept
for the International Linear Collider. It adopts a second metallization layer
to route signals from strips to the bump-bonded \KPIX ASIC and uses a wire-bonded
flex cable for the connection to the DAQ and the power supply system. This
arrangement eliminates the need for a dedicated hybrid PCB. Its performance
was tested for the first time in this project. The system has been evaluated
at the \DIITBF in several test-beam campaigns and has demonstrated an average
single-point resolution of \SI{7.07}{\micro\meter}.
}

\begin{document}
\maketitle
\flushbottom
\section{Introduction}\label{sec:intro}
Beam telescopes have become the workhorse for many campaigns at test beam facilities.
Typically, they consist of two stations or arms, each made up of a set of tracking sensors with a configurable spacing in between, where the Device-under-test (DUT) is usually placed.
They provide very precise reference tracks
and may also provide a common trigger, DAQ integration and even common reconstruction packages.

For over a decade, \EUDET-style pixel beam telescopes~\cite{jansen2016} have been a success-story with a wide-ranging user community
and strongly supported by the EU-funded \EUDET~\cite{eudet}, \AIDA~\cite{aida} and \AIDAII~\cite{aida2020} projects.
\EUDET-style telescopes have six planes of \MIMOSA~\cite{baudot2009,huguo2010} monolithic active pixel sensors providing high-precision beam tracking of around \SI{2}{\micro\metre} in a small active area of 2$\times$\SI{1}{\square\centi\metre}.
The telescope comes in a full package including a common trigger logic unit, the \EUDET TLU~\cite{tlu}, a DAQ framework called \EUDAQ~\cite{Ahlburg_2020} allowing an easy integration of the user DAQ, and a complete reconstruction software suite called the \EUTELESCOPE package~\cite{Bisanz:2020rfv}.

Seven copies of \EUDET-style pixel beam telescopes have been made and are installed at CERN (PS and SPS), at the \DIITBF~\cite{desytb2018}, ELSA in Bonn, and at ESTB~\cite{Fieguth:2011be} at SLAC.
A large number of different user setups have been integrated into \EUDAQ and benefited from the \EUDET TLU~\cite{tlu} and used \EUTELESCOPE for the data reconstruction.
The user community at the various test beam facilities has been very satisfied with the \EUDET-style pixel beam telescopes
but there were two requests which could not be met:
having a telescope for large DUTs inside the \PCMAG \SI{1}{T} solenoid~~\cite{pcmag:yamamoto}
located at the \DIITBF and a large-area tracking coverage for e.g.\ calorimeter setups.
The former is suffering from the fact that the \EUDET-style telesopes have each sensor plane housed in a \SI{1.5}{\centi\metre} thick aluminium case with a dedicated cooling pipe and an auxiliary board.
While both the former and the later are limited by the small active area of the \EUDET-style telescopes.

The design, construction and installation of a large active area, compact telescope at the \DIITBF became one of the deliverables~\cite{lycoris-D15.2} of \AIDAII.
This telescope, named \LYCORIS\footnote{Large Area YX COverage Readout Integrated Strip Telescope}, was from the beginning based on six planes
of silicon-strip sensors oriented in a small-angle-stereo configuration, to meet the desired coverage of at least 10$\times$\SI{10}{\square\centi\metre}.

Within \AIDAII, there were strong activities on developing a successor of \EUDET-TLU called the \AIDA-TLU~\cite{Baesso_2019} and a successor of \EUDAQ called \EUDAQII~\cite{Liu_2019}.
\AIDA-TLU distributes a common clock to all DUTs and timestamps the triggers, thus it enables time-stamp based event reconstruction and removes the readout bottlenecks.
\EUDAQII has upgraded significantly from \EUDAQ with a lot extended protocols and synchronisation modes.
Both upgrades altogether enable new data taking modes beyond the classical ``trigger-busy'' approach of the \EUDET-TLU, where the system trigger rate has been limited by the device with the slowest readout, see details in~\cite{Baesso_2019}.
All \EUDET-style pixel telescopes have already upgraded to \AIDA-TLU with \EUDAQII or, are in the process of doing so.
Hence it was decided for \LYCORIS to use \EUDAQII and the \AIDA-TLU right from the beginning.

This paper is organized as follows: first the design requirements and an overview of the telescope design is given, followed
by a detailed description of the sensor plane, its components and the overall system (Section~\ref{sec:hardware}). Then
the description of the DAQ and the on-line software (Section~\ref{sec:software}) and the performance of the \LYCORIS telescope (Section~\ref{sec:performance}).
Finally, there is a description of how to integrate a DUT (Section~\ref{sec:dut}) followed by an overall
summary (Section~\ref{sec:conclusion}).

\section{The \LYCORIS Telescope System}\label{sec:hardware}
The design of \LYCORIS was driven by two different use-cases: providing a large-area coverage for test beam setups and
providing precision tracking inside the \PCMAG solenoid, where the \LYCORIS arms have to fit in the gap between the
DUT and the inner wall of the \PCMAG field cage. It was quickly concluded that the second use-case imposes much more stringent
constraints and hence the requirements were derived from this use-case.

\subsection{Overall Requirements}\label{sec:hardware:overallreq}

\begin{figure}[htbp]
\begin{center}
\includegraphics[width=0.49\textwidth]{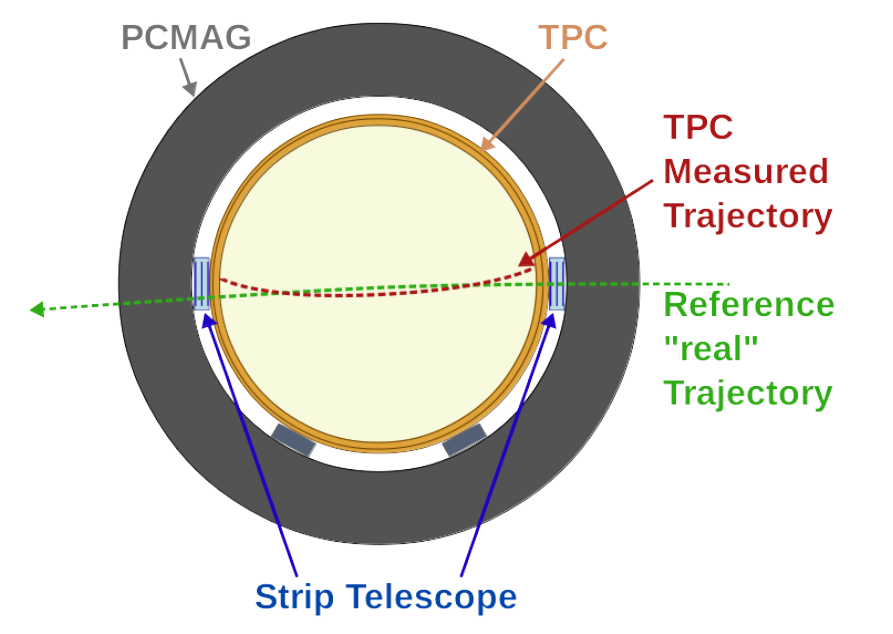}
\includegraphics[width=0.49\textwidth]{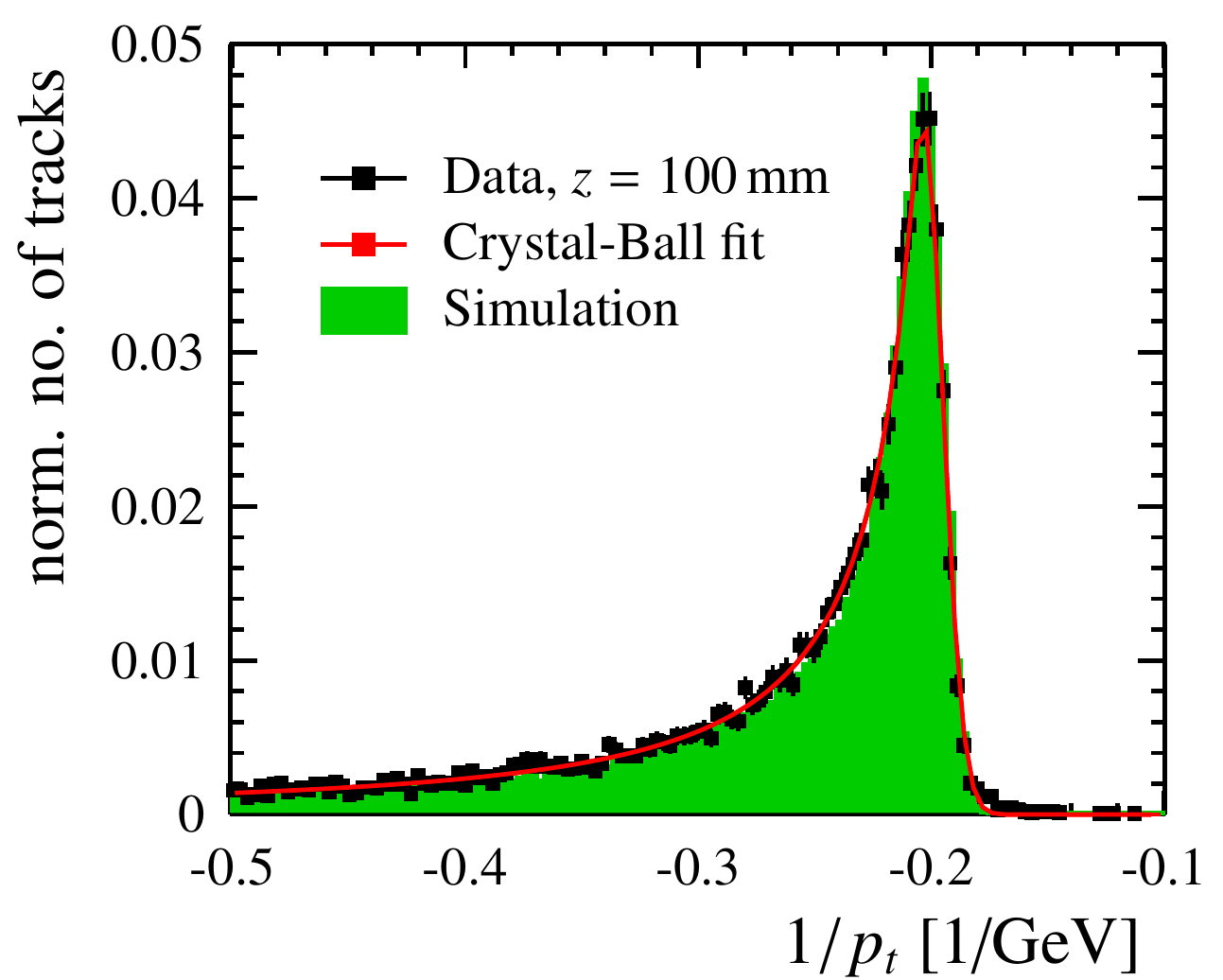}
\caption{\label{fig:hardware:pcmag} Picture of the \PCMAG (left)~\cite{Behnke:2020krd} and the measured momentum spread (right)~\cite{Mueller_2016}.}
\end{center}
\end{figure}

For tests in the \PCMAG solenoid, the DUT (e.g.\ a Time Projection Chamber (TPC)) will be installed inside the \PCMAG with a
small gap on either side, where the \LYCORIS telescope will be installed.
It will then be used to provide a precise measurement of the particle trajectory inside the magnet.
As a result of interactions of the electrons with the magnet wall ($20\%$ radiation length $X_0$), the particles have a large momentum spread when entering
the magnet volume (see Figure~\ref{fig:hardware:pcmag}) compared to the momentum spread of the electron test beam itself.
Effects such as poor momentum resolution due to potential electric field inhomogeneities within a
TPC volume could be studied and corrected with precision information from \LYCORIS.

\begin{figure}[htbp]
\begin{center}
\includegraphics[width=0.49\textwidth]{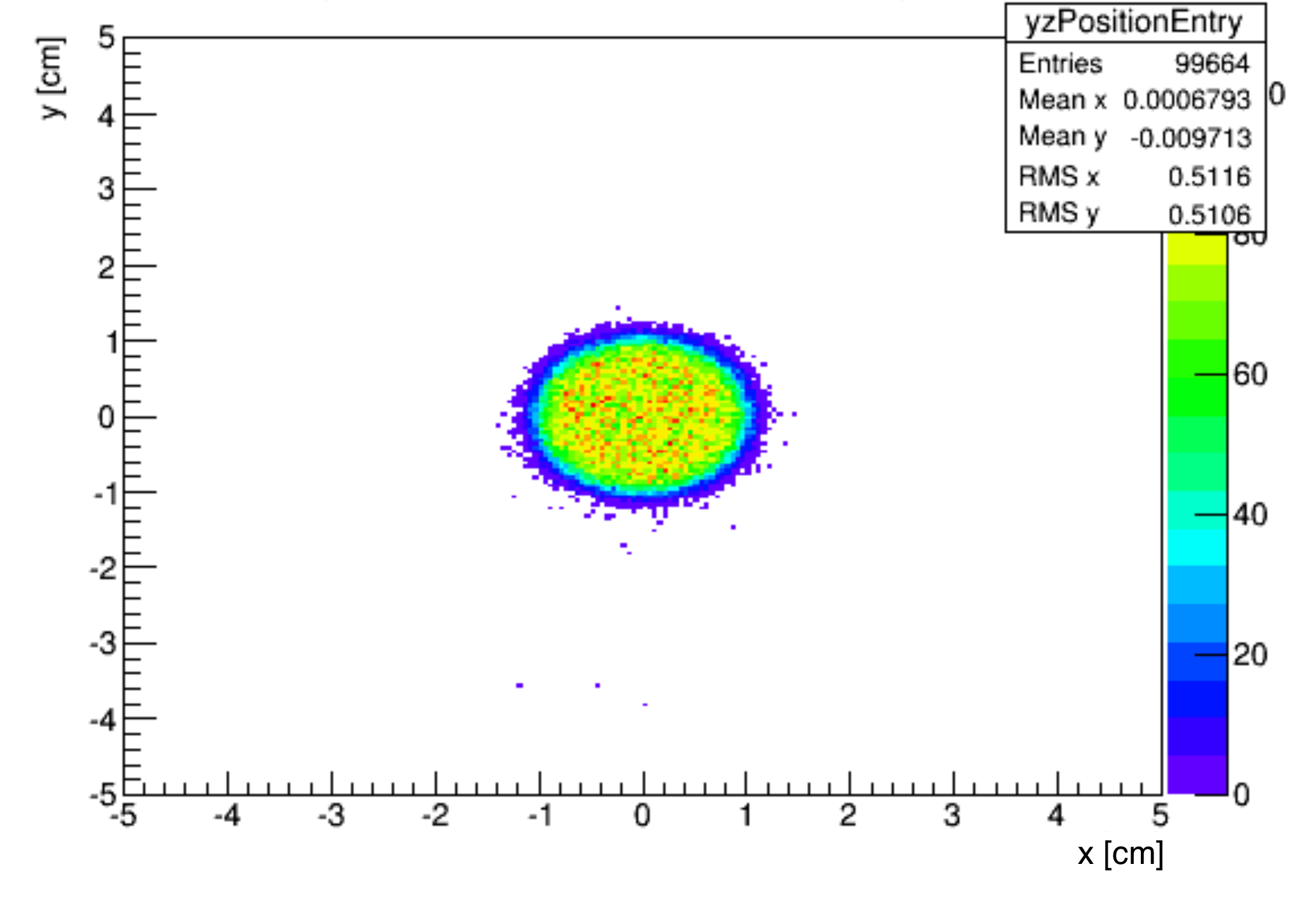}
\includegraphics[width=0.49\textwidth]{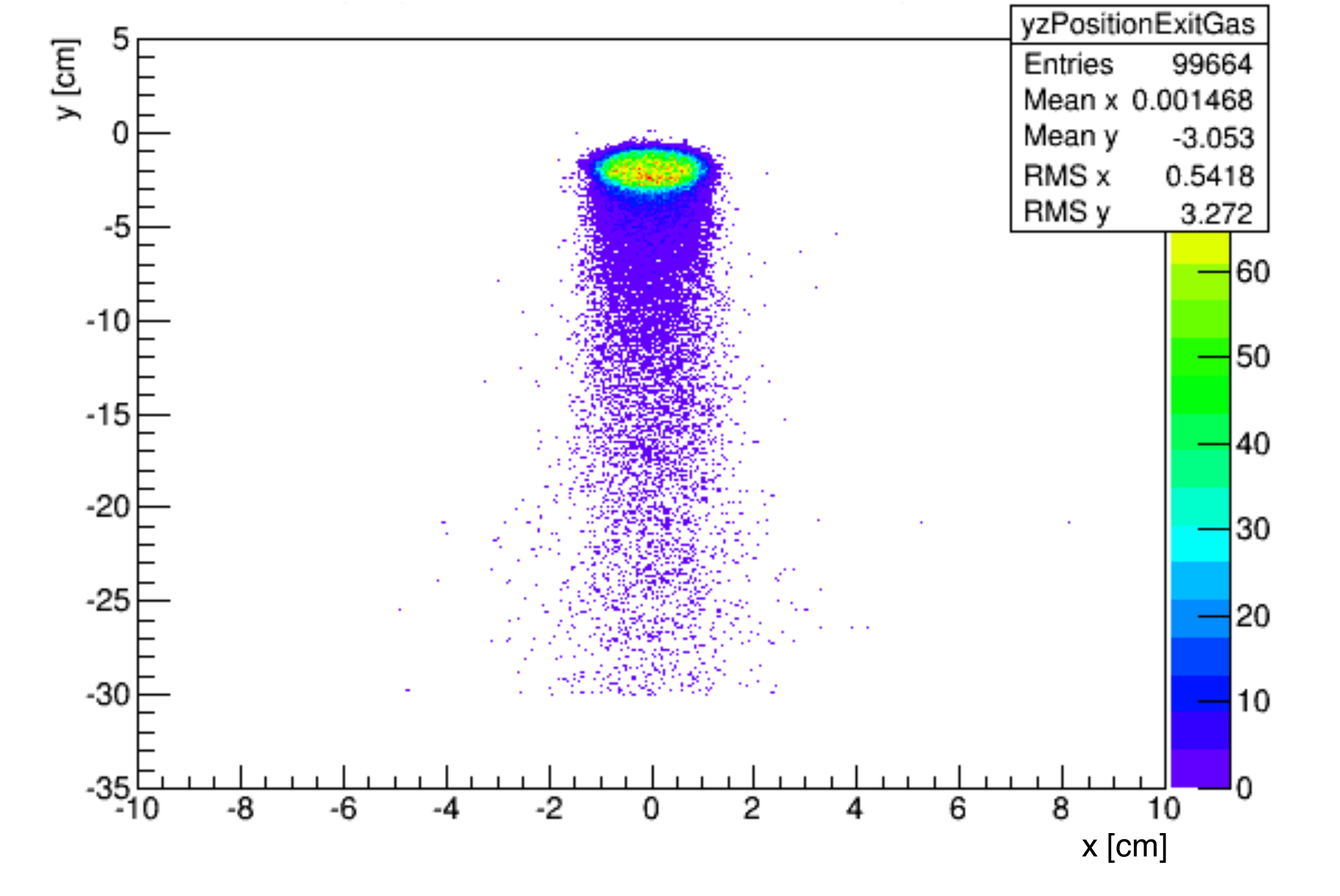}
\caption{\label{fig:hardware:G4simu} Simulated spread of \SI{6}{\GeV} beam at the silicon sensor plane before the TPC field cage (left) and at the last sensor plane (right).
The coordination origin of the simulation is set to be the center of the \PCMAG geometrical center.
The beam spot is centered at $z=$ \SI{2}{\metre} with radius of \SI{0.5}{\milli\metre} and a spread of \SI{2}{\milli\radian},
in order to well describe the beam status with a \SI{1}{\milli\metre} diameter collimator at \DIITBF.}
\end{center}
\end{figure}

Simulation studies have been performed to determine the general requirements on such a system.
A \GEANT~\cite{Allison:2006ve,Allison:2016lfl} simulation accurately describing the current infrastructure including the material
and dimensions of the \PCMAG and a TPC field cage (a detector taken as an example of one potential DUT) has been performed.
These studies have shown that a spatial resolution of better than \SI{10}{\micro\metre} is required in the bending plane of a homogeneous \SI{1}{\tesla} magnetic field,
in order to provide a comparable momentum measurement to a TPC with resolution of $\sim$\SI{5.0E-3}{\per\GeV}.
The overall active area was required to be at least 10$\times$\SI{10}{\centi\metre\squared} %
in order to cover 96\% of beam particles at \SI{6}{\GeV} according to the simulation results shown in Figure~\ref{fig:hardware:G4simu}.

Two technology candidates -- silicon-strip sensors and silicon-pixel sensors -- have been considered.
While the level of 3D information for a pixel system was superior, the available pixel sensors could not meet the area requirements at an affordable cost.
Thus silicon-strip sensors with a sufficiently small pitch were chosen for this project.
To provide the required 3D point resolution information, six silicon-strip sensors with a small-angle stereo configuration are used.
The limiting factor guiding these choices is the available gap, which is as little as \SI{3.5}{\centi\metre} and therefore provides a rather small lever arm.
From these studies, it was decided to build a telescope with six layers in two stations of three sensors each
and foresee configurations with 10$\times$\SI{10}{\centi\metre\squared} and 10$\times$\SI{20}{\centi\metre\squared} coverage,
requiring six or twelve sensors in total.
The key design requirements are summarized in Table~\ref{tab:requirements}.

\begin{table}[htbp]
\begin{center}
\begin{tabular}{l l r} \toprule
Resolution in bending plane                 & $\sigma_y$& $<$ \SI{10}{\micro\metre} \\
Resolution orthogonal to the bending plane  & $\sigma_x$&$<$ \SI{1}{\milli\metre} \\
Area coverage                               & $A_{xy}$  &   10$\times$  \SI{10}{\centi\metre\squared} \\
Thickness of single station                 & d         & $<$ \SI{3.5}{\centi\metre} \\ \bottomrule
\end{tabular}
\caption{\label{tab:requirements} The key design requirements for developing the \LYCORIS telescope.}
\end{center}
\end{table}

Moreover, this design also meets the need to operate the telescope outside of the \PCMAG, where several user groups
have indicated their interest to make use of the large area coverage provided by \LYCORIS, e.g. from the \textsc{CALICE} Collaboration.

\subsection{Overall Telescope Design}

The resolution requirements described above imply a strip pitch of about \SI{25}{\micro\metre} or smaller.
Most silicon-strip sensors used in particle physics or nuclear physics experiments do not provide
the required spatial resolution and/or the area coverage requirements for this telescope.
For example, silicon-strip sensors developed for the HL-LHC mainly use a pitch of $\approx$~\SI{75}{\micro\metre}~\cite{Collaboration:2017mtb,Collaboration:2272264}.
The silicon-strip sensor designed for the main tracker of the \SID detector~\cite{Aihara:2009ad,Behnke:2013lya} for the ILC~\cite{Behnke:2013xla}
was found to have the needed properties and was chosen for \LYCORIS.

\begin{figure}[htbp]
\begin{center}
\includegraphics[width=0.75\textwidth]{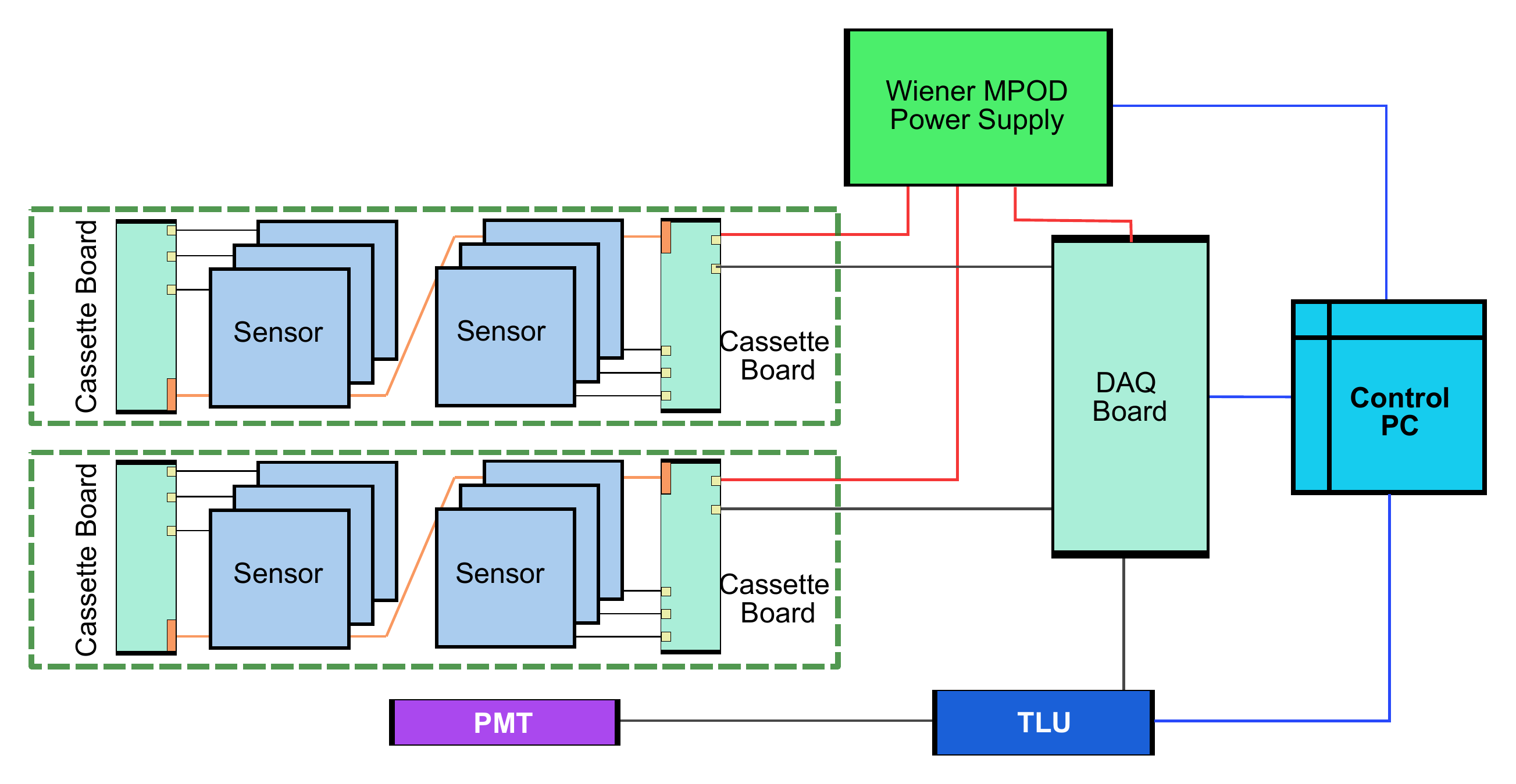}
\caption{\label{fig:hardware:lycorisoverall} The overall design of the \LYCORIS telescope.
The cassettes are indicated by the green dashes and host up to six sensor modules and two cassette boards.
The High Voltage(HV)/Low Voltage(LV) connections are indicated in red, while the connections for the DAQ are indicated in dark grey,
the Ethernet connections using TCP/IP in blue and the connections using the Kapton-flex in orange.}
\end{center}
\end{figure}

The overall architecture of \LYCORIS is shown in Figure~\ref{fig:hardware:lycorisoverall}. The two separate stations
are called ``cassettes'' and provide the mechanical support for up to six sensor planes in total. Three sensor units -- the modules -- are
controlled by one cassette board, which acts as the interface to the power supply crate (Wiener MPOD)
and the DAQ using the SLAC \KPIX DAQ board. The cassette board can be daisy-chained to a secondary cassette board
to control six sensor planes. The DAQ board supports up to four cassette boards
and interfaces with a control PC via Ethernet and with the \AIDA-TLU~\cite{Baesso_2019} via an HDMI connector.
Triggering is provided by a coincidence from a set of scintillators~\footnote{Usually scintillators are placed two before and two after the telescope
in the test beam, however, when telescope is located inside a magnet, scintillators are placed outside the magnet right after the beam.} read out by photo-multiplier tubes (PMT).
The individual telescope components are described in detail in Section~\ref{sec:hardware:module}.

\begin{figure}[htbp]
\begin{center}
\includegraphics[height=4.30cm]{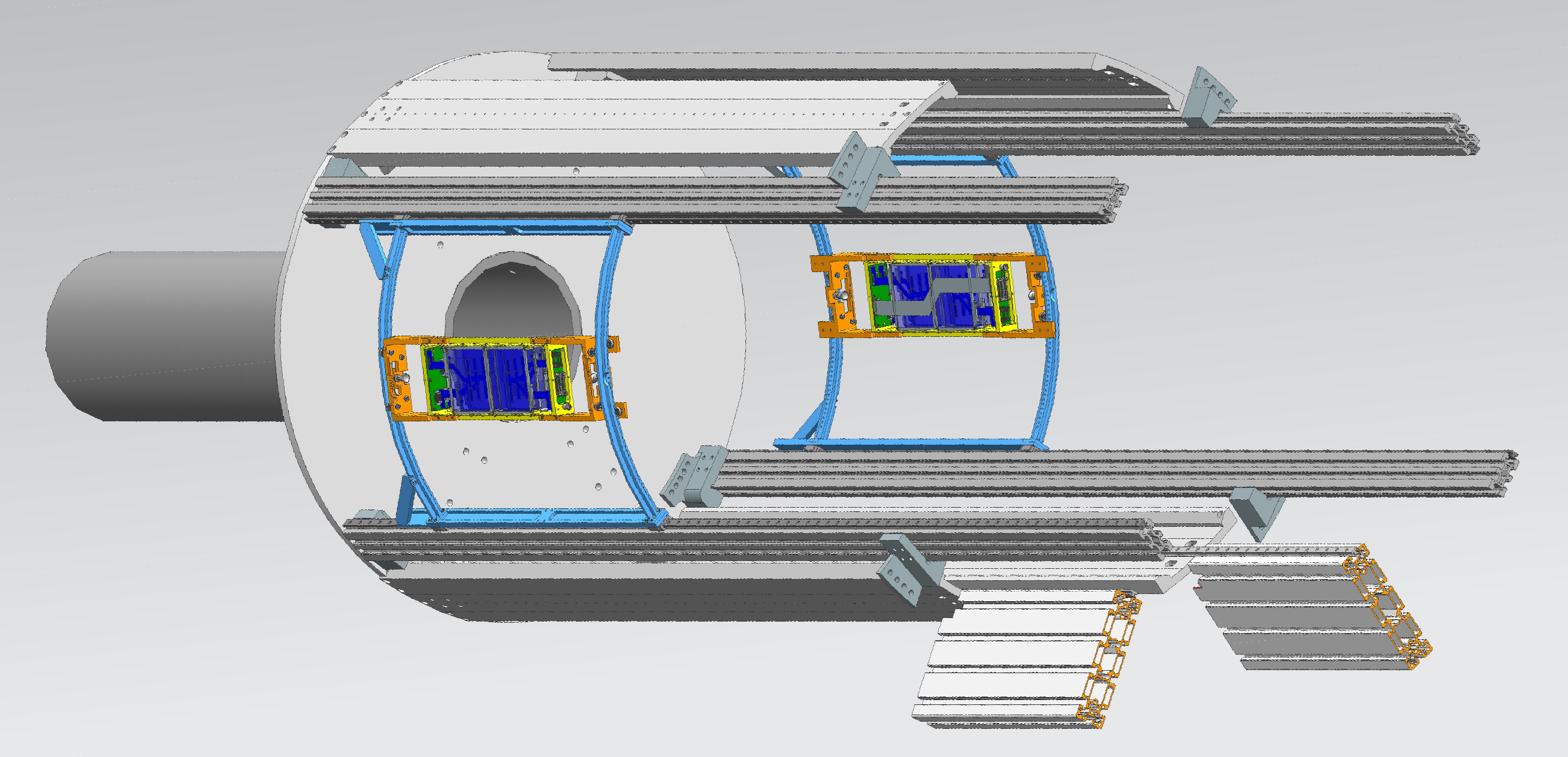}
\caption{\label{fig:installation:pcmag}CAD drawing of the mechanical support to integrate \LYCORIS telescope inside the \PCMAG, where the rails marked in light-blue are used by the two Lycoris cassettes.
}
\end{center}
\end{figure}

The most stringent mechanical requirements are due to operating \LYCORIS inside the \PCMAG solenoid.
The available gap between the current TPC and inner wall of the \PCMAG allows merely a cassette thickness of \SI{3.5}{\centi\metre}. The \PCMAG provides a rail
system to install a DUT, so it can be moved in and out with high accuracy.
The cassettes are mounted on a separate rail system, so they can slide in and out independently of the DUT itself. This simplifies both
installation and operation. A CAD drawing of the actual setup is shown in Figure~\ref{fig:installation:pcmag}.

\subsection{The \LYCORIS Silicon Sensor Module\label{sec:hardware:module}}
\begin{figure}[htbp]
\begin{center}
\includegraphics[width=0.5\textwidth]{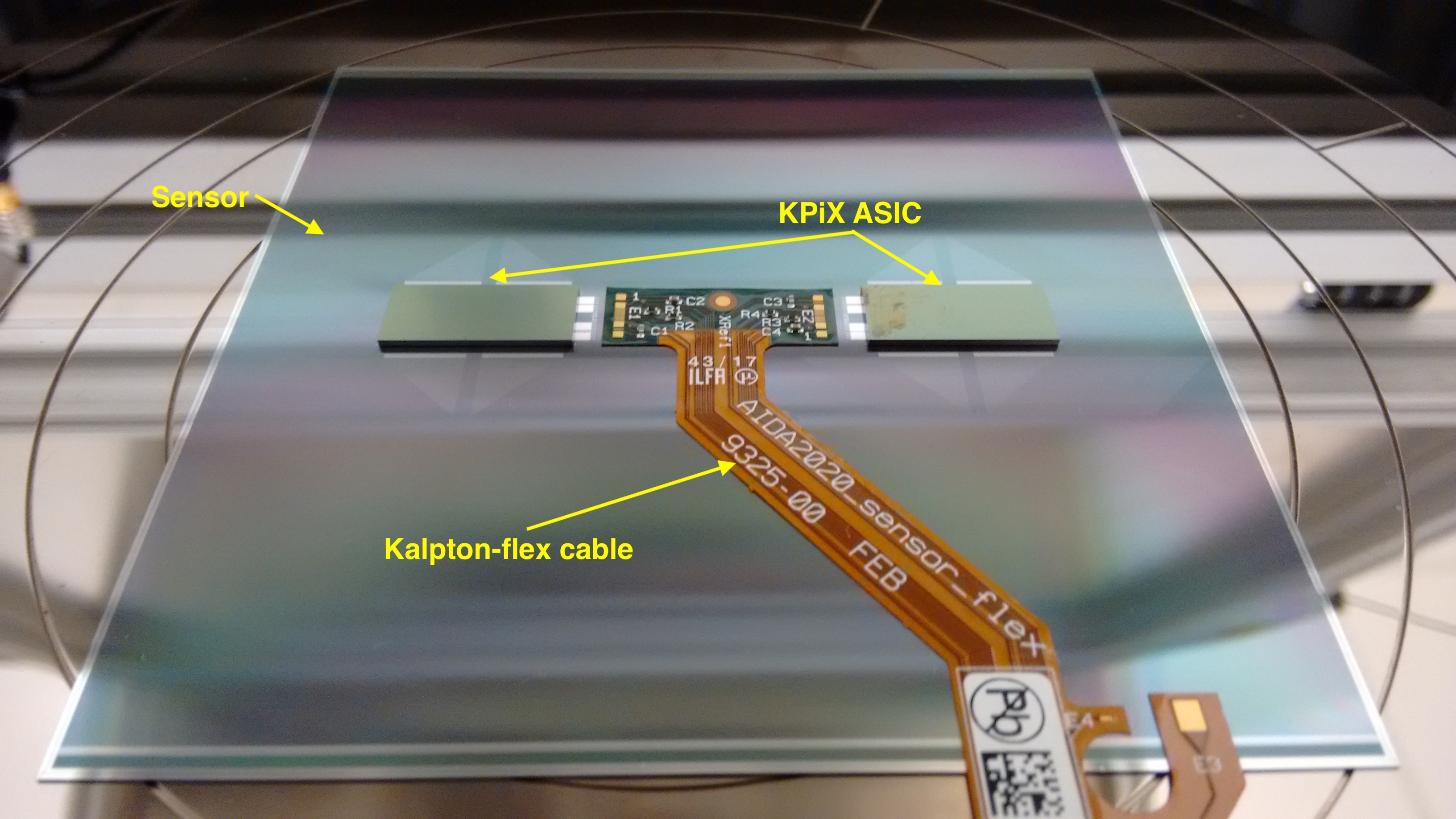}
\caption{\label{fig:hardware:sidsensor} The \LYCORIS module composed of one \SID tracker sensor, two \KPIX ASICs and a Kapton-flex readout cable.}
\end{center}
\end{figure}

The central building block of \LYCORIS is the silicon sensor module shown in Figure~\ref{fig:hardware:sidsensor}
which is composed of one \SID tracker sensor, two \KPIX ASICs~\cite{kpix} and a Kapton-flex cable.
As the \KPIX ASICs are bump-bonded directly on top of the silicon-sensor, there is no need for a hybrid PCB.

\subsubsection{The \SID Tracker Sensor}\label{sec:hardware:sensor}

\begin{figure}[htbp]
\begin{center}
\includegraphics[width=0.5\textwidth]{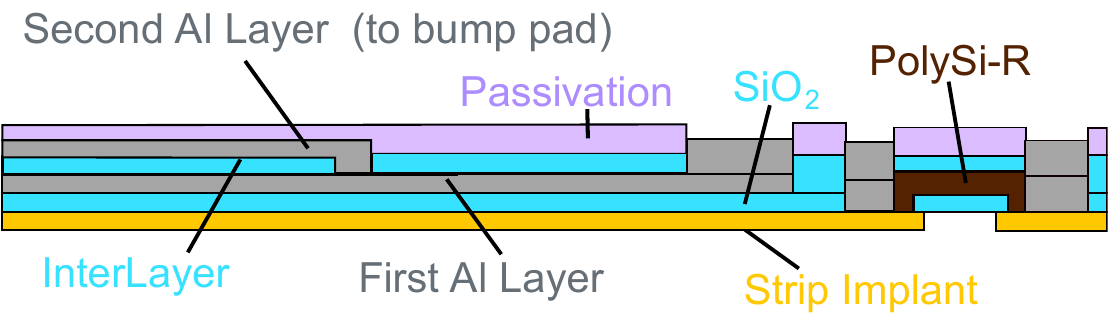}
\caption{\label{fig:hardware:sensor} Schematic cross-section of the \SID tracker sensor.}
\end{center}
\end{figure}

The sensor designed for the \SID tracker, called SiD-OUTER-SSSD 6972, is an n-type silicon micro-strip sensor
with \SI{25}{\micro\metre} pitch (read out on every other strip) manufactured by Hamamatsu~\cite{hamamatsu-www}. It provides a spatial resolution of
$\sim$\SI{7.2}{\micro\metre}~\footnote{This theoretical number is calculated using the simple approach of single strip signals with binary readout}.
The sensor is designed with two aluminum metallization layers, as shown in the schematic cross-section in Figure~\ref{fig:hardware:sensor},
which enables the hybrid-less design.
Table~\ref{tab:hardware:sensor} gives the detailed properties of this sensor.

The second metallization layer routes signals of each strip to a bump pad that is bonded directly to the \KPIX readout ASIC.
This layer is also used for powering the \KPIX ASIC and transmitting other digital signals such as Command, Clock and Trigger.
This approach to a hybrid-less strip sensor design has been tested for the first-time in this project.

\begin{table}[htbp]

\centering
\begin{tabular}{l l}
   & SiD-OUTER-SSSD 6972\\
  \toprule
  Sensor size & $93.53\times$\SI{93.53}{\square\milli\metre}\\
  Thickness & 320$\pm$\SI{15}{\micro\metre} \\
  Crystal Orientation & <100> \\
  Wafer type & n-type\\
  Strip implant & p+ type\\
  Strip p+ width   & \SI{8}{\micro\metre}\\
  Strip Al width   & \SI{9}{\micro\metre}\\
  Readout Al width & \SI{4}{\micro\metre}\\
  Strip bias resistance & 10-\SI{50}{\mega\ohm}/strip \\
  Strip readout coupling & AC \\
  Strip Pitch (readout)  & \SI{25}{\micro\metre} (\SI{50}{\micro\metre})\\
  Strip Multiplicity (readout) & 3679 (1840)\\
  \bottomrule
\end{tabular}
\caption{The \SID tracker sensor (SiD-OUTER-SSSD 6972) properties.}
\label{tab:hardware:sensor}
\end{table}

\subsubsection{The \KPIX ASIC}\label{sec:hardware:kpix}
The \KPIX readout chip was designed by SLAC as a multi-purpose read-out ASIC for the \SID detector concept for the ILC. It was manufactured
using a \SI{250}{\nano\metre} mixed-mode CMOS process from TSMC~\cite{tsmc-www}.
The chip consists of a 1024 channel fully digital readout with a 13-bit ADC and a timing generator controlling the data acquisition.
Each channel consists of a charge amplifier, a shaper and a discriminator. The charge amplifier can be operated either in a high-gain
or low-gain mode and additionally offers a dynamically switchable gain range.
If the default range of \SI{400}{\fC} has been exceeded, which is detected in the range control circuitry, a \SI{10}{\pico\farad} capacitor is automatically added to
extend the range to \SI{10}{\pico\coulomb}.
For a bare \KPIX, the noise floor is about 1000 electrons in the low-gain mode and 700 electrons for the high-gain mode
For a full assembled sensor,  a noise floor of 5\% of a Minimum Ionizing Particle (MIP) in the \SI{320}{\mu\metre} thick silicon tracker sensor was achieved.
The dynamic range switching then extends the available range to 2500 MIPs.

\begin{figure}[htbp]
\begin{center}
\includegraphics[width=\textwidth]{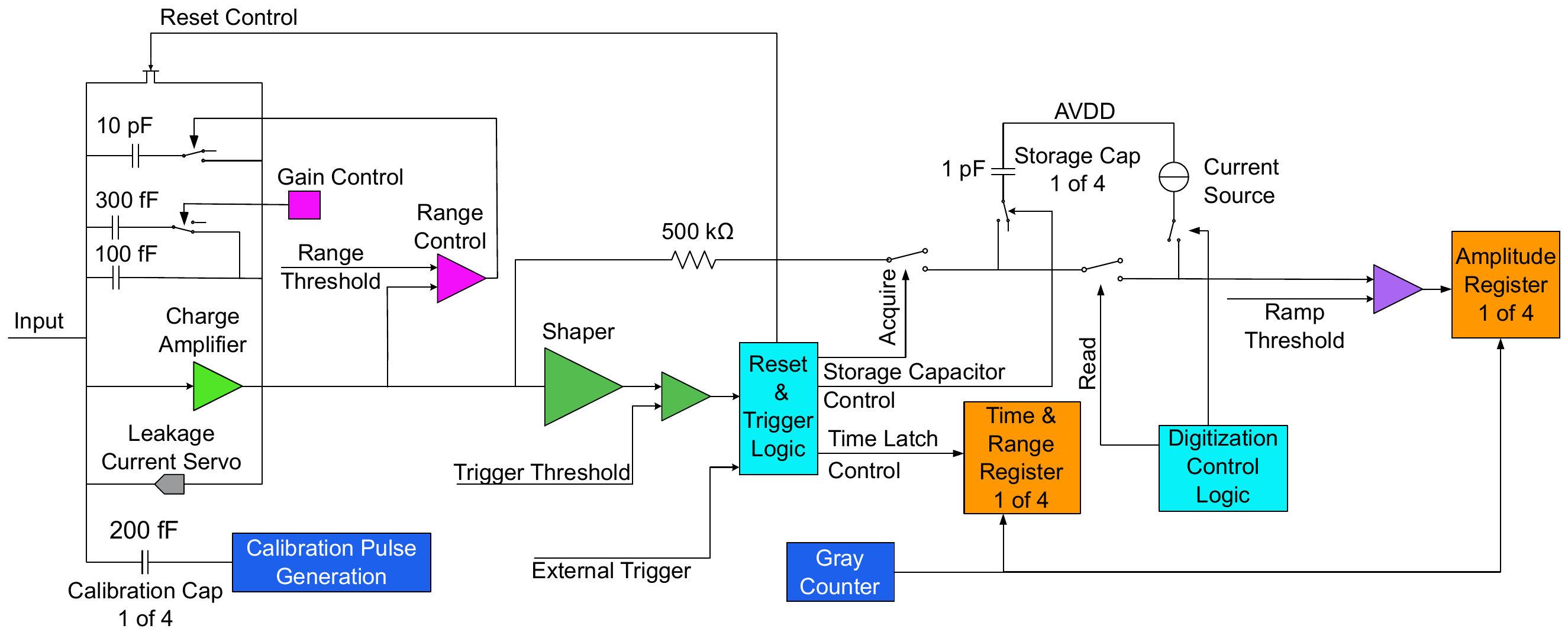}
\caption{\label{fig:hardware:kpixchannel} Simplified block diagram of a single \KPIX readout channel (out of 1024 channels in total).
Shown are the charge-amplifier in light green with its gain control and dynamic range-control circuitry (shown in pink) and the shaper (green).
The main digital blocks are indicated in light and dark blue with the memory buckets shown in orange.}
\end{center}
\end{figure}

Figure~\ref{fig:hardware:kpixchannel} shows the block diagram of a single \KPIX readout channel.
Each channel has four sample and hold capacitors and four timing registers to permit four separate measurements
of the signal amplitude and the time of threshold crossing. This results in four memory buckets for each channel.
In normal operation, the charge amplifier is synchronously reset after each bunch-crossing.

\begin{figure}[htbp]
\begin{center}
\includegraphics[width=0.75\textwidth]{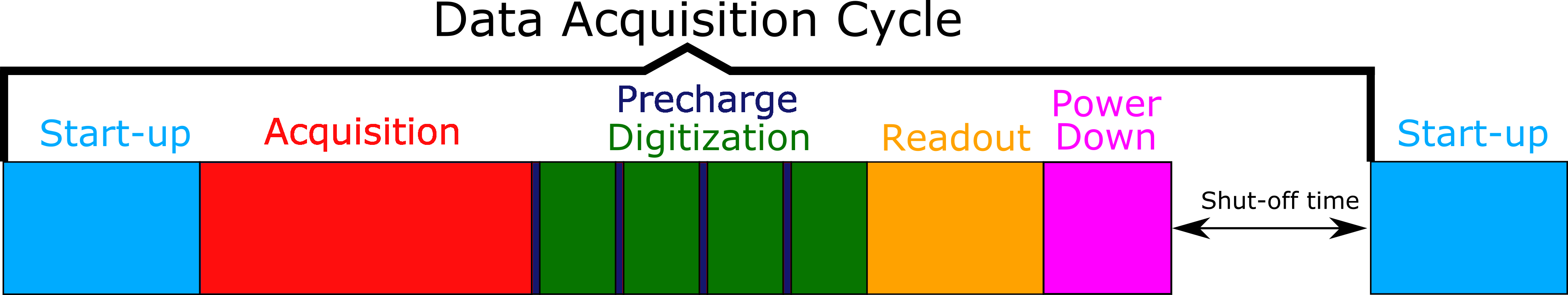}
\caption{\label{fig:hardware:kpixtiming}The five operational states of \KPIX data acquisition cycle followed by the between beam bunches shut-off idle phase (relative time periods for states not to scale).}
\end{center}
\end{figure}

Each channel has a calibration system, leakage compensation, ``DC'' reset and polarity inversion.
\KPIX supports both AC and DC-coupled signals and in DC mode the leakage is compensated by a servo circuit.
The total amount of leakage is determined without any signal being present during the data taking period.
The ``DC'' reset works for asynchronous operation for cosmic-ray runs.
\KPIX can switch the input polarity for use with e.g.\ GEM detectors~\cite{White:2010zza,Yu:2011ij}.
The noise floor is about 1000 electrons in low-gain mode (\SI{0.15}{\femto\coulomb}) and the maximum signal charge is \SI{10}{\pico\coulomb} with the full
dynamic range corresponding to 17 bits.
\KPIX uses two separate power supply lines for its analog (AVDD) and digital (DVDD) circuits,
requiring nominally \SI{2.5}{\volt} and \SI{2.0}{\volt} respectively. \KPIX has been designed to achieve an average power
consumption lower than \SI{20}{\mu\watt\per channel} for applications at the ILC realized by power-pulsing.
This is realized by powering the analog front end only when beam is present, i.e.\ during the signal acquisition period.
This power-pulsing cycle is called the \KPIX data acquisition cycle.
It starts with an external \textit{Acquisition Start Signal} followed by five different operational states and a shut-off idle phase.
The sequence of the five states and the idle phase is shown in Figure~\ref{fig:hardware:kpixtiming}.

An externally provided DAQ board is able to adjust the period of the \KPIX clock separately for each state, in \SI{5}{\nano\second} increments with a minimum period of
\SI{10}{\nano\second}. This allows a wider acquisition and pre-charge window to be achieved, while at the same time reducing the time spent on digitization and readout.
Table~\ref{tab:hardware:kpix-timing} shows the typical clock periods to operate \KPIX at each state, including the five data acquisition operational states and the shut-off idle phase.
These unique properties of the \KPIX ASIC make it ideal for the ILC environment. In addition to its use in the \SID
tracking system, as discussed in Section~\ref{sec:hardware:sensor}, \KPIX is also the proposed readout strategy for the \SID Silicon-tungsten sampling electromagnetic
calorimeter~\cite{Steinhebel:2017qze,Barkeloo_2019}, as the many readout channels read out between beam bunch crossings through
a power-pulsing cycle allow for a highly granular and compact calorimeter design with a minimal cooling system.

\begin{table}[htbp]
\centering
\begin{tabular}{lr}
Clock           & Typical clock period (ns)  \\ \toprule
\KPIXTAcq       &  320\\
\KPIXTDigi      &  50\\
\KPIXTPreC      &  6000\\
\KPIXTReadout   &  200\\
\KPIXTIdle      &  200\\
\bottomrule
\end{tabular}
\caption{The \KPIX clock domains and the typical clock periods to operate \KPIX at each state.}
\label{tab:hardware:kpix-timing}
\end{table}

The five operational states of the \KPIX data acquisition cycle (shown in Figure~\ref{fig:hardware:kpixtiming}) are described in the following.

\begin{description}
\item[Start-up:] \KPIX powers up after receiving the \textit{Acquisition Start Signal}.
The start up phase duration is configurable via the TimeBunchClkDelay setting and during this phase the \KPIX uses the acquisition clock \KPIXTAcq. \KPIX operation during this
phase includes the ramp up of analog and digital current to operation values and the resetting of internal \KPIX registers and the Gray counter.
\begin{equation*}
	t_{\mathrm{Start-up}} = T_{\mathrm{Acquisition}} \cdot TimeBunchClkDelay
\end{equation*}

\item[Acquisition:]
The acquisition phase is the period when the chip records data and the clock used is the acquisition clock \KPIXTAcq.
During this phase, the chip is able to store per channel up to four events in the analog buffers and timestamp each event using a 13-bit
counter called BunchClockCount (BCC) which increments by every eight clock pulses and is cleared every cycle.
The opening time for this phase $t_{\mathrm{acquisition}}$ is configurable by setting the maximum BCC ($n_{BCC,max.}$) thus calculated as
\begin{equation*}
t_{\mathrm{Acquisition}} = n_{BCC,max.} \cdot 1\;BCC
         = n_{BCC,max.} \cdot  8\; T_{\mathrm{Acquisition}} \,\,.
\end{equation*}
There is a small configurable pause between readout and digitization which is used to prepare the analog buffers for the upcoming digitization.

\item[Digitization:]
The analog information stored in four storage capacitors for each of the 1024 channels is now digitized in
four cycles, doing 1024 analog buckets in parallel. The digitization is handled by a Wilkinson ADC with a 13-bit resolution for each channel.
A current mirrored into each of the 1024 channels runs down the charge in the storage capacitor for each cycle.
The duration of the so-called pre-charge period is $2\cdot$ \KPIXTPreC per individual storage capacitor or $8\cdot$\KPIXTPreC in total. During the pre-charge
the analog bus connecting the storage capacitors to the Wilkinson converters is charged to a known level as it can contain charge from a previous event
as such it has to be done before connecting to the next storage capacitor. In addition, the read-out bus is pre-charged to high as the read-out can only pull the bus low.
A ramp-threshold discriminator then detects the transition through zero, and the content of a common Gray counter is then stored
in the corresponding memory bucket together with the corresponding timestamp determined by BCC and the range identifier bits.
The total time required for the digitization of a single bucket, excluding the time for the pre-charge is
\begin{equation*}
t_{\mathrm{Single-Digitization}}= (8192+18)\cdot T_{\mathrm{Digitization}}
\end{equation*}
with the total digitization time being given by
\begin{equation*}
t_{\mathrm{Digitization}}= 4 \cdot (8192+18)\cdot T_{\mathrm{Digitization}} \,\,.
\end{equation*}
\item[Readout:]
After digitization is complete, all data is sequentially read out using the \KPIXTReadout clock.
A \KPIX~{\textit{word}} corresponds to 13~bits, corresponding to the ADC resolution and the BCC.
The readout is organized in rows of 32 channels, which corresponds to 416 bits equivalent to $416\cdot$\KPIXTReadout.
Reading out each shift register is preceded by the parallel loading
of all the shift registers which takes $300\cdot$\KPIXTReadout.
In total the readout of one row takes $716 \cdot$ \KPIXTReadout.
For each channel there are nine words in total and \KPIX has 32 rows of 32 channels in total.

Overall the entire readout of \KPIX takes $32\cdot9\cdot716\cdot$\KPIXTReadout$= 206208 \cdot$\KPIXTReadout.

\item[Power-down:]
After the readout has completed, the current is pulsed down while maintaining the supply voltages.
\end{description}

Additional details on the various \KPIX trigger modes, the calibration circuitry and the interconnects are given below.

\paragraph{\KPIX Trigger Modes}\label{sec:hardware:kpix:trigger}
\KPIX can take data in two different operation modes, corresponding to two different triggering schemes:
\begin{description}
\item[Self triggering:] If the charge generated within a channel rises above a user
defined threshold, it will be recorded by \KPIX.
\item[External triggering:] When an external trigger arrives, \KPIX records the generated charge
from all channels simultaneously.
\end{description}
In both modes a maximum of four triggers per channel can be stored by \KPIX.
Given the preferred beam telescope operation and the usual need of an external trigger by DUT setups,
the \KPIX external trigger mode is the preferred running mode for the \LYCORIS telescope.

\paragraph{Calibration}\label{sec:hardware:kpix:calib}
In order to calibrate the ADC response, each of the 1024 readout channels is equipped with a calibration module which injects
a series of user defined calibration amplitudes through an 8-bit DAC for verifying the ADC response.
The ADC has a linear response to the different calibration values and its slope is consequently used in the following studies to convert the raw ADC value to
a charge in fC. The calibration data-taking is operated using the external triggering mode with a software-controlled start signal.
All the four memory buckets for every channel can be calibrated individually to correct for potential pedestal and offset differences.
The first calibration pulse can be selected for a high calibration range, so that the system can be tested up to the full-scale range of \SI{10}{\pico\coulomb}.
For negative-polarity signals, an inverter is inserted after the charge amplifier so the polarity of the calibration signal is reversed as well.

\begin{figure}[htbp]
\begin{center}
\includegraphics[height=4cm]{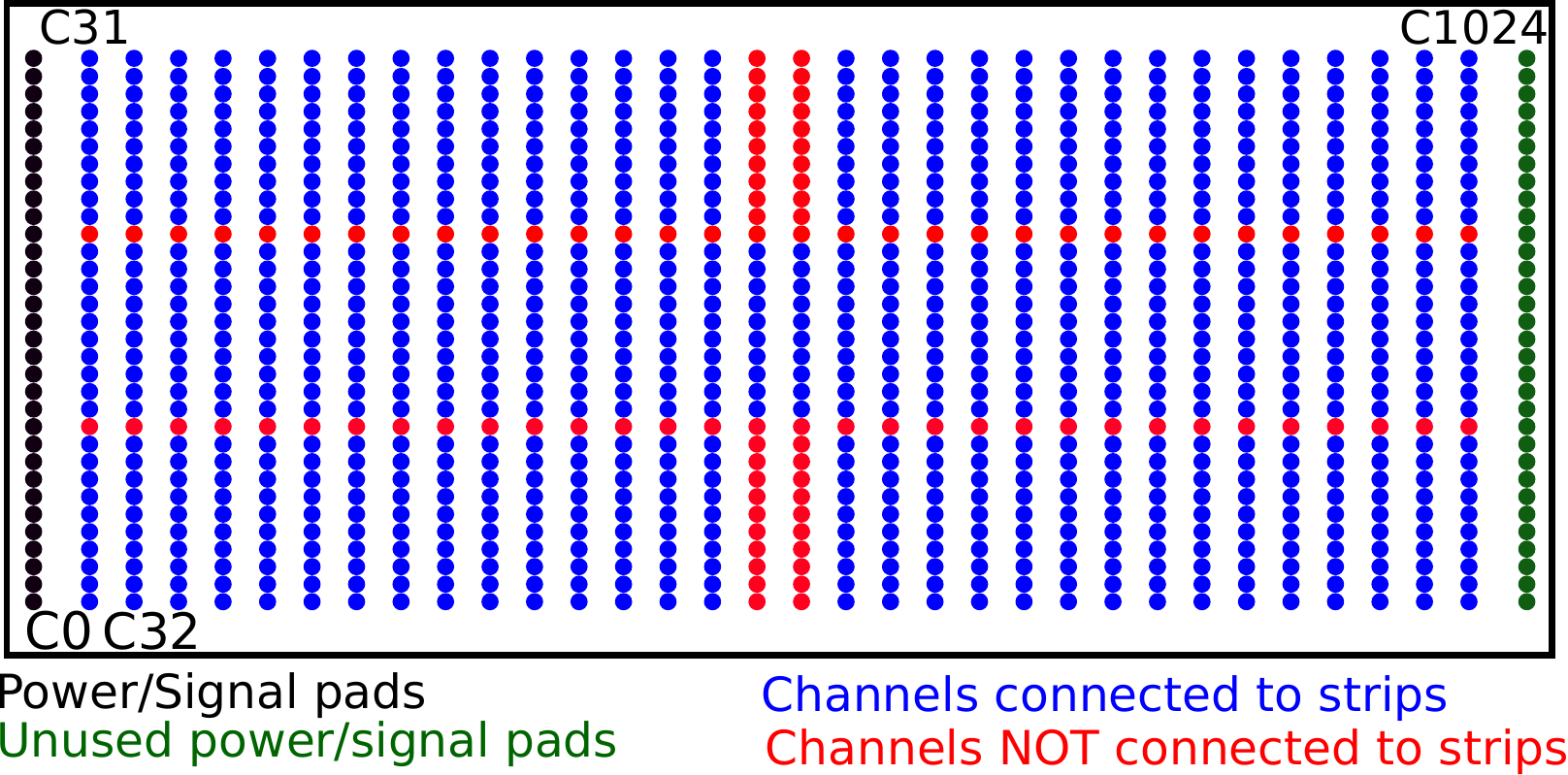}
\includegraphics[height=4cm]{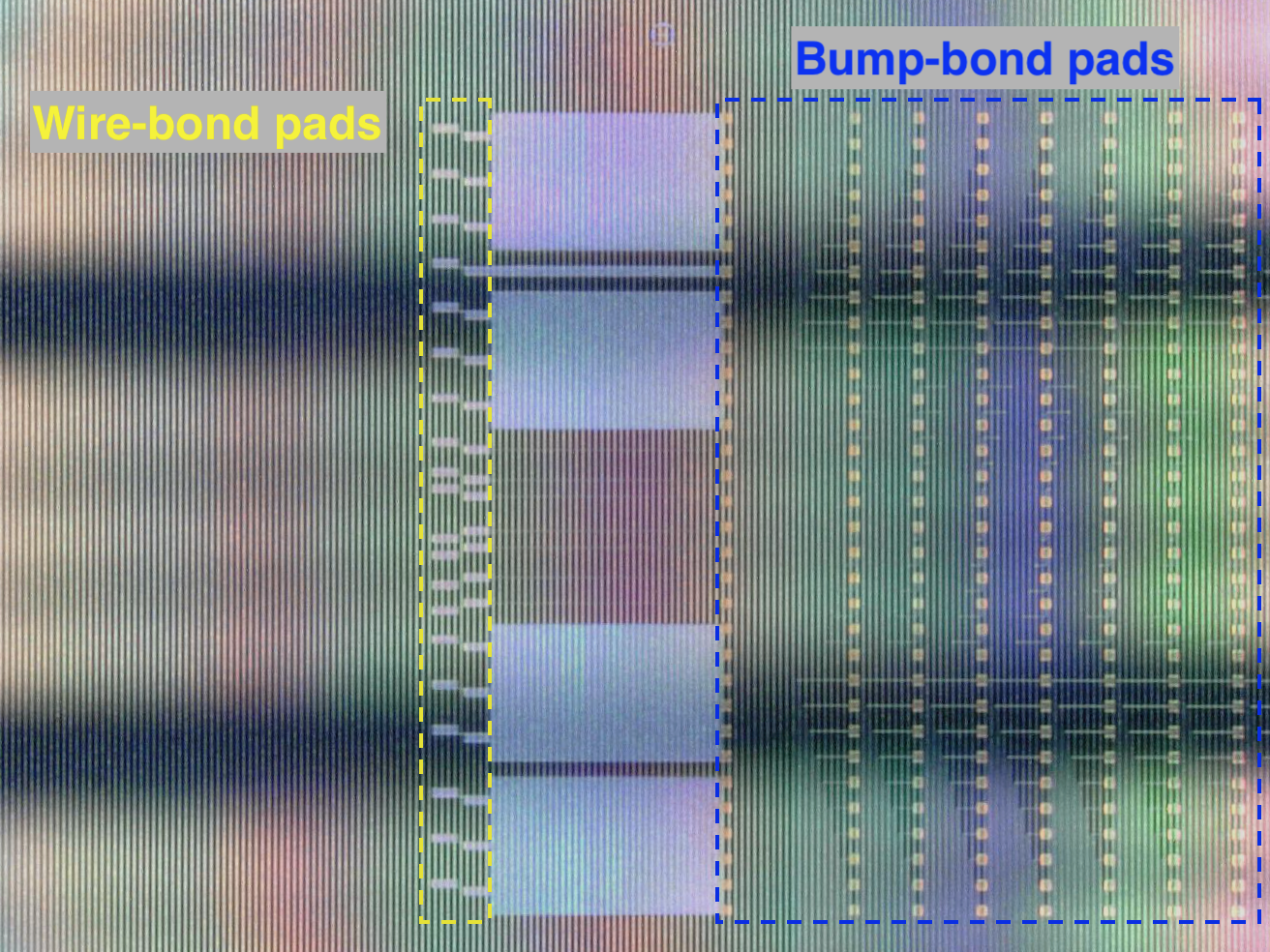}
\caption{\label{fig:hardware:kpix:bumpbond-map} A schematic diagram of \KPIX bump-bonding array (left) with the 920 connected channels in (blue).
The two outermost columns (black/green) on each side route the AVDD/DVDD power to \KPIX, as well as exchange digital signals.
A microscope photo (right) of a section of the pad grid on the sensor surface.
The bump-bonding pads (blue) connect to the used Power/Signal column (black on left sketch) of one \KPIX and some of its neighboring pads.
The wire-bonding pads (yellow) connect the Power/Signal column through the sensor's second metallization layer to outside (via Kapton-Flex).
}
\end{center}
\end{figure}

\paragraph{Interconnects}
 \KPIX has been designed to be bump-bonded to various Silicon sensors. The eutectic Sn-Pb bumps on a $200 \times $\SI{500}{\micro\metre} lattice are placed in wells on each \KPIX pad.
Figure~\ref{fig:hardware:kpix:bumpbond-map} (left) shows a schematic diagram of the \KPIX bump bonding array,
where the center bump matrix (blue/red) provides connections to the \KPIX readout channels while the two outermost columns (black/green)
route the AVDD/DVDD power, the digital input signals including the clock, control, external trigger and the outgoing digital data stream.
Two \KPIX chips are necessary to read out the 1840 strips of one \SID tracker sensor, but only 920 (blue) out of the 1024 readout channels of a single \KPIX are connected
to a strip while the others (red) are bonded to the sensor bump pads and are left floating.
The bump pads on each side of \KPIX are designed with the same functionality.
The \SID tracker sensor is designed to connect only one side (green) of the \KPIX power/signal bump-pads, thereby the other side (black) is left open.
Figure~\ref{fig:hardware:kpix:bumpbond-map} (right) shows a microscope photo of the pads on the sensor surface for bump-bonding and wire-bonding,
where the marginal bump-bonding pads to provide \KPIX power and signals are routed to the wire-bonding pads via the second metallization layer.

\subsubsection{The Kapton-Flex Cable}

\begin{figure}[htbp]
\begin{center}
\includegraphics[width=0.6\textwidth]{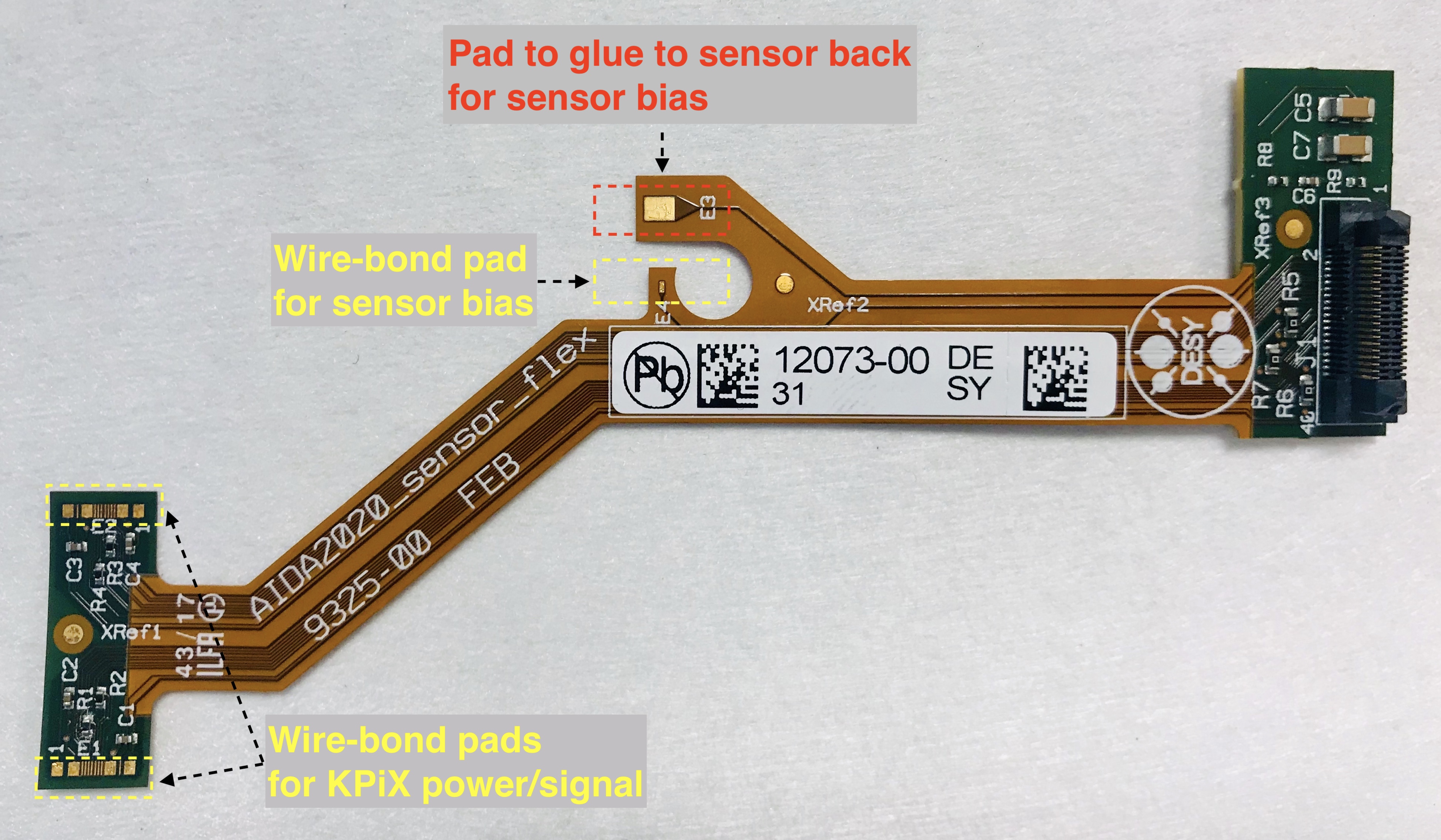}
\caption{\label{fig:hardware:flex-cable}The Kapton-Flex cable used on the \LYCORIS module.}
\end{center}
\end{figure}

The Kapton-Flex cable shown in Figure~\ref{fig:hardware:flex-cable} is designed to route both HV and LV power
and digital traces between the sensor module and the DAQ system.
Both HV and LV lines are further filtered on the Kapton-Flex to further reduce the noise due to power supplies.
It feeds the sensor bias voltage through wire-bonding from the HV bonding pad to the sensor bias ring,
and the bias return is made by connecting the HV return to the sensor back plane using conductive silver epoxy paste.
For the two \KPIX chips on the same sensor, their low voltage traces (AVDD and DVDD) as well as their grounds, are tied together.
For the same \KPIX, the AGND and DGND are tied together through a \SI{0}{\ohm} resistor.
The bias voltage is AC coupled to the \KPIX AVDD through a bypass capacitor; \KPIX is measuring
through its analogue voltage instead of its analogue ground to avoid any mismeasurement from grounding differences.
The clock and the trigger signals are Low Voltage Differential Signals (LVDS), of which the termination resistors are placed on this Kapton-Flex cable.

\subsubsection{Module Assembly}\label{sec:hardware:moduleassembly}

Assembling a \LYCORIS module is a four-step process:
\begin{enumerate}
  \item The bump-bonding between the two flipped \KPIX chips and the silicon sensor;
  \item Gluing the Kapton-Flex cable onto the silicon sensor;
  \item Connecting the pads on Kapton-Flex cable with the corresponding pads on the silicon sensor using wire-bonding;
  \item Silver epoxy glue the HV return pad on the Kapton-Flex cable to the sensor back plane.
\end{enumerate}

\begin{figure}[htbp]
  \begin{center}
  \includegraphics[width=0.8\textwidth]{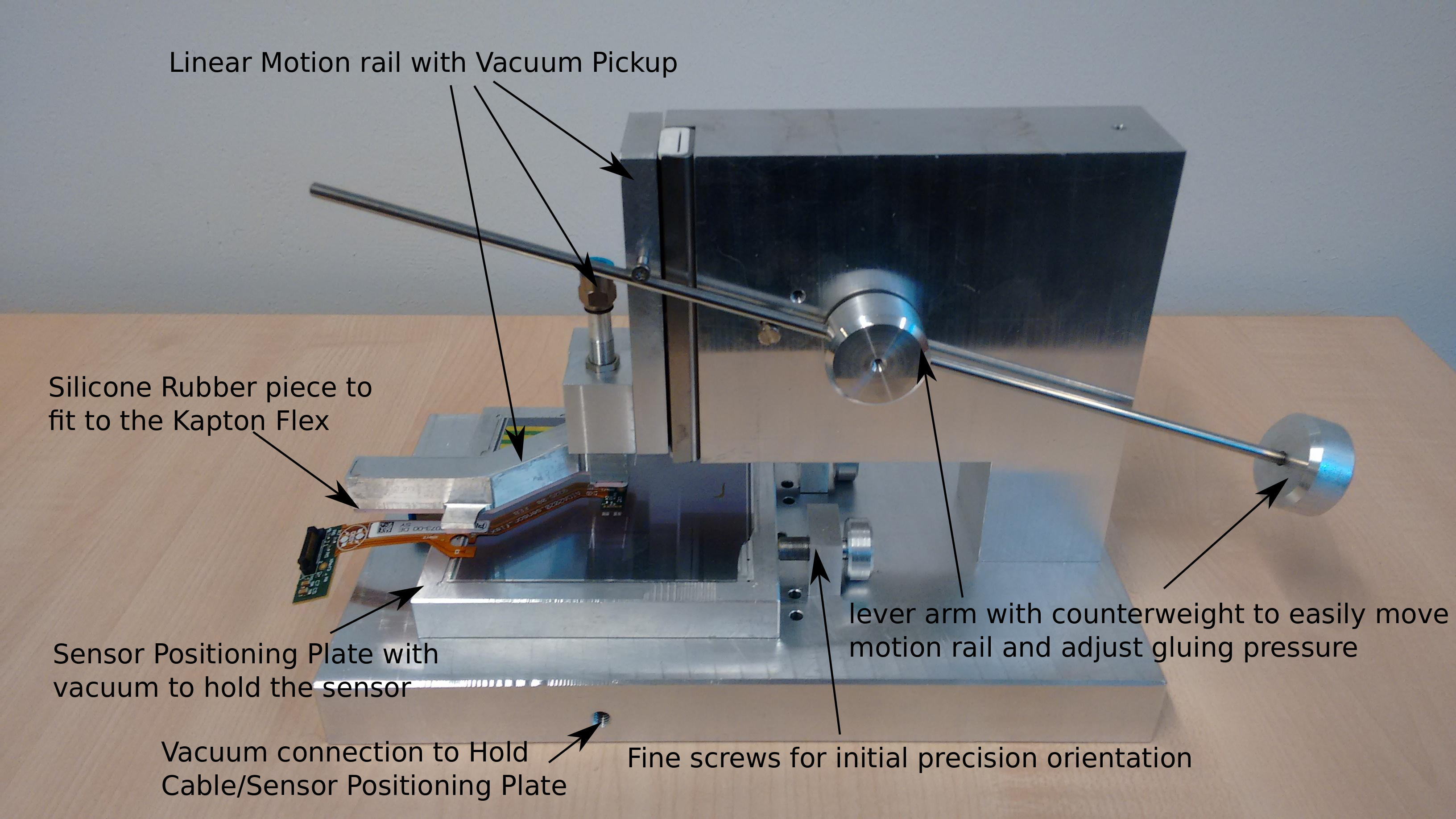}
  \caption{\label{fig:hardware:gluetool}The custom gluing tool used to assemble the \LYCORIS modules.}
\end{center}
\end{figure}

Both bump-bonding and wire-bonding are well-established industry standard processes and have excellent track-records in assembling
sensor modules for the LHC experiments.
The bump-bonding were performed at Fraunhofer IZM. The bumps has been pre-applied on the \KPIX by its manufacturer and the under bump metalization has also been pre-applied on the sensor by its manufacturer. The wire-bonding step were performed in-house at DESY.
Placing a Kapton-Flex cable accurately on the active sensor surface using glue, however, is a highly customized process.
Therefore, a dedicated gluing tool was designed to glue the Kapton-Flex to the sensor (see Figure~\ref{fig:hardware:gluetool}).
The non-conductive epoxy paste adhesive \texttt{Araldite 2011} requiring 12~hours curing has been used based on the experiences of ATLAS~\cite{Abdesselam:2006wt} and
the gluing procedure has been validated to ensure it meets the following requirements:
\begin{itemize}
  \item No damage to the sensor when applying an even pressure on the glue area to obtain an optimal cure;
  \item The cable is positioned precisely to the area given by the edge of the three bonding areas and stays steady during the cure;
  \item No extra glue split over the wire-bonding pads which directly connects to the second metallization layer.
\end{itemize}

Among the 29 sensors produced, 24 have been successfully assembled, while two of them have not yet been fully assembled for future upgrades to the flex cable. The rest five sensors failed to be assembled for different reasons: one sensor was grinded down to validate the bump-bonding quality, two sensors failed to be wire-bonded due to aged bonding pads on the flex cable, and the rest two sensors were damaged by a high pressure during the gluing process.
The quality of the 24 assembled sensor modules are further controlled by IV measurements. All the sensors show a typical development as shown in Section~\ref{sec:performance:calib}, except for two sensors that cannot deplete anymore~\footnote{The reason is not identified though many inspections have been conveyed, one promising cause is micro-level damages in the sensor bulk during the gluing step.}.

\subsection{The Cassette}
\begin{figure}[htbp]
\begin{center}
\includegraphics[height=4.5cm]{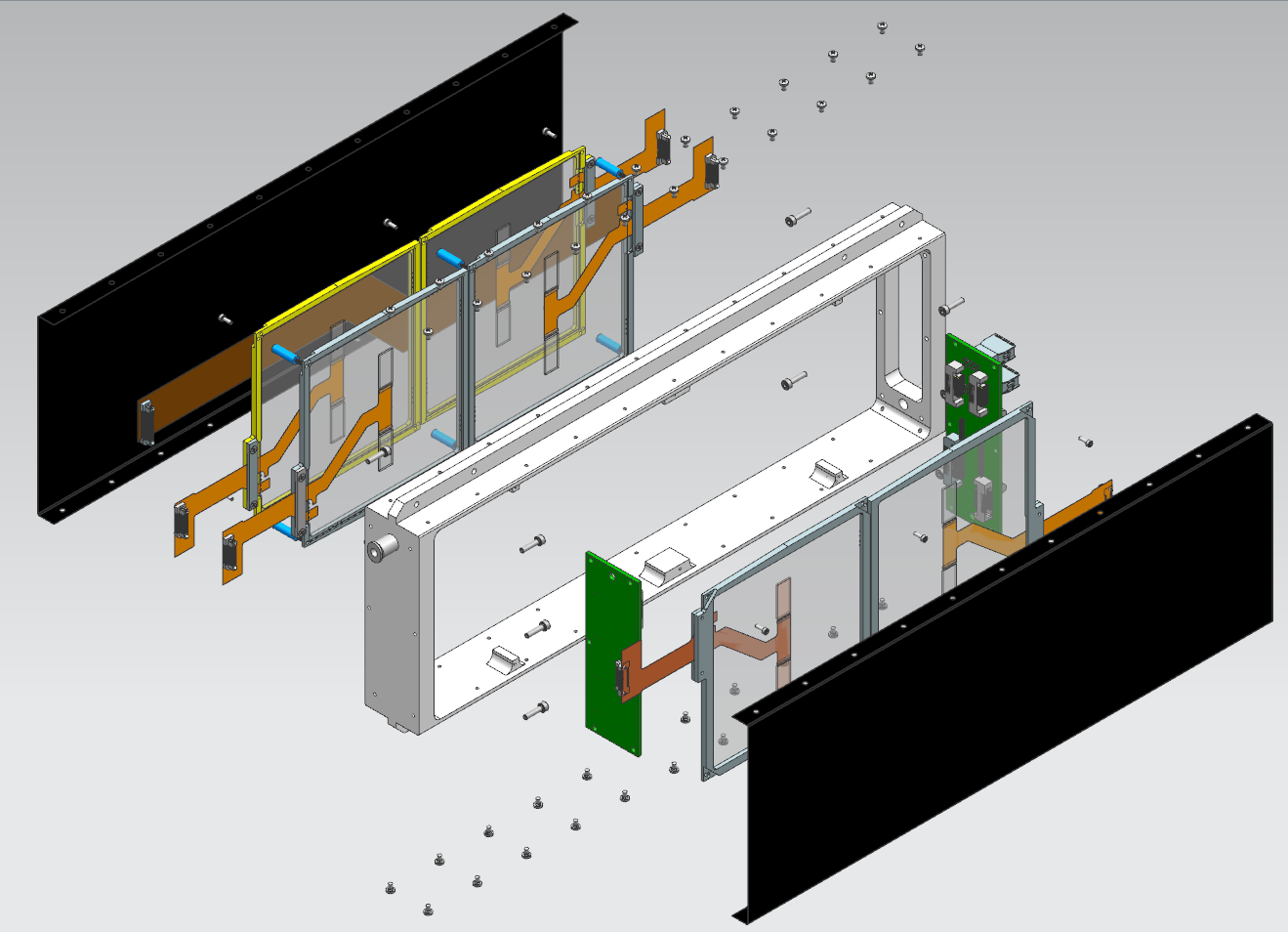}
\includegraphics[height=4.5cm]{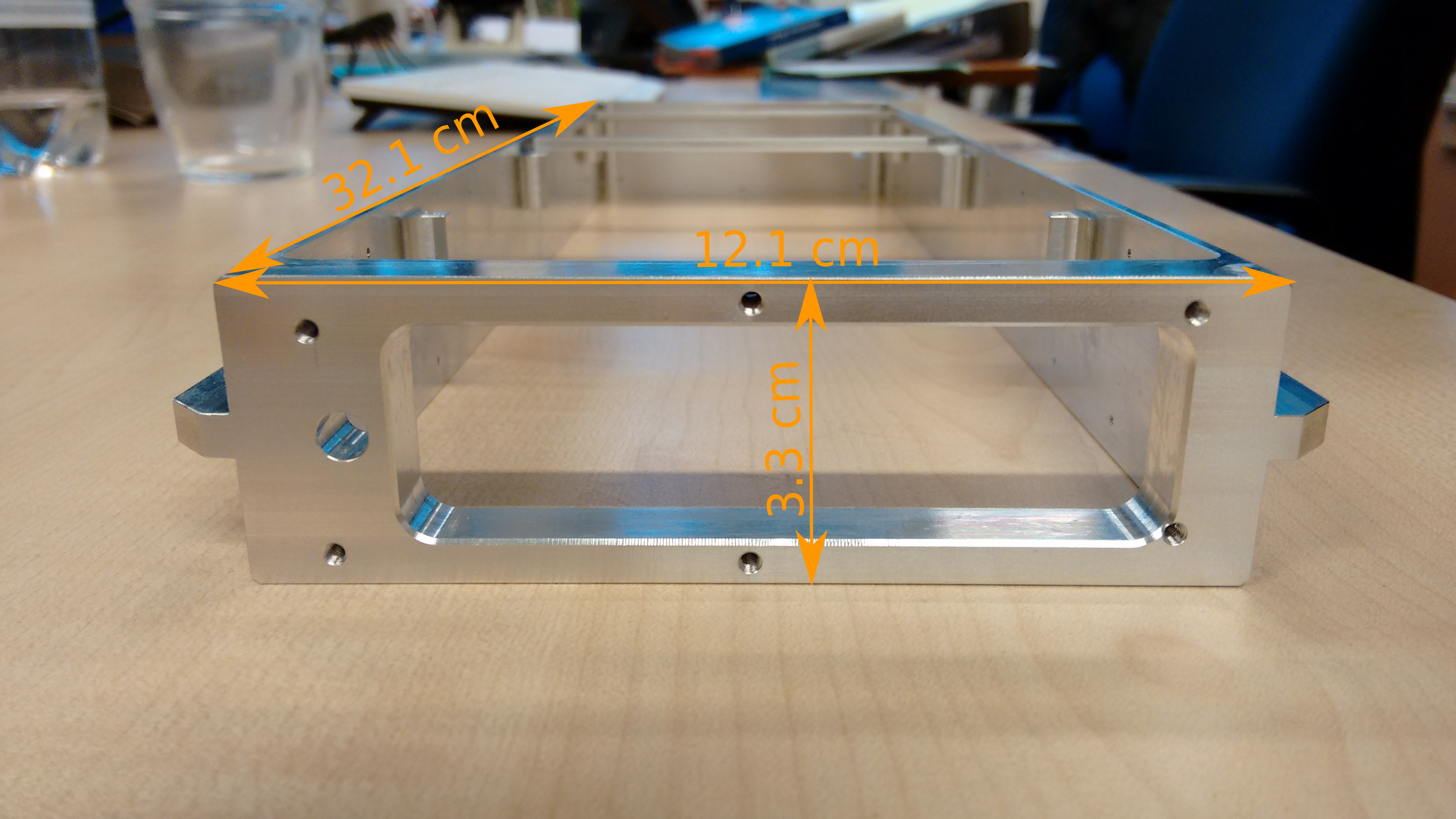}
\caption{\label{fig:hardware:cassette}
The Cassette: An exploded view of one \LYCORIS cassette holding two stacks of three sensors and two cassette boards (left) and a photo of the cassette frame with
dimensions (\SI{3.3}{\centi\metre} tall, \SI{12.1}{\centi\metre} wide and \SI{32.1}{\centi\metre} deep) (right).}
\end{center}
\end{figure}

The \LYCORIS cassette, which holds up to two stacks of three sensors, consists of an aluminum frame and two cassette boards and is
covered by two carbon fiber windows as shown in Figure~\ref{fig:hardware:cassette}.
The cassette is \SI{3.3}{\centi\metre} tall, \SI{12.1}{\centi\metre} wide and \SI{32.1}{\centi\metre} deep.
Each individual sensor module is glued to a frame which is designed together with the cassette so that the frame can be
installed with three (+2, -2, 0 degree) different orientations.
The frame is made out of Torlon, which is non-conductive but of similar strength as aluminum, so that it can be milled using standard techniques.

\subsubsection{The Cassette Board}
The cassette board serves as a transition board between the DAQ board and the modules. It is designed to drive up to three sensor modules in a cassette.
It distributes Low Voltage (LV) power to the connected \KPIX chips through linear regulators with an input voltage of \SI{3.0}{\volt},
connects each High Voltage (HV) power line from the power supply to each connected sensor,
and distributes all digital signals between the \KPIX chips and the DAQ board.
Each output line of the linear regulator is filtered separately for noise.
AVDD (2.5V) is separately regulated for the two \KPIX chips on the same sensor,
while the digital power DVDD (2.0V) is shared by all \KPIX ASICs from a single regulator.

\begin{figure}[htbp]
\begin{center}
\includegraphics[height=6cm]{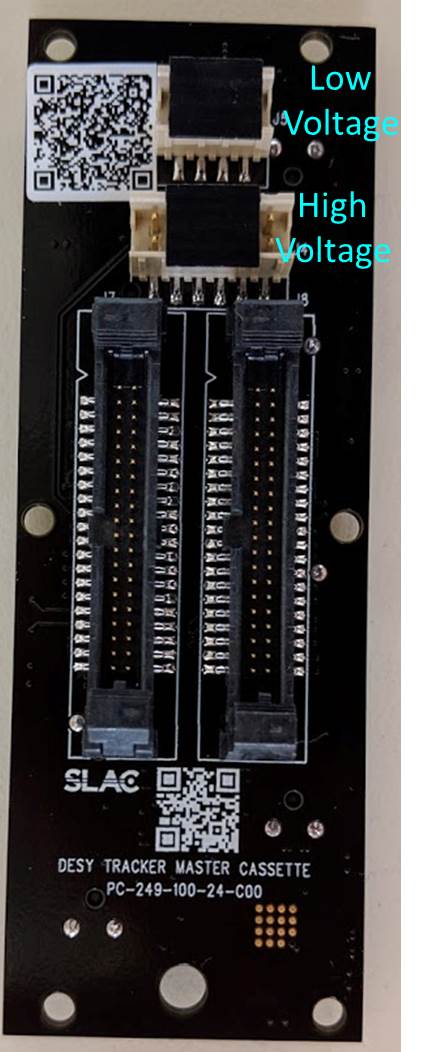}
\includegraphics[height=6cm]{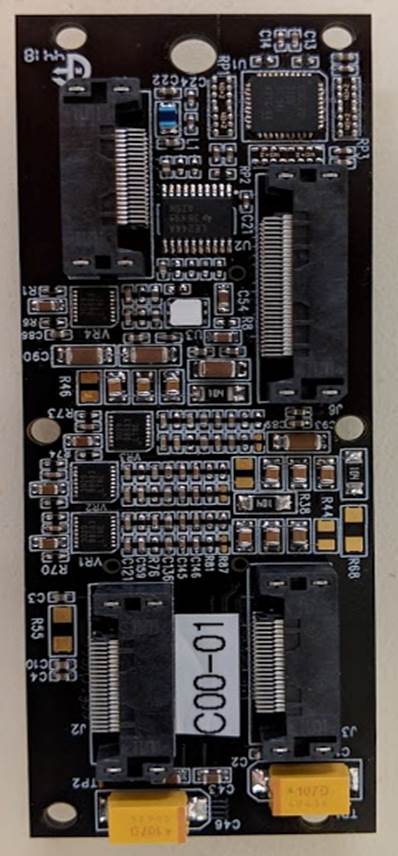}
\includegraphics[height=6cm]{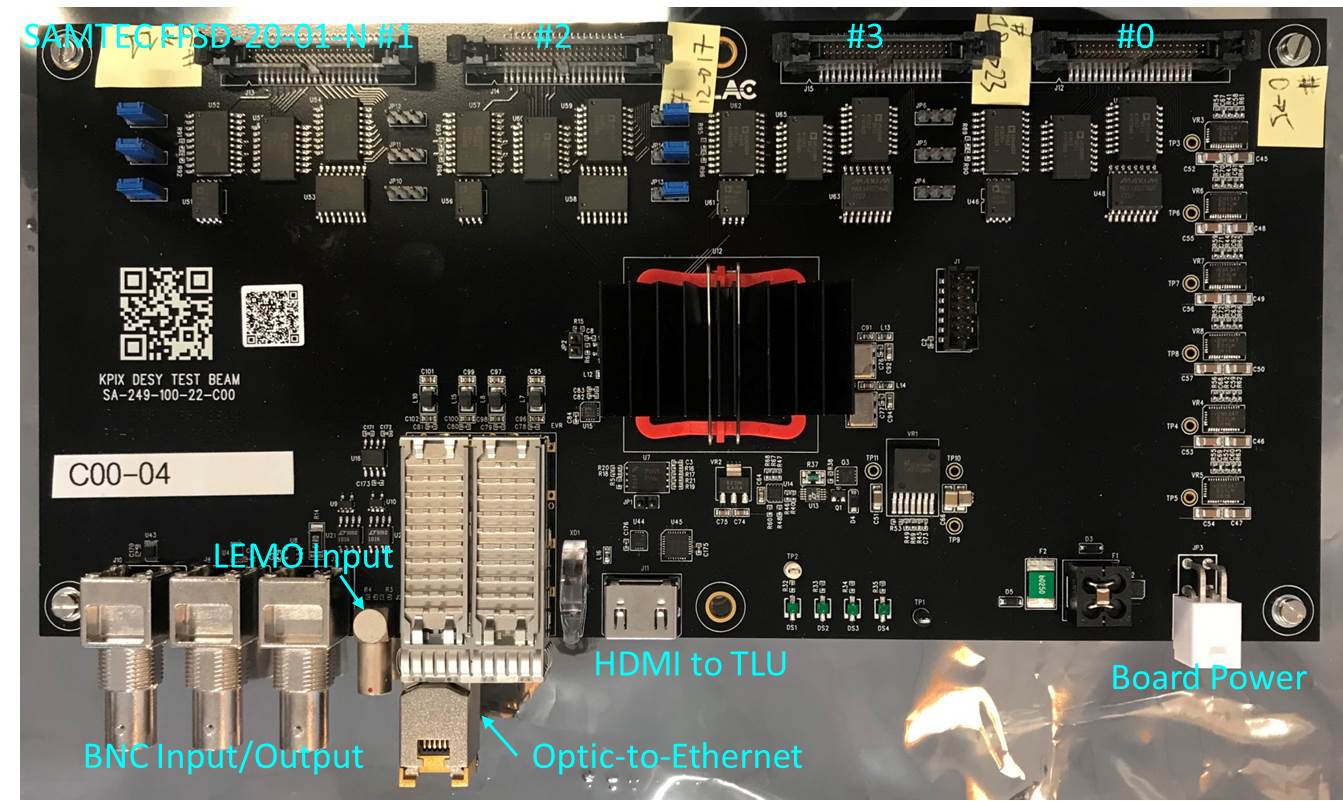}
\caption{\label{fig:hardware:boards}The primary cassette board (outer side - left and inner side - middle) and the DAQ board used by \LYCORIS (right).}
\end{center}
\end{figure}

Two versions of the cassette board exist and can be daisy-chained via a custom Kapton-Flex cable.
One is installed at the front of the cassette and acts as the primary board while the other is installed at the other end of the cassette and serves as the secondary.
They drive the connected sensor modules in the same way, but the primary cassette board has two extra functions.
First, its outer side as shown in Figure~\ref{fig:hardware:boards} (left)
hosts the interfaces with the DAQ board and the power supply for the sensors.
Second, it has one \ITWOC temperature and humidity sensor equipped on its inner side as shown in Figure~\ref{fig:hardware:boards}
(middle) to monitor the environment inside the cassette.

\subsection{Trigger and DAQ Hardware}
The DAQ board used by \LYCORIS is shown in Figure~\ref{fig:hardware:boards} (right).
It requires \SI{12}{\volt} / \SI{0.5}{\ampere} for operation.
It connects to the user control PC through an SFP port, allowing for flexibility in the physical interface used.
Currently, the RJ-45 module is used for ease in connecting to standard 1000Base-T networking equipment, but it is possible to
upgrade to optical Ethernet or a custom optical interface in the future if needed.
The board is equipped with a Xilinx Kintex7 FPGA to provide I/O interfaces and event building.
It has been designed with the following features:
\begin{itemize}
  \item All signal interconnects with the connected cassette boards are communicated via isolation buffers, so that the DAQ board and
  all the connected \KPIX chips are electrically isolated;
  \item Multiple input connections to provide acquisition start signals and external triggers for the \KPIX data taking: two LEMO ports, two BNC ports and one HDMI port;
  \item An Ethernet interface using SFP+ plugins to connect to a PC running the DAQ software.
\end{itemize}

\LYCORIS is designed to synchronize with other devices using the \AIDAII Trigger Logic Unit (TLU).
This has been implemented by correlating timestamps on triggers recorded respectively by \LYCORIS and by \AIDAII TLU.
The DAQ board is able to record a global timestamp on each incoming trigger using a 64-bit counter which is incremented with
every clock pulse of the \SI{200}{\mega\hertz} FPGA system clock and is reset by an external start signal at the beginning of each data run.
In order to correlate the \LYCORIS trigger timestamps to the TLU trigger timestamps,
the DAQ board can use the \SI{40}{\mega\hertz} TLU clock to generate its system clock
and the 64-bit counter can be reset by the start signal $T_0$ sent out by the TLU.
Both of the TLU clock and the $T_0$ signals are received through the HDMI input.

Besides the clock and the trigger signals, the TLU is also able to send out a so-called shutter signal to indicate the
presence of the beam. This signal is generated with a configurable delay on the $E_{min}$ signal from the \DESYII
accelerator (see Section~\ref{sec:performance:desyii} for further details)
and serves as the \KPIX \KPIXAcqstart signal.

The latency of the trigger timestamps registered by \LYCORIS and by the TLU is
important for the offline analysis in order to synchronize events between
different devices. This latency has been measured and verified to be very
stable for the same port of the same TLU unit. The variation among ports of
different TLU units is very small resulting a typical latency of \SI{105}{\nano\second}~\footnote{If the trigger comes at the very edge of one clock cycle, then the latency is extended for one additional clock cycle (\SI{25}{\nano\second}).}.

\subsubsection{DAQ Board FPGA Firmware}
\begin{figure}[htbp]
\begin{center}
\includegraphics[width=0.6\textwidth]{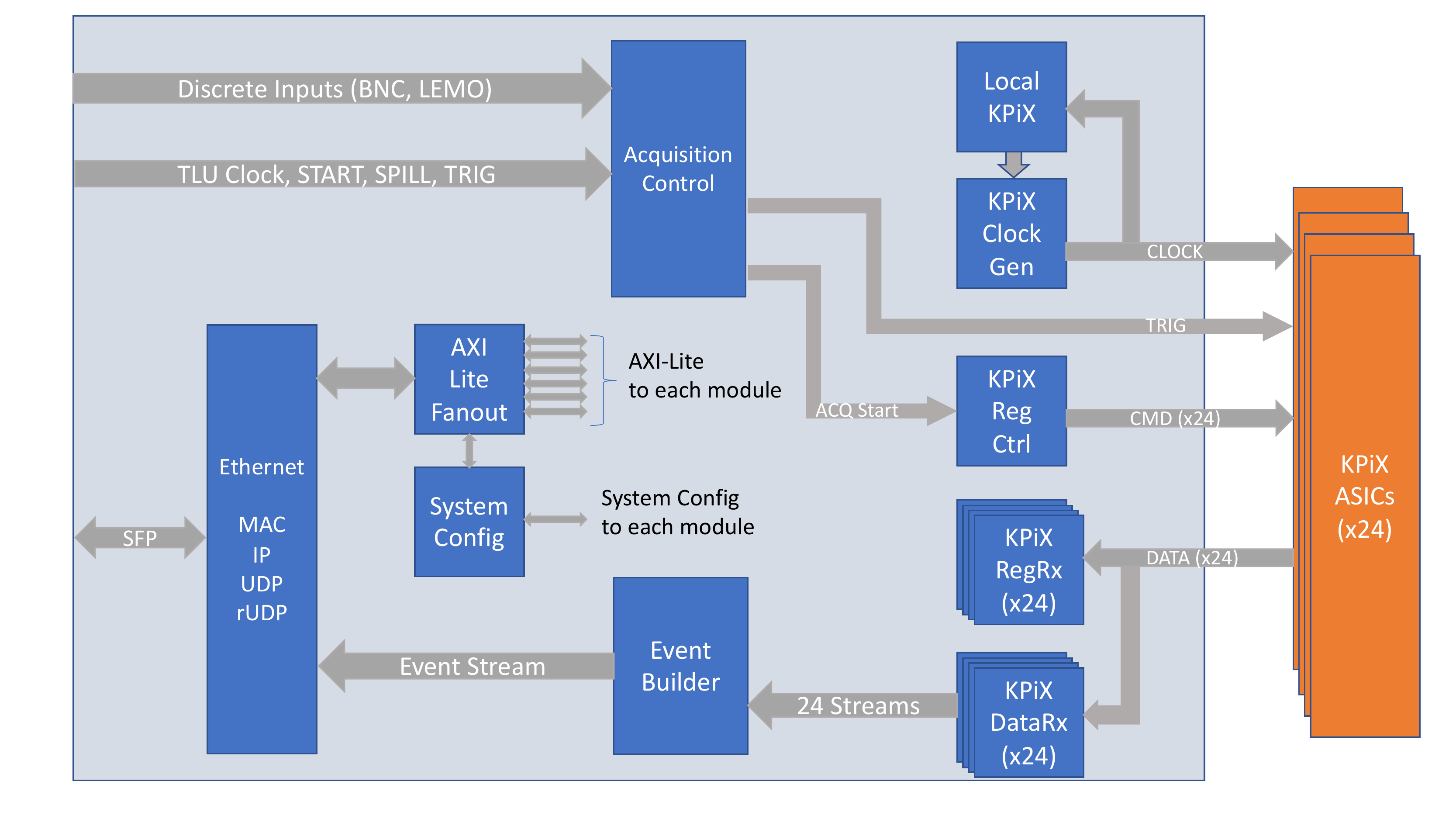}
\caption{\label{fig:hardware:firmware}Block diagram of the DAQ firmware design.}
\end{center}
\end{figure}

The FPGA firmware serves as a bridge between the DAQ software running on
commodity PC hardware, and the interfaces required by the \KPIX ASICs and the
TLU timing system. The firmware allows for configuration and acquisition control
of up to 24 attached \KPIX chips. Configuration registers allow for acquisition
to be driven by the TLU interface or a combination of DAQ software commands and
TTL inputs. The later configuration is typically used for taking baseline and
calibration data. Figure~\ref{fig:hardware:firmware} shows the block diagram of
the FPGA firmware. It reveals what is behind the digital traces between the DAQ
board and the connected \KPIX chips.

The interface to the DAQ software consists of an Ethernet/IP/UDP stack, with an
additional Reliable UDP (rUDP) layer on top to ensure reliable transmission of
UDP packets. This entire stack actually accounts for over half of the firmware
logic utilization of the design. The firmware uses an AXI-Lite bus internally
for handling register transactions from software.

The \KPIX ASIC digital interface consists of four signals: CLK, CMD, RSP, TRIG.
Register read and write commands are sent serially on the CMD line. Register
read responses arrive from the \KPIX on the RSP line. Requests to start a new
acquisition cycle are also sent on the CMD line, and the acquired data arrives
serially on the RSP line once the acquisition is complete. Register access is
not allowed during a \KPIX acquisition. The firmware logic handles the
serialization and deserialization of CMD and RSP, respectively. It also
translates register access requests and acquisition start commands into the
serial format expected by \KPIX. During acquisition data readout, it reformats
the serial data stream into 64-bit ``samples", with each sample representing the
ADC and timestamp data for a single channel/bucket in the system.

The firmware allows for the clock speed to change dynamically during each phase
of the \KPIX acquisition cycle. To accomplish this, the firmware runs a local
copy of the \KPIX digital logic, so that it can snoop on the current state of the
\KPIX. All \KPIX chips in the system run in lockstep with each other, so the state
of the local \KPIX is the same as the chips on the sensors. The clock period to
use during each phase is set via configuration registers prior to the start of
each run.

For each acquisition, the Event Builder combines the 64-bit samples from each of the (up to) 24 attached \KPIX ASICs into a single large data frame.
Each \KPIX has 1024 channels, and each channel has four buckets, so a data frame from an acquisition may contain up to 1024$\times$4$\times$24 samples. The
data frame also contains a 4 $\times$ 64-bit word header containing timestamp and run count metadata.


\subsection{Power Supplies}
The power supply system used is a commercially available MPOD Mini system from Wiener which is mounted into a 19-inch rack.
The \texttt{WIENER LV MPV8016I} provides all the LV (DAQ, AVDD and DVDD) and the one for HV is \texttt{WIENER/ISEQ EHS F205x\_106HV}.
This system can be remote-controlled via Ethernet using the SNMP protocol, thus providing online monitoring of the current and the voltage of each power line.

The power supply LV lines for the sensors in a cassette are common.
They connect to a set of linear regulators on the cassette board and are then distributed to the sensors.
Two LV lines serve each cassette board, cleanly separating analog and digital power provided to the \KPIX chips.
Each sensor has its a dedicated HV line, which allows the fine-tuning of depletion voltage for each sensor individually.

\section{DAQ, Monitoring and Reconstruction Software}\label{sec:software}
\begin{figure}[htbp]
  \begin{center}
    \includegraphics[width=0.85\textwidth]{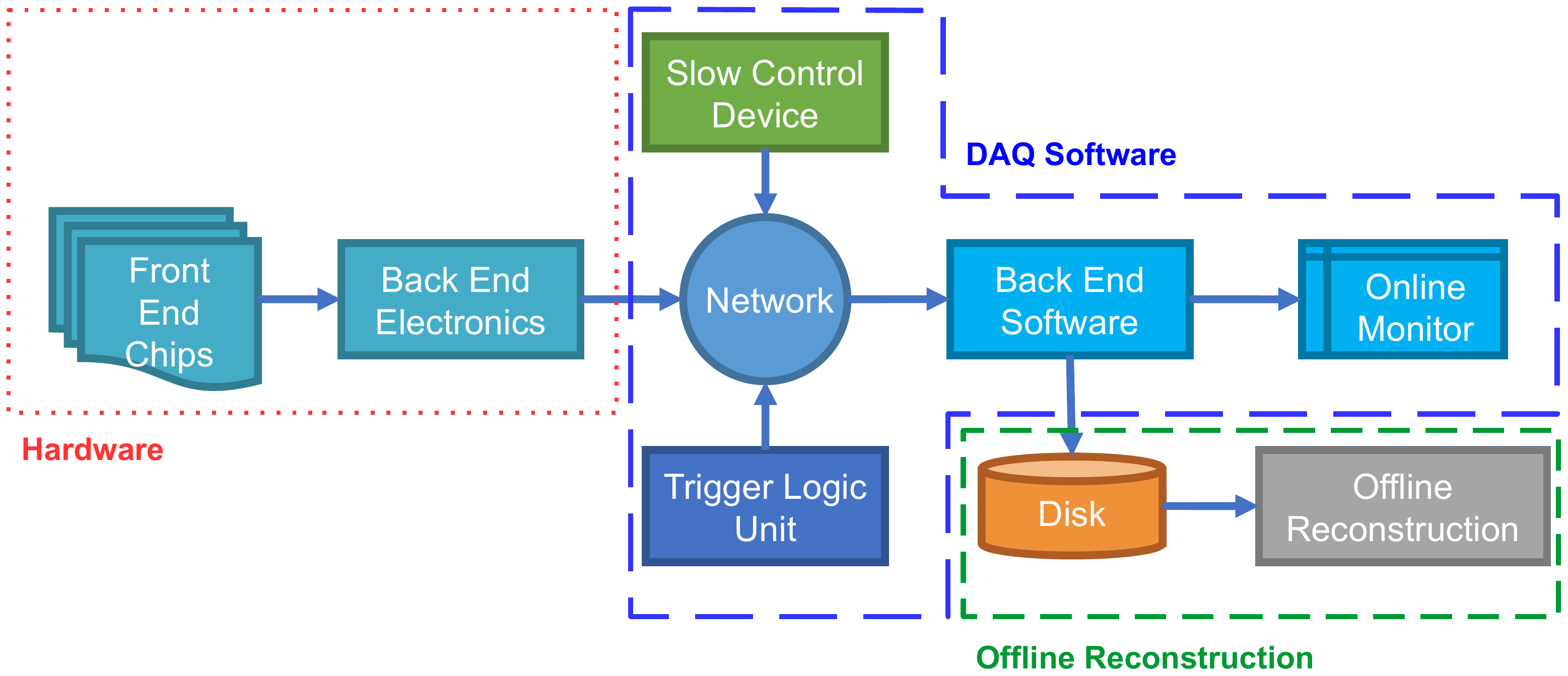}
    \caption{Flow chart highlighting the data flow in the DAQ back-end system for \LYCORIS.The components in the red box have been discussed in the Section~\ref{sec:hardware}, while
    the components inside the blue and green boxes are discussed in Section~\ref{sec:software}.}
    \label{fig:daqsw:general}
  \end{center}
\end{figure}

In this section the DAQ and Monitoring software packages required to operate \LYCORIS are described,
together with the reconstruction software packages used in this paper.
These packages are all part of the complete software suite, that is provided to the users, so they can easily analyze
their test beam data taken with \LYCORIS. The overall data flow from the \LYCORIS hardware to disk and the offline reconstruction is shown in
Figure~\ref{fig:daqsw:general}.

The back-end DAQ software runs on a Control PC, polling all the connected devices.
It is responsible for collecting all the data and writing it to disk for the offline track reconstruction.
Additionally, it provides online monitoring to provide a quick feedback on the data-quality during data-taking.
This DAQ software for \LYCORIS consists of two separate layers: the layer communicating direct to the \LYCORIS hardware uses the \Rogue DAQ framework while
the interfacing of \LYCORIS with the \AIDA TLU and the user DAQ is handled by \EUDAQII.

\subsection{\Rogue DAQ}
The software employed to communicate with the DAQ board and \KPIX chips is based on the \Rogue platform~\cite{roguewebsite}. \Rogue was created by SLAC to facilitate development of
software interfaces to hardware in DAQ systems.
It has been designed with easy-to-understand mechanisms to connect independent management and data processing modules together and uses a set of
well-defined and easy-to-understand interfaces. It has been developed using both \CPLUSPLUS and \PYTHON and the interfacing between the \CPLUSPLUS and \PYTHON parts is done using \BOOSTPYTHON.
\Rogue provides a properly acquired and released global interpreter lock which enables true multi-threading and predefined handlers to access the data buffer pipelines and
hardware interfaces, which \EUDAQII is using to establish full control of the data taking with \KPIX and \LYCORIS.

\subsection{\EUDAQII integration}
The integration of \Rogue into \EUDAQII has been done using the \PYTHON interfaces of both packages. The \LYCORIS strip telescope has used two distinct versions of the \KPIX DAQ system during its development history.
The initial version was able to only synchronize \LYCORIS to an external device by merely counting the incoming triggers,
while the final version can additionally synchronize the incoming triggers by timestamping with a centrally provided clock.
\LYCORIS is using the new \KPIX DAQ system which enables full synchronization with other devices such as the \EUDET telescope via the \AIDA TLU.

The \KPIX \Producer, which is a part of the \EUDAQII package, is a \PYTHON-based \Producer to operate \LYCORIS using the imported \KPIX DAQ libraries.
In the \EUDAQII framework, the \Producer acts as the interface between the user DAQ and the \EUDAQII core components including the \RunControl and the \DataCollector.
The \EUDAQII \RunControl is used to issue all commands to the connected \Producers and the underlying DAQ systems including \LYCORIS.
The \EUDAQII \DataCollector uses the \KPIX binary data format libraries to store the incoming \LYCORIS data-stream and the \DATACONVERTER
module converts \KPIX binary format to the common format used by \EUDAQII. This is necessary to make the online monitoring in \EUDAQII available.

\begin{figure}[htbp]
  \begin{center}
    \includegraphics[width=0.85\textwidth]{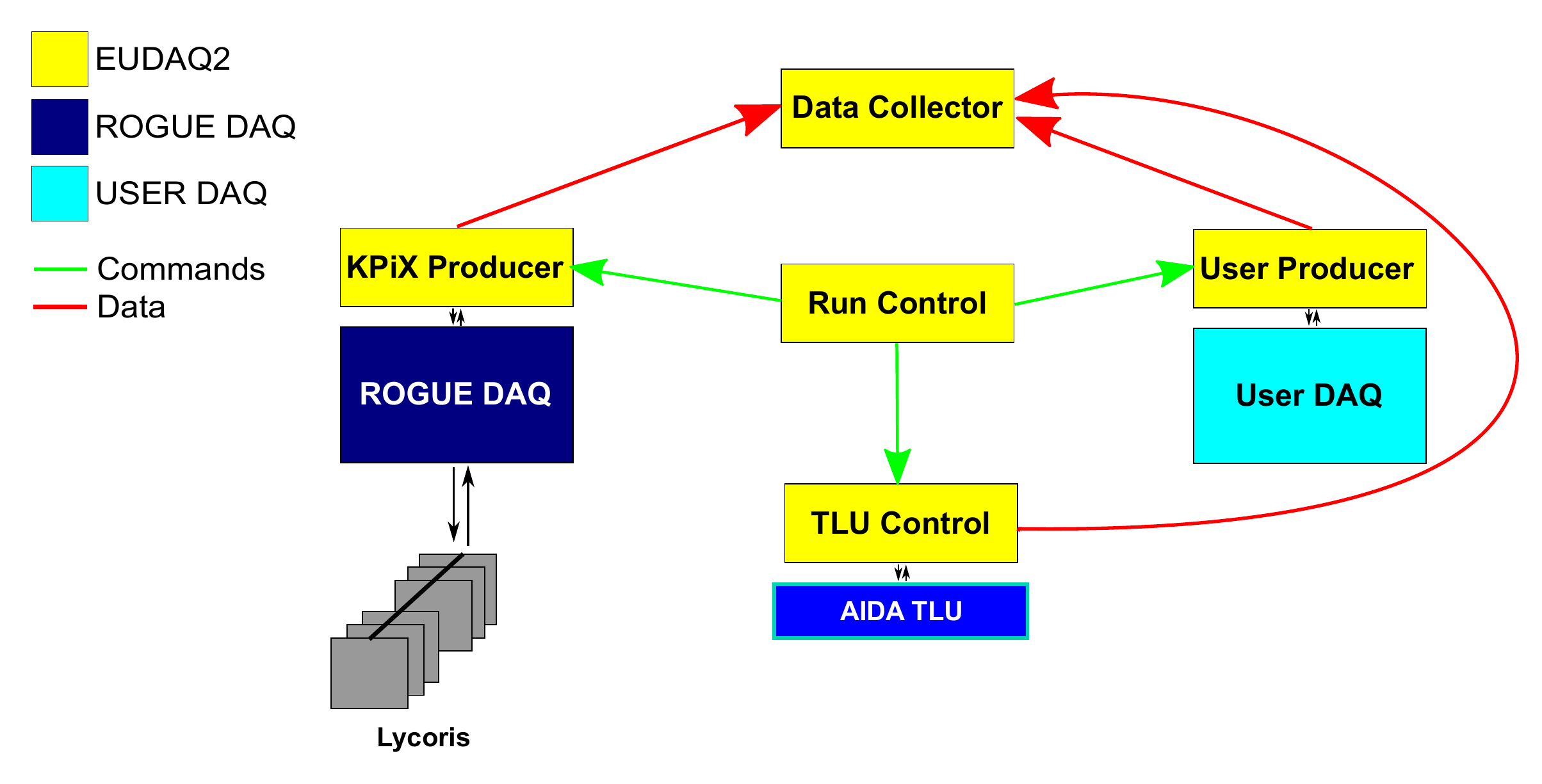}
    \caption{The interplay between the various \EUDAQII producers and the \EUDAQII \DataCollector with the \LYCORIS DAQ system.}
    \label{fig:daqsw:eudaq2flow}
  \end{center}
\end{figure}

Furthermore, the \KPIX user module contains a customized GUI, which provides more detailed information like the data rate during a run,
and an analysis executable called \texttt{lycorisCliConverter}, which produces a set of basic plots, e.g.\ the ADC distribution of each channel,
and stores them into a \ROOT
file. Figure~\ref{fig:daqsw:eudaq2flow} show the interplay of the various DAQ software components.

\subsection{Online Monitoring \& Slow Controls}
\subsubsection{Online Monitoring}
\begin{figure}[htbp]
  \begin{center}
    \includegraphics[width=0.8\textwidth]{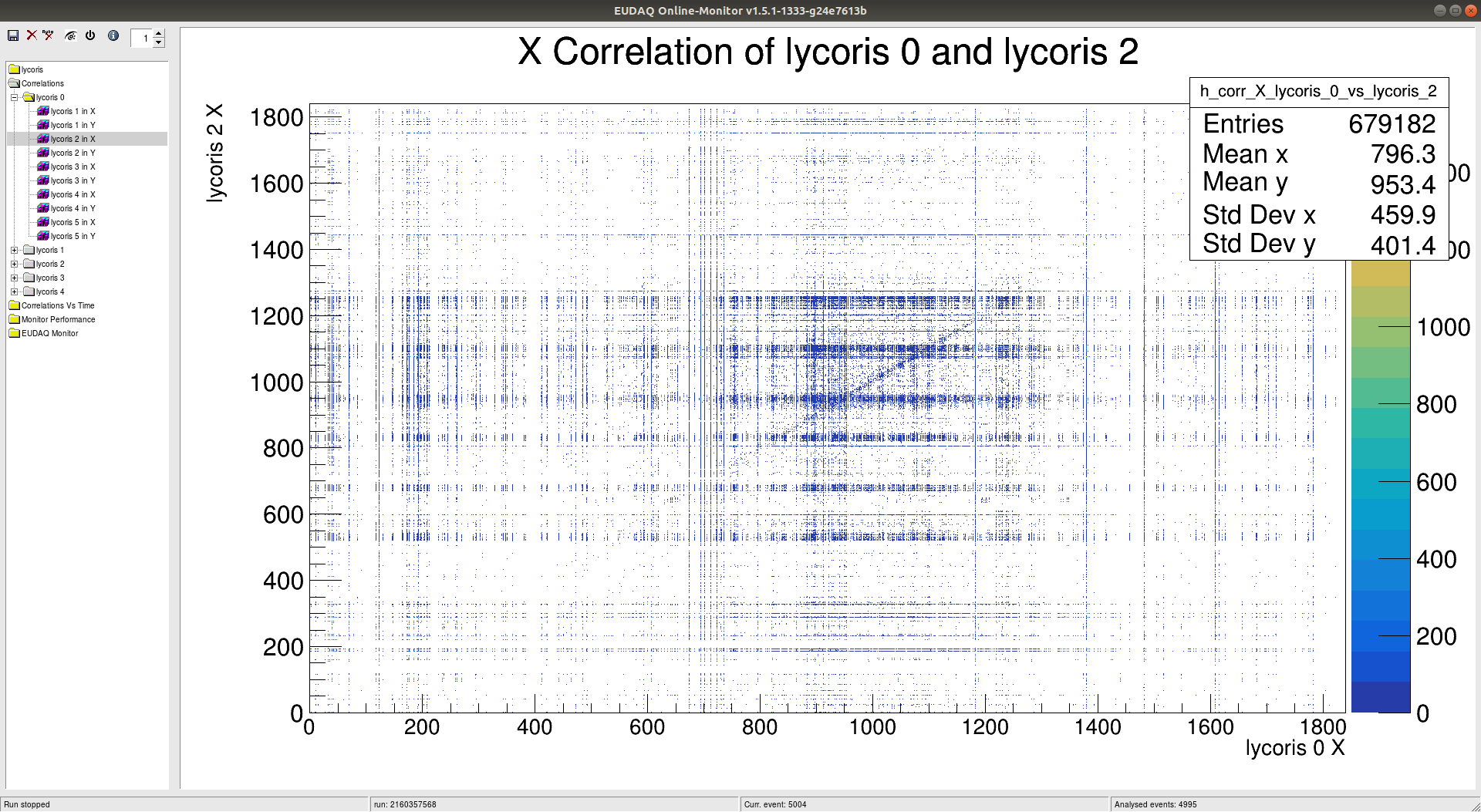}
    \caption{Online monitor showing spatial correlations of two \LYCORIS sensor planes
    from one example data taken outside the \PCMAG with the \EUDAQII and the \AIDA TLU at the \DIITBF,
    where the x and y axes are in unit of the strip readout pitch (\SI{50}{\micro\metre}).
    This shows a good alingment between the two sensor planes and a beam spot of \SI{20}{\milli\metre}$\times$\SI{20}{\milli\metre} (same size as the used collimator)
    located in the center of the sensor.
    }
    \label{fig:daqsw:onlinemon}
  \end{center}
\end{figure}

The online monitor for data taking with the \LYCORIS telescope has been developed using the \EUDAQII online monitor module \StdEventMonitor.
It provides a list of 2D histograms showing spatial correlations of hitted strips from two different sensor planes,
which can be used to control the data quality, beam alignment and the beam spot position.
Figure~\ref{fig:daqsw:onlinemon} is a screenshot of the online monitor during \LYCORIS data taking, where the shown example histogram shows a clear spatial correlation between the first and the third \LYCORIS telescope planes.
Online monitor data streams are converted by the \DATACONVERTER, including data format decoding, ADC to charge conversion and baseline noise subtraction to remove pedestals and the common-mode noise, see details in Section~\ref{sec:software:reco}.

\subsubsection{Slow Controls}
The \LYCORIS slow control system monitors the humidity and temperature inside the cassette units and the bias voltages for each sensor.
The humidity and temperature are monitored by an \ITWOC sensor located on the master cassette board, which is polled by the \KPIX DAQ.
The HV modules of the WIENER MPOD system are controlled and monitored remotely through an SNMP~v2c (Simple Network Management Protocol) compliant protocol.
It is integrated into \EUDAQII with a \Producer to poll the measured current and voltage, and with one dedicated \DataCollector the collected data can
be stored directly in the format used by \EUDAQII.

\begin{figure}[htbp]
  \begin{center}
    \includegraphics[width=0.49\textwidth]{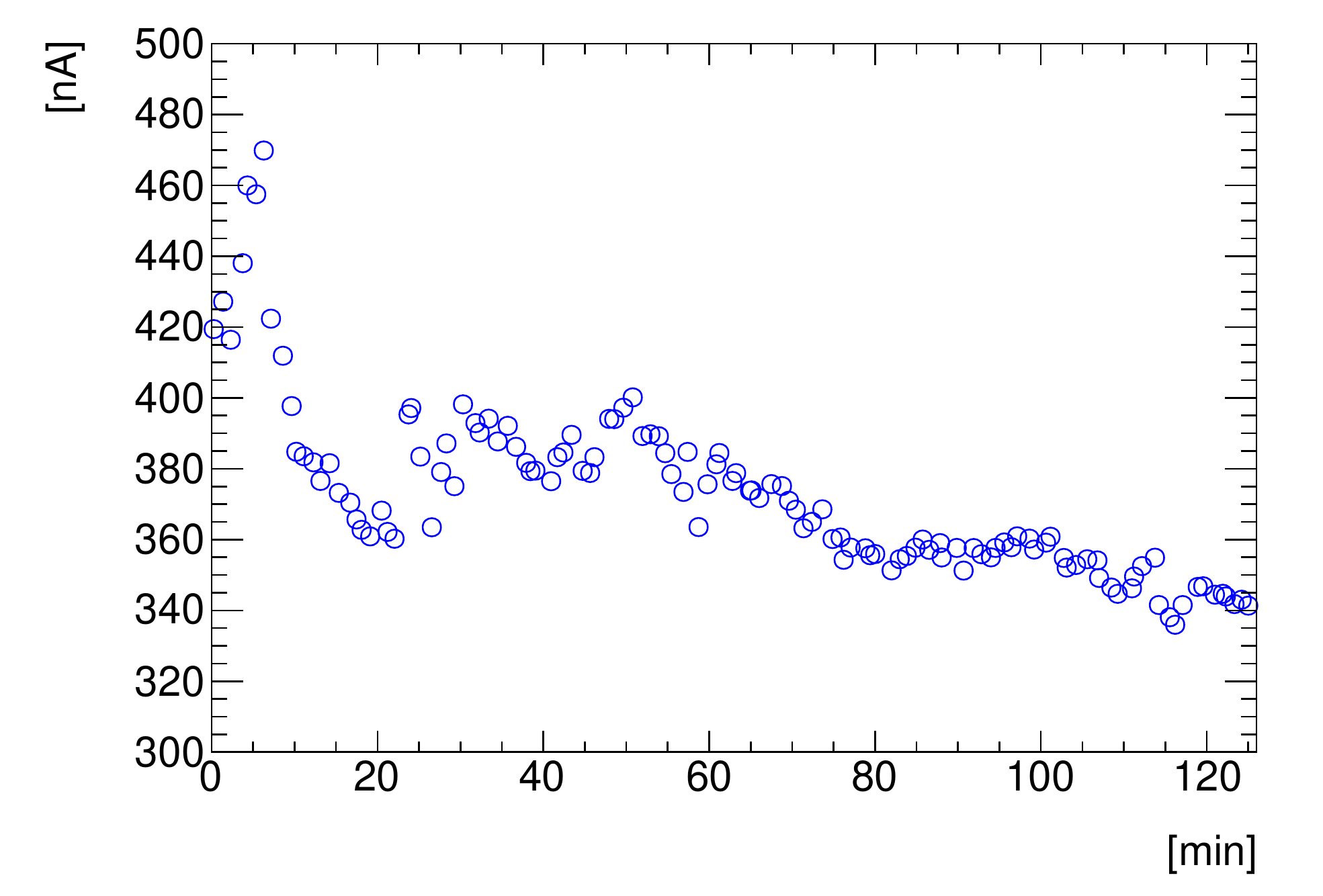}
    \caption{The bias current in nA vs. time in minute over two hours of sensor plane 0
    from example data taken with the \EUDAQII and the \AIDA TLU at the \DIITBF.
    }
    \label{fig:daqsw:sc}
  \end{center}
\end{figure}

Figure~\ref{fig:daqsw:sc} shows the bias current in \SI{}{\nano\ampere} of a sensor measured over two hours during data taking, providing fast feedback on the status of each sensor.

\subsection{Event Reconstruction}\label{sec:software:reco}
\begin{figure}[htb]
  \begin{center}
    \includegraphics[width=0.95\textwidth]{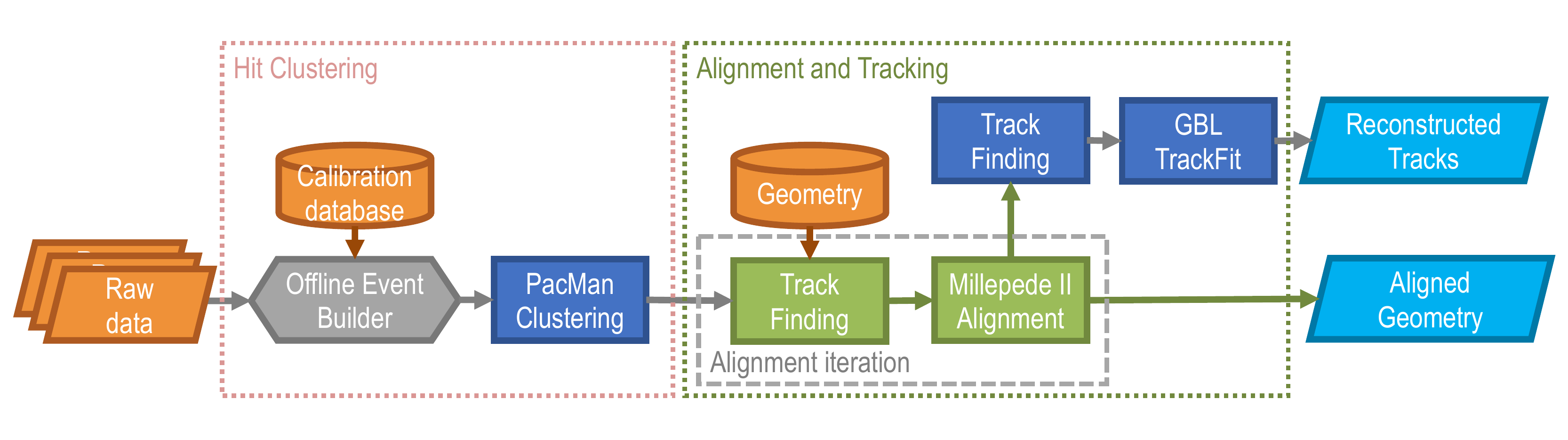}
    \caption{Flow chart of the tracking reconstruction algorithms.}
    \label{fig:daqsw:reco}
  \end{center}
\end{figure}

Two software packages are used for the track reconstruction; the first package loops over the raw data provided by \LYCORIS and produces hit clusters for the tracking algorithms,
while the second one does the overall alignment, the track finding and fitting and produces both a fully aligned geometry and the fitted tracks.
Figure~\ref{fig:daqsw:reco} shows a simplified block diagram of data processing for track reconstruction with \LYCORIS.

\subsubsection{Hit Clustering}\label{sec:software:hitclustering}

The Hit Clustering analysis package reads in the raw data, applies the necessary corrections and then groups strips into hit clusters.
The package is written in \CPLUSPLUS and built with \CMAKE~\cite{cmake}. There are two main steps in this package,
the offline event builder and the PacMan clustering algorithm.
\paragraph{The Offline Event Builder}

The first step converts the raw binary data into events suitable for reconstruction.
The definition of an event depends on whether \LYCORIS is operated in external trigger mode or internal trigger mode (see Section~\ref{sec:hardware:kpix:trigger}).
In external trigger mode, one event is a snapshot of the full telescope, i.e.\ one event consists of data from the same bucket
of all 1024 channels of all \KPIX chips in the same acquisition cycle.
For the internal trigger mode, every readout channel is triggered individually but one event can still be defined
by matching an internal trigger with an external trigger using timestamps.
Therefore, one event in this case may contain data from different \KPIX readout channels at different memory buckets of all
the \KPIX chips in the same acquisition cycle. As already explained in Section~\ref{sec:hardware:kpix:trigger}, \LYCORIS
prefers to be operated using the external trigger mode and the all the following steps are targeted for this trigger mode.

The following is the baseline noise subtraction, which includes subtractions of the pedestals and the common-mode noise.
The pedestal subtraction is performed separately for each memory bucket of each \KPIX readout channel,
and defined as the median of the charge response over a series of acquisition cycles in order to suppress the impact of outliers.
The common-mode noise subtraction is applied on an event-by-event basis for each individual \KPIX to ensure
that data recorded in different memory buckets during different acquisition cycles are comparable.
It is calculated as the median of the pedestal-subtracted charge responses of one \KPIX chip for one event.
Therefore, the calibration constants (see Section \ref{sec:hardware:kpix:calib}) have to be applied to convert the raw ADC
value to a charge in \si{\femto\coulomb} beforehand so that charge response is comparable from one channel to another when
calculating the common-mode noise.

At the end of the baseline noise subtraction, all \KPIX readout channels that are not connected to a strip are removed,
so every \KPIX channel represents one readout strip on the sensor, and thus is referred to be the strip identity.

Before grouping strips into clusters, we apply the following quality criteria to remove strips with faulty ADC readings:
\begin{itemize}
  \item The calibration slope has to be non-zero;
  \item The calibration needs to show a good linearity by requiring its Pearson correlation coefficient to be greater than $0.85$;
  \item The Median Absolute Deviation (MAD) of the measured charge of each bucket of each strip over all cycles must be non-zero.
\end{itemize}

The intrinsic noise level $N_i$ for one strip at one memory bucket is defined as the width of its noise-corrected charge response when no signal presents.
For one charge measurement $q_i$ of one strip at one memory bucket, its significance is calculated as $q_i/N_i$.
The results of the complete noise-removal procedure and the noise performance will be discussed in Section~\ref{sec:performance}.

\paragraph{PacMan Clustering Algorithm}
The next step in the chain is the clustering algorithm which follows right after the baseline noise removal.
A relatively loose selection is applied on all strips before the clustering by requiring $S/N > 2$ to ensure a sufficient cluster purity while retaining
suitable statistics at this stage.
The clustering algorithm then searches iteratively for the most significant strip as the \textit{cluster seed}
and groups its neighbor strips to form a \textit{cluster}.
For each cluster, the group process stops if the next-neighbor strip has a higher significance in order to keep clusters isolated.

The attributes assigned to each generated cluster are the charge $Q$,
the significance (signal-to-noise ratio) and the cluster size (strip multiplicity).
The cluster charge $Q$ is defined as the sum of the charges of all associated strips.
Its noise $N$ is then calculated by adding the noise of all its associated strips in quadrature.
As a result, its significance can be calculated as
\begin{equation}
\frac{Q}{N} = \frac{\sum^{n}_{i}q_i}{\sqrt{\sum^{n}_{i}N_i^2}},
\end{equation}
where $n$ is the number of strips in the cluster, and $q_i$ and $N_i$ represent the signal charge and noise level for each associated strip.
The position of each cluster perpendicular to the strip direction (the y-axis) is obtained by charge weighting
\begin{equation}
y_{cluster} = \frac{\sum^{n}_{i}q_i\cdot y_i}{\sum^{n}_{i}q_i}.
\end{equation}
The results from the clustering procedure are called \textit{hits} throughout the remainder of this section.

\subsubsection{Alignment and Tracking}
The alignment and tracking algorithms are part of a highly modular \PYTHON package.
The three major algorithms of this package are the alignment code using Millepede~II~\cite{millepede2:2006},
the track-finding algorithm developed specifically for \LYCORIS
 and the track (re)fitting using the General Broken Lines (GBL) package~\cite{claus2012}.

\begin{figure}[htbp]
 \begin{center}
    \includegraphics[width=0.65\textwidth]{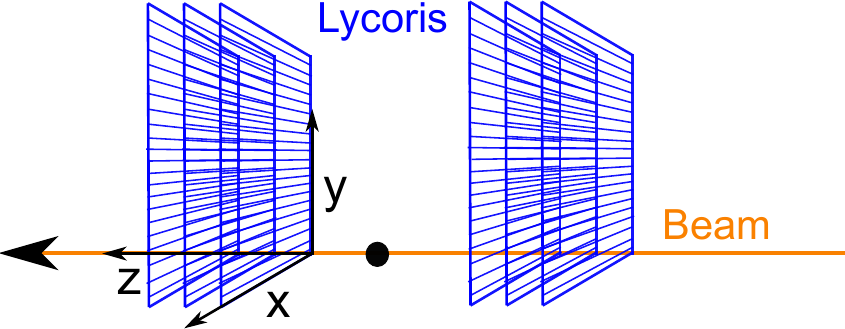}
    \caption{The global coordinate system used by the \LYCORIS Tracking and Alignment tool. The z-axis is along the beam axis,
    orthogonal to the \LYCORIS sensor planes. The x-axis is defined to be parallel to the silicon-strips on the axial sensor
    and the y-axis is perpendicular. The origin is located in the center of the telescope as indicated by the black dot.}
    \label{fig:daqsw:globalcoordinates}
  \end{center}
\end{figure}

For the reconstruction, the following global coordinate system is used throughout the paper:
The z-axis is along the beam axis, orthogonal to the \LYCORIS sensor planes.
The x-axis is defined to be parallel to the silicon-strips on the axial sensor and the y-axis is perpendicular
to the silicon strips (See Figure~\ref{fig:daqsw:globalcoordinates}).

\paragraph{Pre-selections}
There is a series of pre-selection cuts that are applied to the raw hits before the alignment and tracking algorithm.
First an event filter is applied to remove abnormal noisy events with at least 50 hits before any track finding algorithm is deployed.
Subsequently, hits from the three layers in each cassette need to be correlated
within a \SI{1}{\milli\meter} window based on an initial geometry description.
Hits that can not be correlated to others or hits with more than $5$ strips are all removed.

\paragraph{Alignment}

The alignment uses the Millepede~II algorithm which has already been used successfully by H1~\cite{Blobel:2002ax,Blobel:2006zz,Kleinwort:2006zz},
CMS~\cite{Flucke:2008zzb,Chatrchyan:2009sr}, BELLE-II~\cite{Bilka:2019tnt} and others.
Millepede~II is responsible for relative position alignment, meaning a 3-D reference frame is always needed as input; which is usually given by fixing the first and the last plane of the whole tracking system.
It builds on the case where least-squares fit problems with a very large number of parameters can be subdivided into two classes, global and local parameters.
Millepede~II then solves this to determine the alignment constants, where the global parameters are independent of the number of local parameters.

\paragraph{Track Finding}
Two track finding algorithms are designed to reconstruct stand-alone tracks with \LYCORIS hits.
The two algorithms can be used in any order and hits associated to one track candidate are not used for the subsequent algorithm to avoid double counting.
The track model is a straight line. In the case that a magnetic field is present, a coordinate transformation is applied to tracking parameters to account for the impact of the B-field based on the beam energy.

\begin{description}
  \item[\striplet finder:]
  It constructs \striplets in each cassette then matches them to form track candidates.
  The seed of \striplet is a doublet constructed by hits from the two stereo (\ang{2} stereo angle) layers in one cassette.
  By intersecting the two hits a 2D position is obtained in the x-y plane that is later interpolated to the axial layer for a third hit search.
  The third hit will be added to form the so-called \striplet when its distance to the doublet interpolation is smaller than \SI{100}{\micro\metre}.
  The four tracking parameters~\footnote{The offsets in X and Y, and the slopes in the x-z and the y-z planes.}
  of one \striplet are determined based on assumptions on slopes in both x-z and y-z planes for two reasons.
  First the three 1D measurements cannot determine four tracking parameters; second the doublet intersection needs both slopes
  for corrections due to the distance in z-axis between the two stereo layers.
  Assumptions on slopes are given by the beam direction with a correction if a magnetic field is present.
  Once all the \striplets are found in both cassettes, \striplet matching between cassettes starts.
  First their slopes on the y-z plane are matched by requiring the curvature difference to be smaller than \SI{0.01}{\radian}.
  Then they are extrapolated to the center between two cassettes for position matching by requiring the deviation to be
  smaller than ($<$\SI{10}{\milli\meter}) in x and ($<$\SI{1}{\milli\meter}) in y.
  To increase the purity, the matched \striplets only become a valid track candidate if they are the unique match to each other.

  \item[Strip road search:]
  It first forms track roads using four hits from the seed layers, then adds in one to two test hits from other test layers
  to finalize the search. The track road needs four hits to provide the full spatial description of a track.
  The default choice is to use the four stereo layers as the seed layers while the two axial layers are used as test layers to maximize the x-z plane description.
  An event will be rejected if there are fewer than five layers with hits or any layer possesses more than eight hits.
  Once all track roads are found, they are interpolated to the test layers and only the unique hits within a distance of \SI{200}{\micro\metre} are taken.
  The distance from the test hits to the interpolation is then used to calculate a $\chi^2$ to estimate the goodness of this track candidate.
  At the end, only the best track candidate is selected based on the number of associated hits and the goodness of fit $\chi^2/ndf$.
\end{description}
The combination of the high purity/low efficiency \striplet finder and low purity/high efficiency strip road search allows to reach a high
overall track-finding efficiency to account for some possible inefficient telescope plane.

\paragraph{Track Fitting}
After finding of all possible track candidates a re-fit of all valid track candidates using GBL is performed,
taking into account all measurements and scattering materials in the path of the track.
For a track with an initial trajectory from a pre-fit of the measurements
(internal seed) or an external prediction (external seed) the description of
multiple scattering is added by offsets in a local system. Along the initial
trajectory points are defined which can describe a measurement or a (thin)
scatterer (insensitive material) or both. Measurements are arbitrary functions of the local track
parameters at a point (e.g.\ 2D: position, 4D: direction+position). The re-fit
provides corrections to the local track parameters (in the local system) and the
corresponding covariance matrix at any of those points. Non-diagonal covariance
matrices of measurements will be diagonalized internally. Outliers can be
down-weighted by use of M-estimators.
A single measurement point can be omitted from the refit in order to calculate
unbiased residuals.

\section{Performance of the \LYCORIS telescope}\label{sec:performance}

This section presents all the results obtained using \LYCORIS starting from the bare sensor performance to the momentum resolution.
Firstly the test-beam setups used are described, which are the basis for most results through this section.

All the shown performance studies use only data stored in the first memory bucket of each \KPIX readout channel as no good-quality data have been collected using other three memory buckets.
The observed symptoms for these cells are that most of the \KPIX channels are found bad in either their ADC response or in their baseline noise level.
The root-cause for this problem is yet to be understood, but studies conducted post data-taking, have indicated, that by increasing the \KPIXTPreC (see \ref{sec:hardware:kpix})
by a factor of two, the data stored in the three other memory buckets can be recovered.
Unless mentioned explicitly, all the data shown are taken in the high-gain mode using the external trigger, which is the default mode for the \LYCORIS telescope.

\subsection{Test Beam Setup}
The system performance of \LYCORIS telescope has been evaluated at the \DIITBF in several test-beam campaigns. They can be categorized into two different categories:
one for performance measurements with an \EUDET-style pixel telescope \AZALEA outside the \PCMAG solenoid; the other one for validating the system performance inside the \PCMAG solenoid with the \AZALEA telescope as DUT.

\subsection{Operation at \DESYII}\label{sec:performance:desyii}
The \KPIX ASIC used in \LYCORIS requires an \KPIXAcqstart, so in order to operate \LYCORIS at the \DIITBF efficiently,
several particular properties of \DESYII have to be taken into account in order to synchronize \LYCORIS with the accelerator.

\DESYII~\cite{Hemmie:1983et,Hemmie:1985uw} is a synchrotron using normal-conducting magnets with a circumference of \SI{292.8}{\m} and a maximum energy of \SI{7.0}{\GeV}.
Its standard operation without any extraction stores a bunch of electrons or positrons for two magnet cycles
$\mathrm{T_{DESY\;Magnet\; Cycle}}$ of \SI{80}{\milli\s} which define the \SI{160}{\milli\s} \DESYII cycle.
One bunch of about 10$^{10}$ electrons or positrons is injected on-axis at $E_{\rm min} = \SI{0.45}{\GeV}$ from the linear accelerator \LINACII (LINear ACcelerator) via
the accumulator storage ring \PIA (Positron Intensity Accumulator) and is accelerated to $E_{\rm max} = \SI{6.3}{\GeV}$.
The beam is typically stored for two magnet cycles or one \DESYII cycle and is then dumped about \SI{160}{\milli\s} after injection, just before the next injection from \PIA.

\begin{figure}[htbp]
\begin{center}
\includegraphics[width=0.9\textwidth]{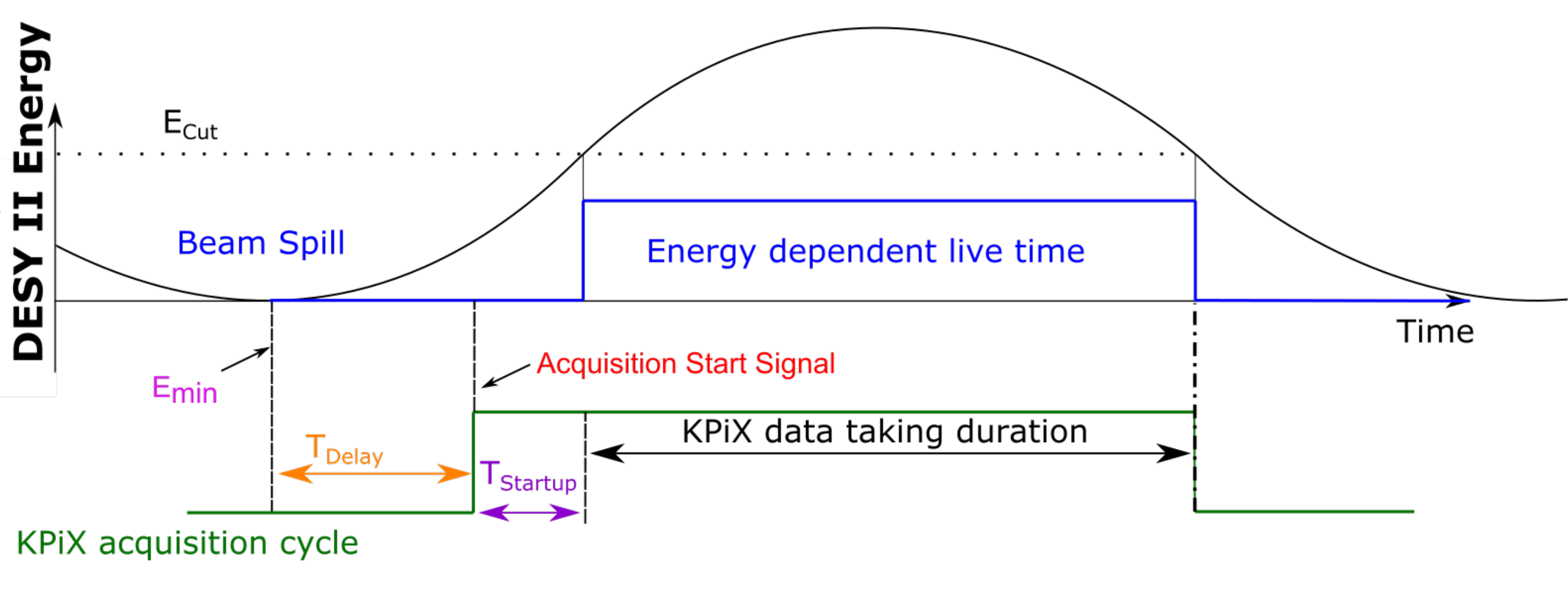}
\caption{\label{fig:installation:desyii} Synchronization of \LYCORIS with the \DESYII accelerator cycle.}
\end{center}
\end{figure}

The beam at the \DIITBF is generated using a carbon fiber, generating bremsstrahlung photons and then making electron-position pairs at a secondary target.
A threshold on the particle energy $\mathrm{E}_{\rm cut}$ can be applied with a dipole magnet. Depending on the selected $\mathrm{E}_{\rm cut}$, beam will only be available if
the \DESYII beam energy is exceeding the selected particle energy~\cite{desytb2018}, as shown in Figure~\ref{fig:installation:desyii}.

Therefore, to take data efficiently with the \KPIX acquisition cycle, the \KPIXAcqstart (as described in Section~\ref{sec:hardware:kpix}) needs to be timed correctly that data
taking commences after the \DESYII beam energy exceeding $\mathrm{E}_{\rm cut}$.
The $\mathrm{E}_{\rm min}$ signal from \DESYII is used as a reference to generate the \KPIXAcqstart by applying a configurable delay timer $\mathrm{T_{Delay}}$,
and the fixed $\mathrm{T_{Startup}}$ time of \KPIX ASICs needs to be extracted in the calculation:
\begin{equation*}
 \mathrm{T_{Delay}(E_{cut})} = \frac{\mathrm{T_{DESY\;Magnet\; Cycle}}}{2\pi} \left(\arcsin\left(\frac{\mathrm{2E_{cut}-E_{max}+E_{min}}}{\mathrm{E_{max}-E_{min}}}\right)+\frac{\pi}{2}\right) -\mathrm{T_{Startup}}
\end{equation*}

This operation ensures that \LYCORIS is always ready for data taking right after particles with energy above the selected energy $\mathrm{E}_{\rm cut}$ start to arrive.
The Data Acquisition period then lasts until the beam energy is below the threshold again, upon which the digitization and readout of all data take place.
A sketch of the complete synchronization scheme is shown in Figure~\ref{fig:installation:desyii}.

\subsubsection*{Setup outside of the \PCMAG} 
\begin{figure}[htbp]
  \begin{center}
    \includegraphics[width=0.475\textwidth]{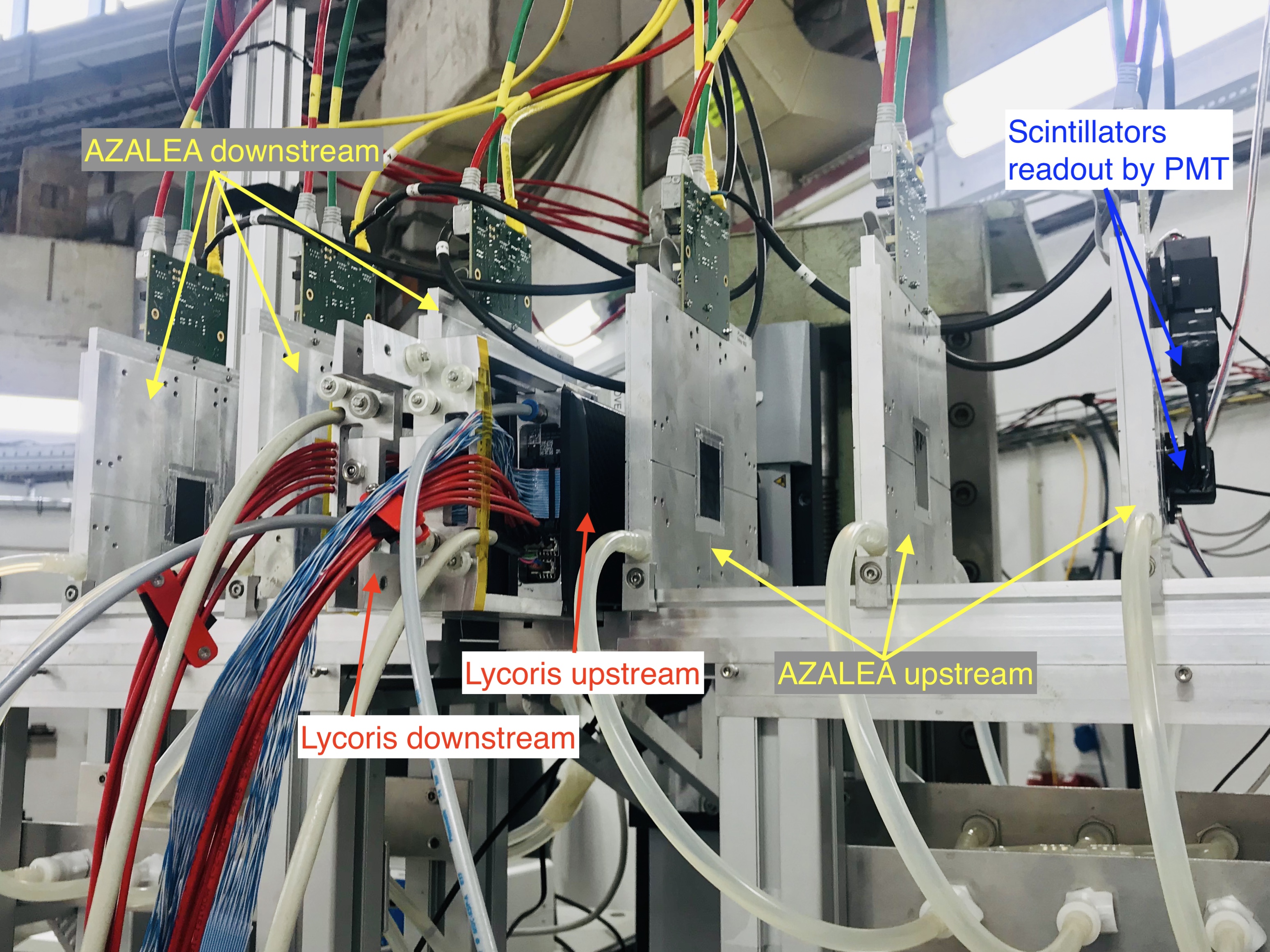}
    \includegraphics[width=0.475\textwidth]{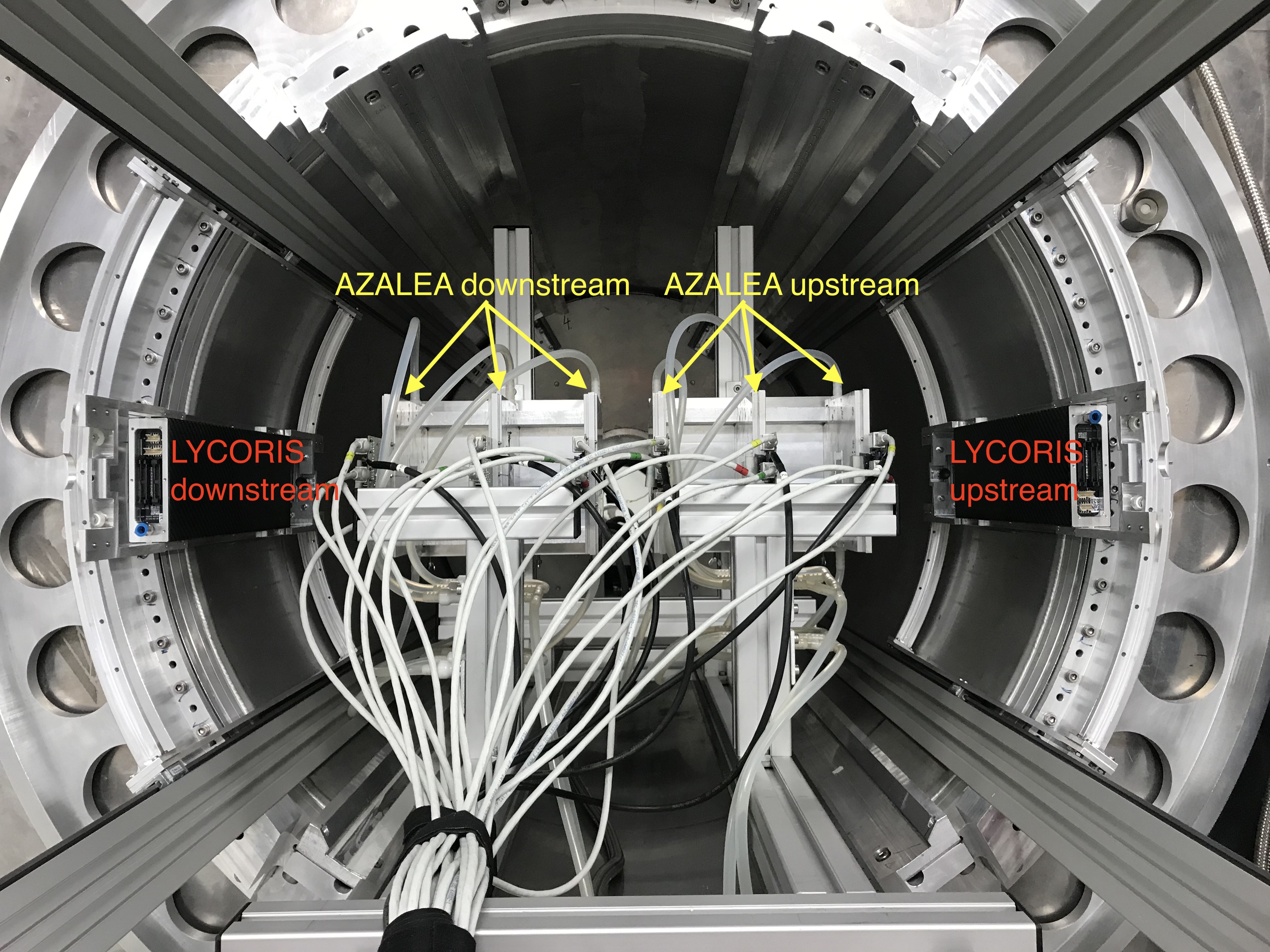} 
  \caption{\label{fig:performance:tbT24} Setup for \LYCORIS tracking performance measurements with the \AZALEA telescope outside the \PCMAG solenoid at \DIITBF in March 2020 (left)
  and inside the \PCMAG with the \AZALEA telescope as a DUT in the center (right).}
\end{center}
\end{figure}

The performance results shown in this paper are based on the latest data collected outside \PCMAG in 2020 March. \LYCORIS was installed next
to each other in between the up- and downstream arms of the \AZALEA telescope, and with a close distance to the nearest \AZALEA plane see
Figure~\ref{fig:performance:tbT24} (left).
Data was taken with the beam energy threshold set at \SI{4.4}{\GeV}; trigger coincidence was provided by two crossing scintillators
located closely in front of the first \AZALEA plane; the beam collimator is set to $x \times y=$\SI{20}{\milli\metre}$\times$\SI{10}{\milli\metre}
and the distance from the beam collimator to the first scintillator is about \SI{220}{\centi\metre}.
The trigger rate under this configuration is high enough to fill all the four memory buckets of each \KPIX channel at every acquisition cycle.

\subsubsection*{Setup inside the \PCMAG} 
\LYCORIS was installed on its dedicated rail system with \AZALEA telescope inside the \PCMAG solenoid in TB24/1, see Figure~\ref{fig:performance:tbT24} (right).
In this configuration, \AZALEA is used as a DUT for \LYCORIS and is installed in between the \LYCORIS up- and downstream cassettes.
Data was taken with the magnetic field set to \SI{0.9}{\tesla} and a beam energy threshold at \SI{4.4}{\GeV}.
Three crossing scintillators are placed right after the beam collimator to provide coincidence, which is about \SI{200}{\centi\metre} away from the solenoid wall.
As TB24/1 is a subsequent area, its beam collimator is about \SI{1157}{\centi\metre} far from the primary beam collimator.
This results in a relatively low particle rate in TB24/1 with trigger coincidences varying from zero to four in one \KPIX acquisition cycle.

\subsection{Sensor Performance}
\begin{figure}[htbp]
  \begin{center}
  \includegraphics[width=0.49\textwidth]{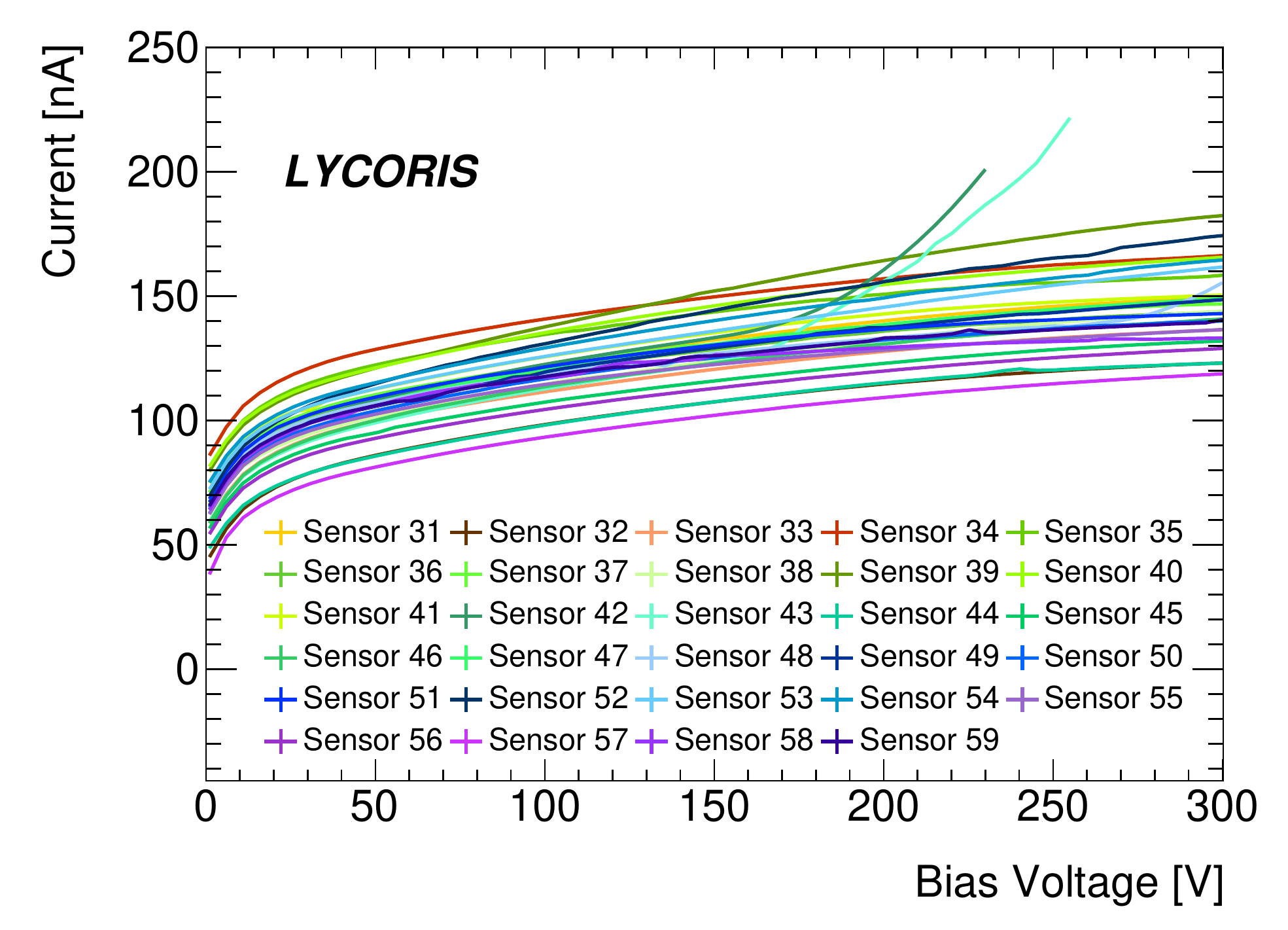}
  \includegraphics[width=0.49\textwidth]{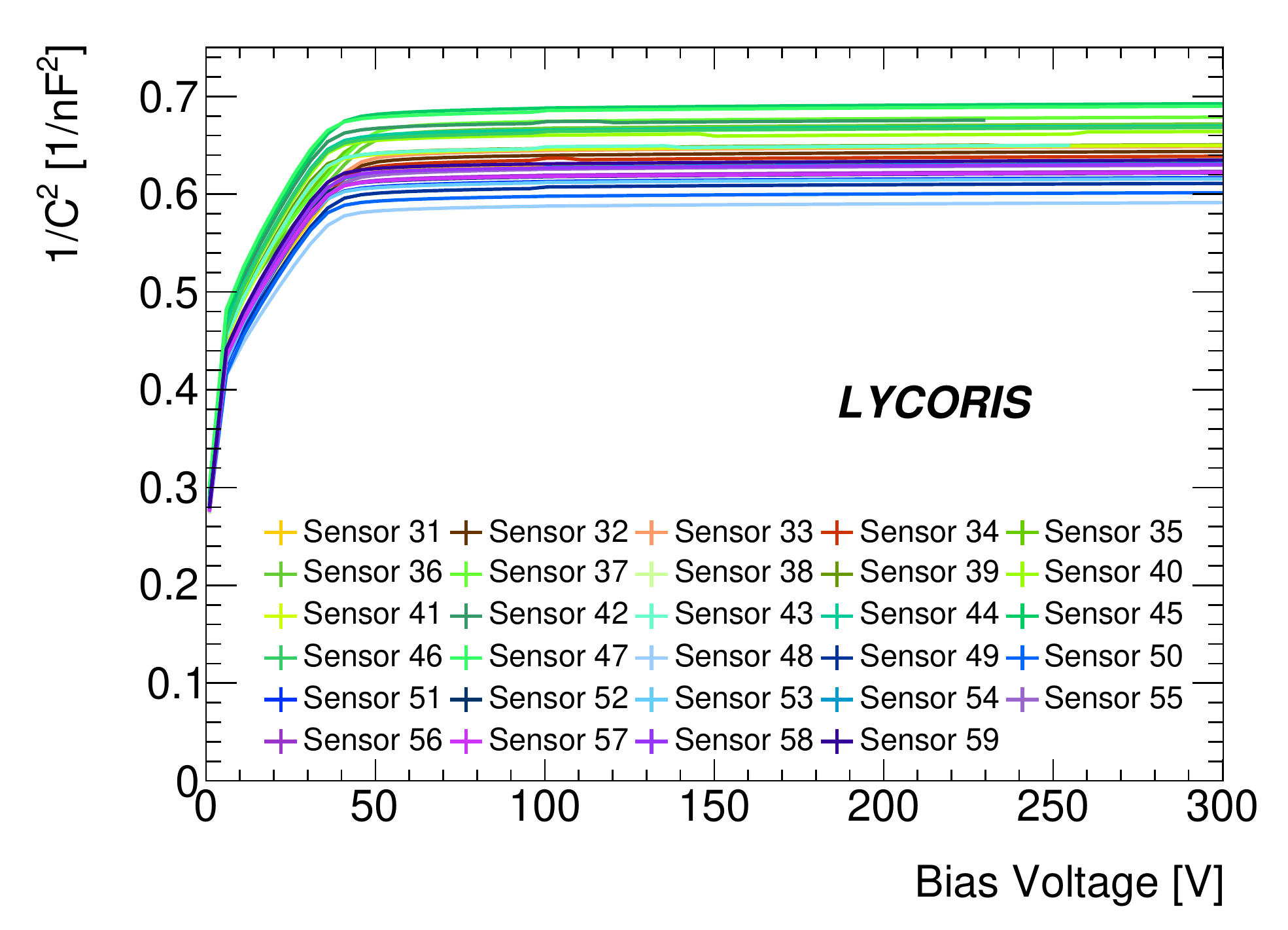}
  \caption{\label{fig:performance:ivcv}The IV (left) and CV (right) curves of all the bare sensors produced.
  6 sensors, Sensor 59, Sensor 43, Sensor 40, Sensor 48, Sensor 47 and Sensor 46, forming an active area of 10$\times$\SI{10}{\square\centi\metre} are used in studies in this paper.
  }
\end{center}
\end{figure}

Figure~\ref{fig:performance:ivcv} shows the measured electric properties of all 29 bare sensors using a probe station.
The measured IV curves (left) show that the dark current level of all sensors is very good, about 100 to \SI{150}{\nano\ampere}.
The backplane capacitance is measured through the bias ring with an LCR meter at \SI{10}{\kilo\hertz} AC frequency and the  measured CV curve (see Figure~\ref{fig:performance:ivcv} (right))
determines the depletion voltage that is at around \SI{50}{\volt} for all the sensors.

\begin{figure}[htbp]
  \begin{center}
  \includegraphics[width=0.49\textwidth]{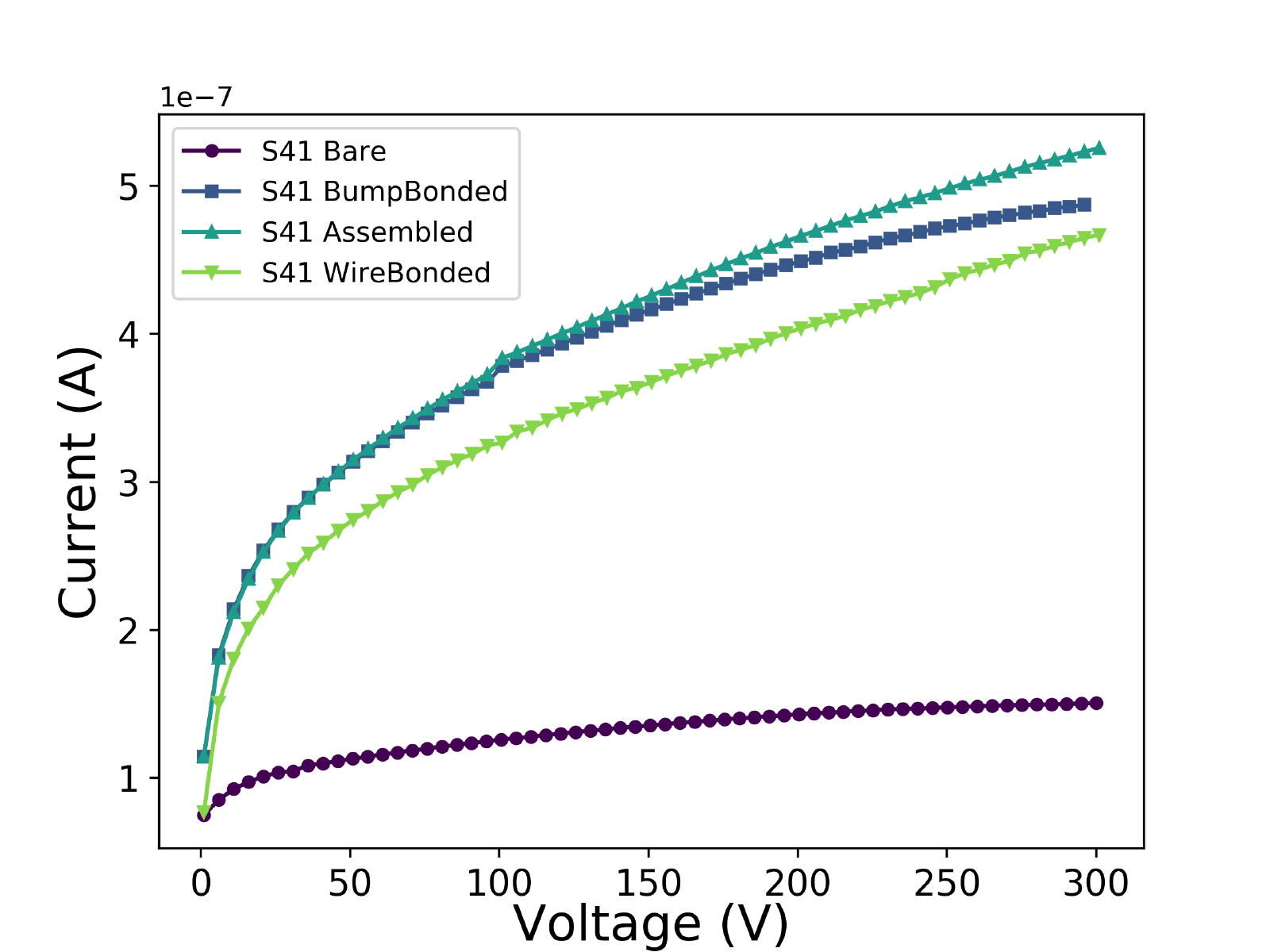}
  \caption{\label{fig:performance:ivassembly}The IV curves of Sensor 41 at various assembly steps, where the "Assembled" refers to the assembly step after gluing the flex onto the sensor surface.}
\end{center}
\end{figure}

The assembly quality is controlled by measuring the IV curve of the sensor after every major assembly step (see Section~\ref{sec:hardware:moduleassembly}),
namely after bump-bonding, after gluing the flex onto the sensor surface, and after wire-bonding.
22 sensors are successfully, fully assembled, and all except for two are showing an expected development.
Figure~\ref{fig:performance:ivassembly} gives one example from the IV curves measured for the Sensor 41 during assembly.
Overall, the dark current at each step stays at the same order of magnitude as the bare sensor, confirming no damage occurred to the sensor during the assembly process.
The behaviors of the curves are also expected: the large increase after bump-bonding comes from impurities added from the treatment of the sensor surface during the bump-bonding procedure;
the slight decrease in dark current from the "Assembled" curve (denoted for after the step of gluing the flex onto the sensor suface) to the curve after wire-bonding is understood due to an
improvement of humidity of sensor storage.

The measured bulk capacitance through bias ring of the entire sensor is \SI{1.26}{\nano\farad} on average when fully depleted.
Given the 3679 sense strips are the same in size and are coupled to the backplane in the same way,
the backplane capacitance of each sense strip is therefore \SI{1.26}{\nano\farad}$/3679=$\SI{0.34}{\pico\farad}.
The AC inter-strip capacitance is measured by the producer Hamamatsu when sensor is fully depleted (\SI{60}{\volt}) and the result
on average is \SI{8}{\pico\farad} per strip.
A MIP creates about 3.5 to \SI{4}{\fC} in \SI{320}{\micro\metre} silicon.
According to the sensor design, $20\%$ of the signal charge generated in the intermediate strip goes
to the backplane and the rest are shared by the neighbor strips for readout~\cite{TNelson}.
Assuming signal loss to the backplane is negligible for readout strips and the possibility to hit on a floating strip or a readout strip
is equal, the expected signal charge on average is then 90\% of the MIP charge thus 3.1 to \SI{3.6}{\fC}.

\subsection{\KPIX Calibration Performance}\label{sec:performance:calib}
\begin{figure}[htbp]
\includegraphics[width=0.49\textwidth]{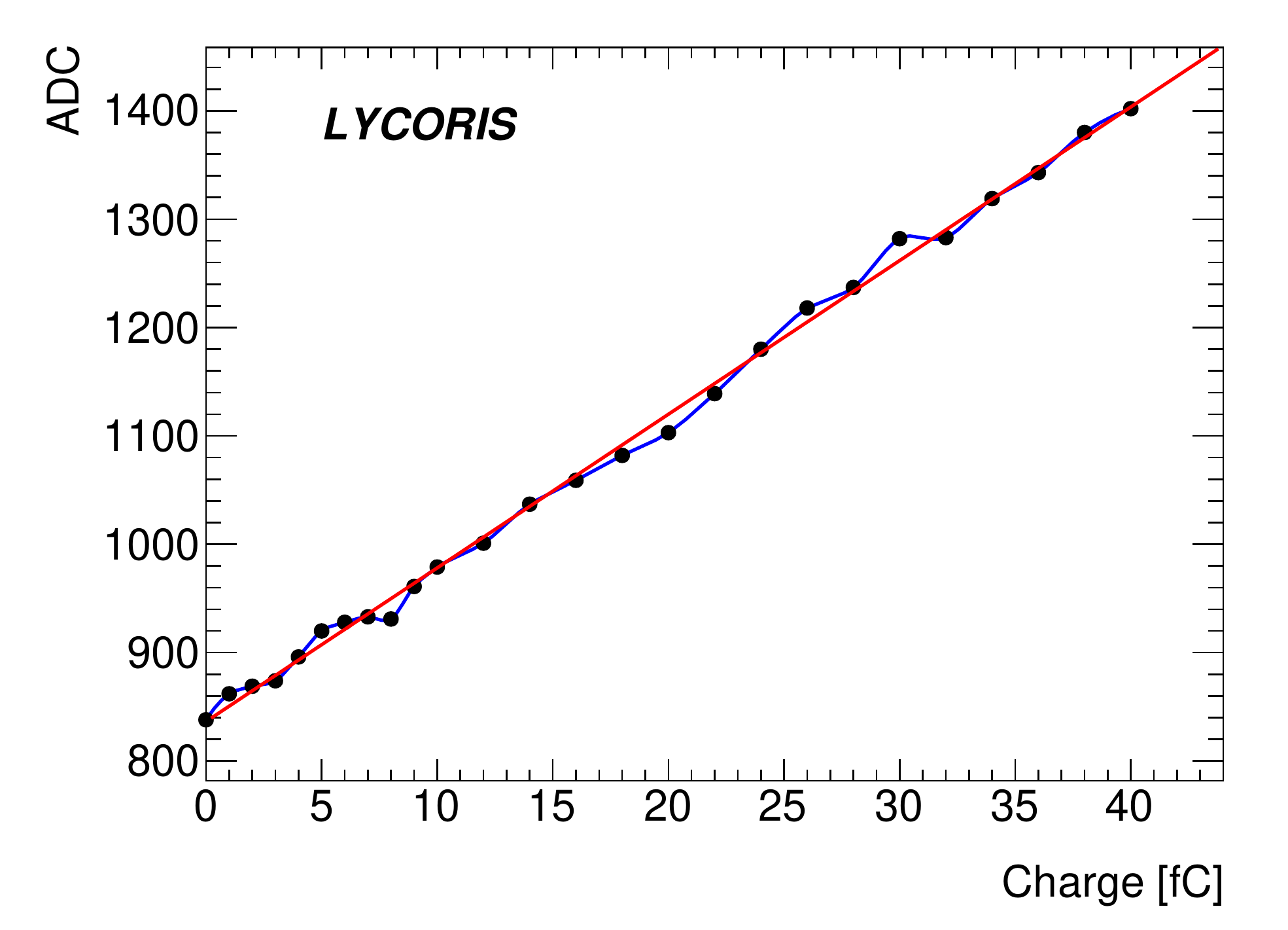}
\includegraphics[width=0.49\textwidth]{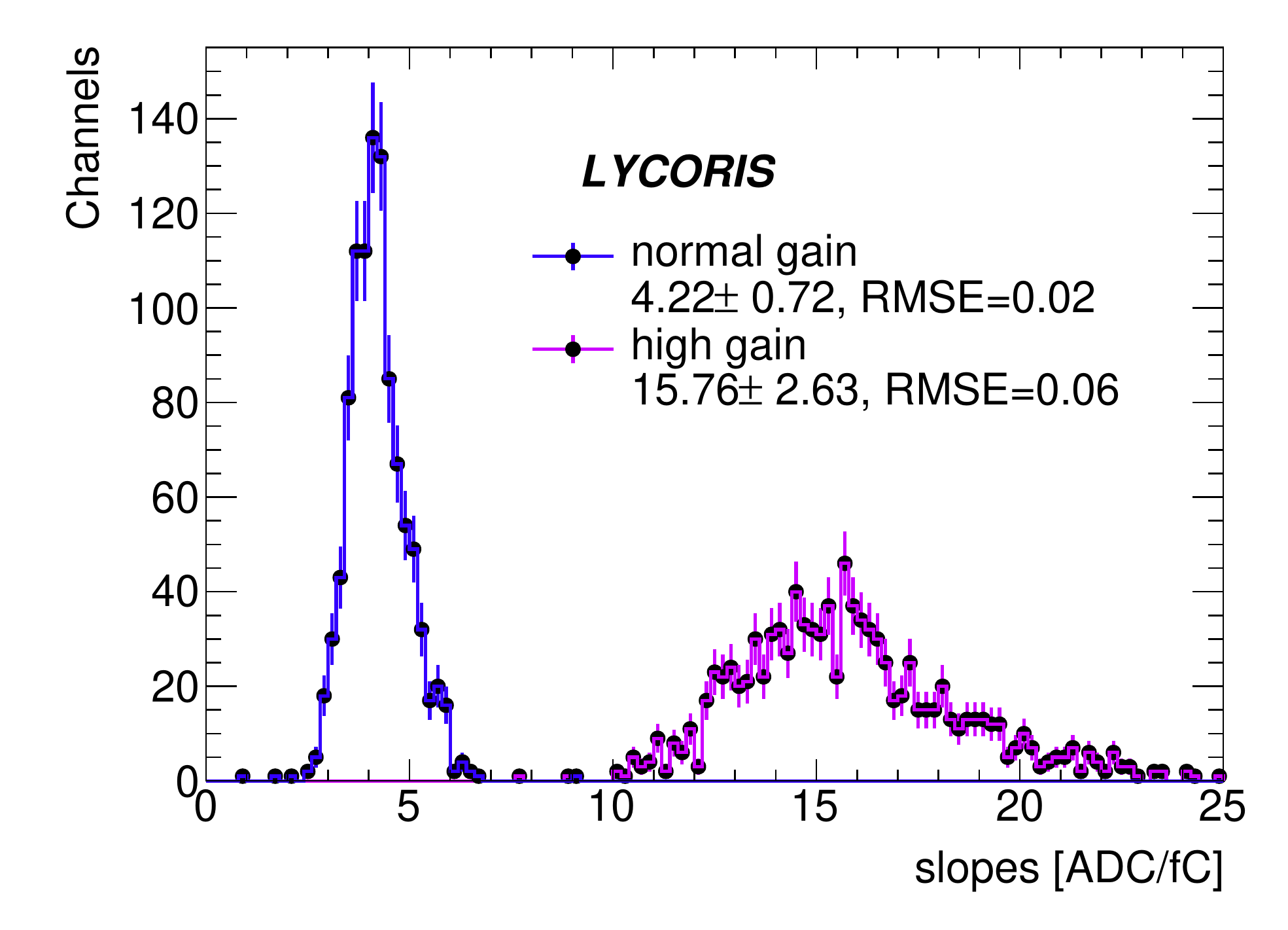}
\caption{\label{fig:performance:calib_1kpix} The ADC calibration curve of one example channel from one \KPIX, a linear fit is performed in red (left).
The slope distribution of the ADC response to the charge injected for all 1024 channels of a \KPIX, the blue curve refers to calibration data in normal-gain mode while
the violet one is using the high-gain mode; the mean value, its RMS and RMS error (RMSE) are given for each curve in the legend.}
\end{figure}

As described in Section~\ref{sec:hardware:kpix:calib}, ADC response of every readout channel is calibrated with an 8-bit DAC.
The calibration data used for studies shown in this paper is taken with the most accurate configuration for the \KPIX calibration system
i.e.\ every available DAC value is measured, corresponding to 26 measurements varying from 0 to \SI{40}{\femto\coulomb} for each channel.
The interval of the 26 measurements is \SI{1}{\femto\coulomb} from 0 to \SI{10}{\femto\coulomb} and later \SI{2}{\femto\coulomb} ($\sim 12500$ electrons) from 10 to \SI{40}{\femto\coulomb}.
The slope required to convert the ADC response to \si{\femto\coulomb} (or number of electrons) can be determined by performing a
linear fit on the ADC response against the DAC input value for each channel, see one example in Figure~\ref{fig:performance:calib_1kpix} (left).
The kinks of single charge injection points reflect natural fluctuations at single measurement level which is not reproducible and expected.
The most important is have the slope reproducible, i.e. stable charge response, which has been confirmed from various calibration data samples.
Figure~\ref{fig:performance:calib_1kpix} (right) shows the distribution of the slopes of all the $1024$ channels from one \KPIX.

\begin{figure}[htbp]
  \begin{center}
  \includegraphics[width=0.45\textwidth]{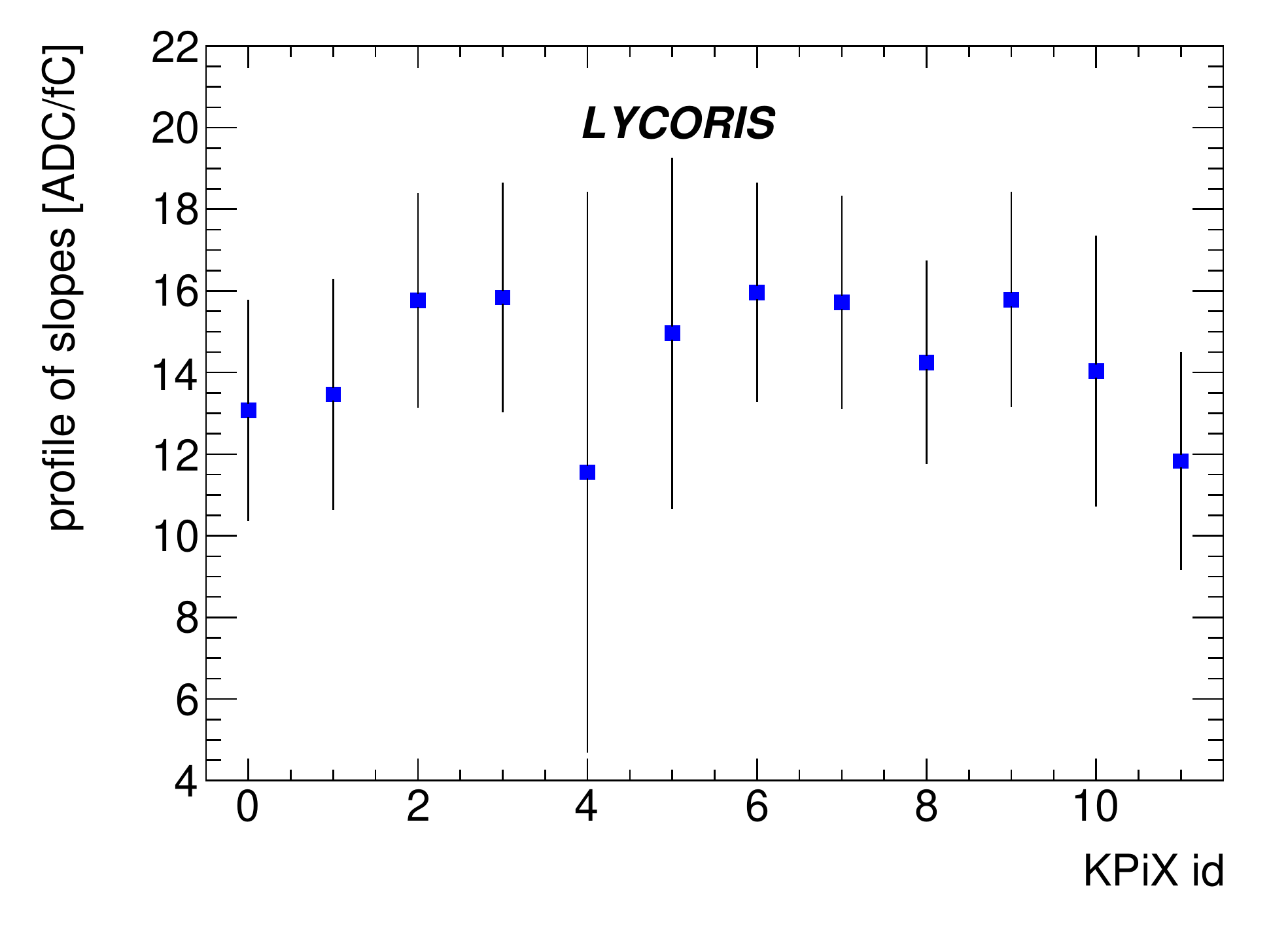}
  \caption{\label{fig:performance:calib_all-kpix} A profile histogram of the slope distribution of all twelve \KPIX ASICs in the high-gain mode.}
\end{center}
\end{figure}

The two distributions shown were taken using the normal-gain and high-gain modes of \KPIX respectively.
The ratio of the mean values of the two curves is around $1:4$ as expected from the designed value of the coupling capacitors shown in Fig~\ref{fig:hardware:kpixchannel}.
The small RMS Error of the slope distributions in Figure~\ref{fig:performance:calib_1kpix} (right) shows a good uniformity of all channels over single \KPIX chip.
Figure~\ref{fig:performance:calib_all-kpix} shows a profile histogram of the slope distribution in the high-gain mode for all twelve \KPIX ASICs used by \LYCORIS,
where eleven out of the twelve chips are showing a very uniform behavior.

\subsection{Module Performance}
\subsubsection{Noise Performance}
In the external trigger mode, \KPIX records the charge response of each channel when the external trigger arrives.
As a result of the short opening window for one trigger and additionally due to the low particle rate at the test beam,
only one to two out of the 1890 readout strips read out signal in every externally triggered event.
As such, the majority of the \KPIX channels are recording random noise.
In this case, the charge distribution of one entire \KPIX chip is expected to follow a Gaussian distribution centered at zero with outliers coming from signal contributions.
The standard deviation of the Gaussian charge distribution defines the chip's noise level
and it is robustly estimated using the MAD: $\sigma = b\cdot MAD$ with $b = 1.4826$ for Gaussian distributions.

\begin{figure}[htbp]
  \begin{center}
  \includegraphics[height=5cm]{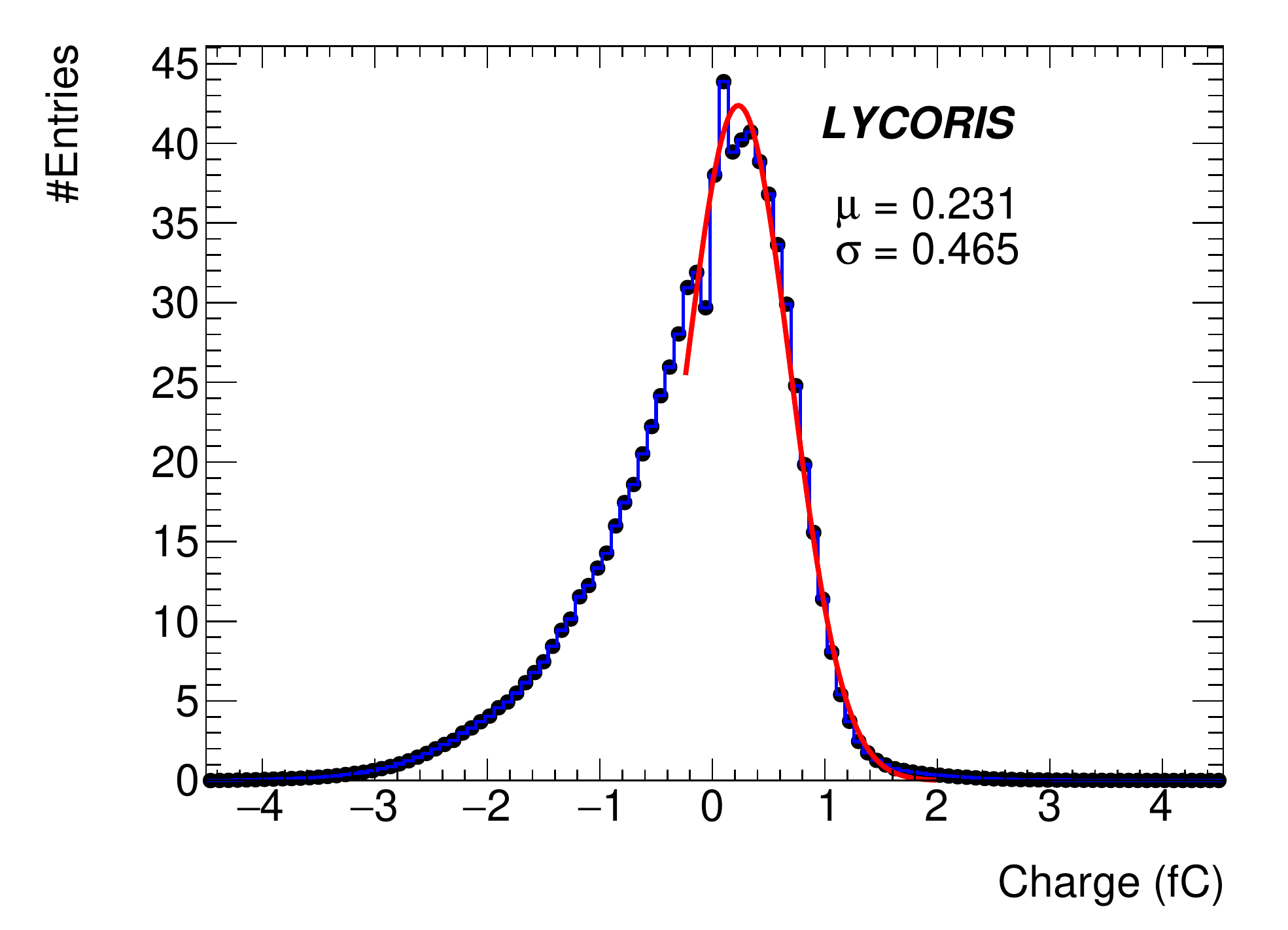} %
  \includegraphics[height=5cm]{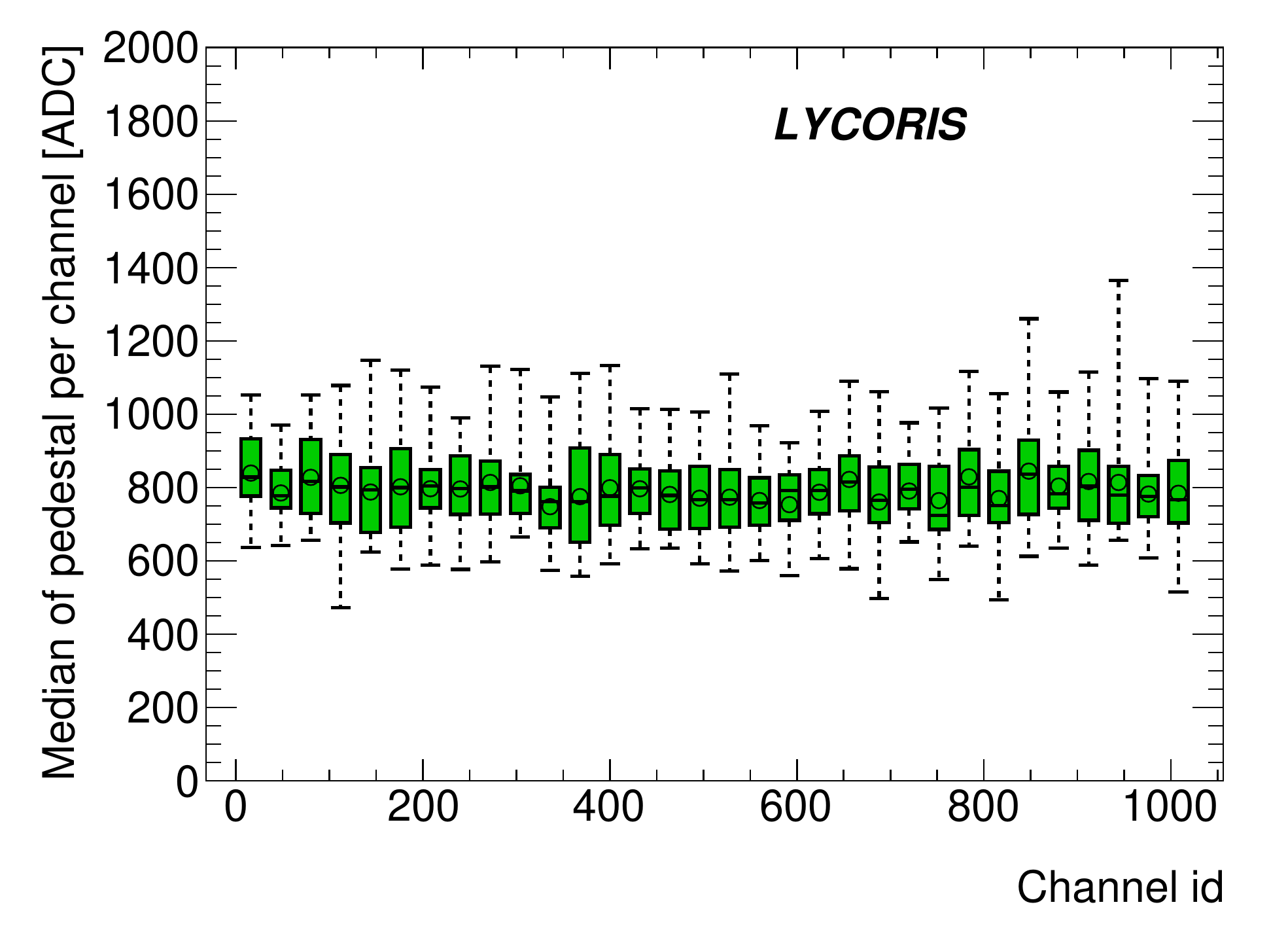}%
  \caption{\label{fig:performance:noise_ped} Charge distribution of one entire example \KPIX chip after the pedestal subtraction (left)
  and its pedestal distribution over its channels in a candle plot (right).}
\end{center}
\end{figure}

As stated in Section~\ref{sec:software:reco}, the baseline noise subtraction involves two steps: the pedestal subtraction and
the common-mode noise subtraction.
Figure~\ref{fig:performance:noise_ped} (left) shows the recorded charge distribution after the pedestal subtraction for one
example \KPIX chip, and its pedestal distribution against the \KPIX channel number is shown using a candle plot in Figure~\ref{fig:performance:noise_ped} (right).
The mean of the charge distribution is close to zero after pedestal subtraction, demonstrating the pedestal subtraction corrects the offset efficiently.
However, the clearly asymmetric shape demonstrates the significant contribution from the event-by-event common-mode noise.

\begin{figure}[htbp]
  \begin{center}
  \includegraphics[height=5cm]{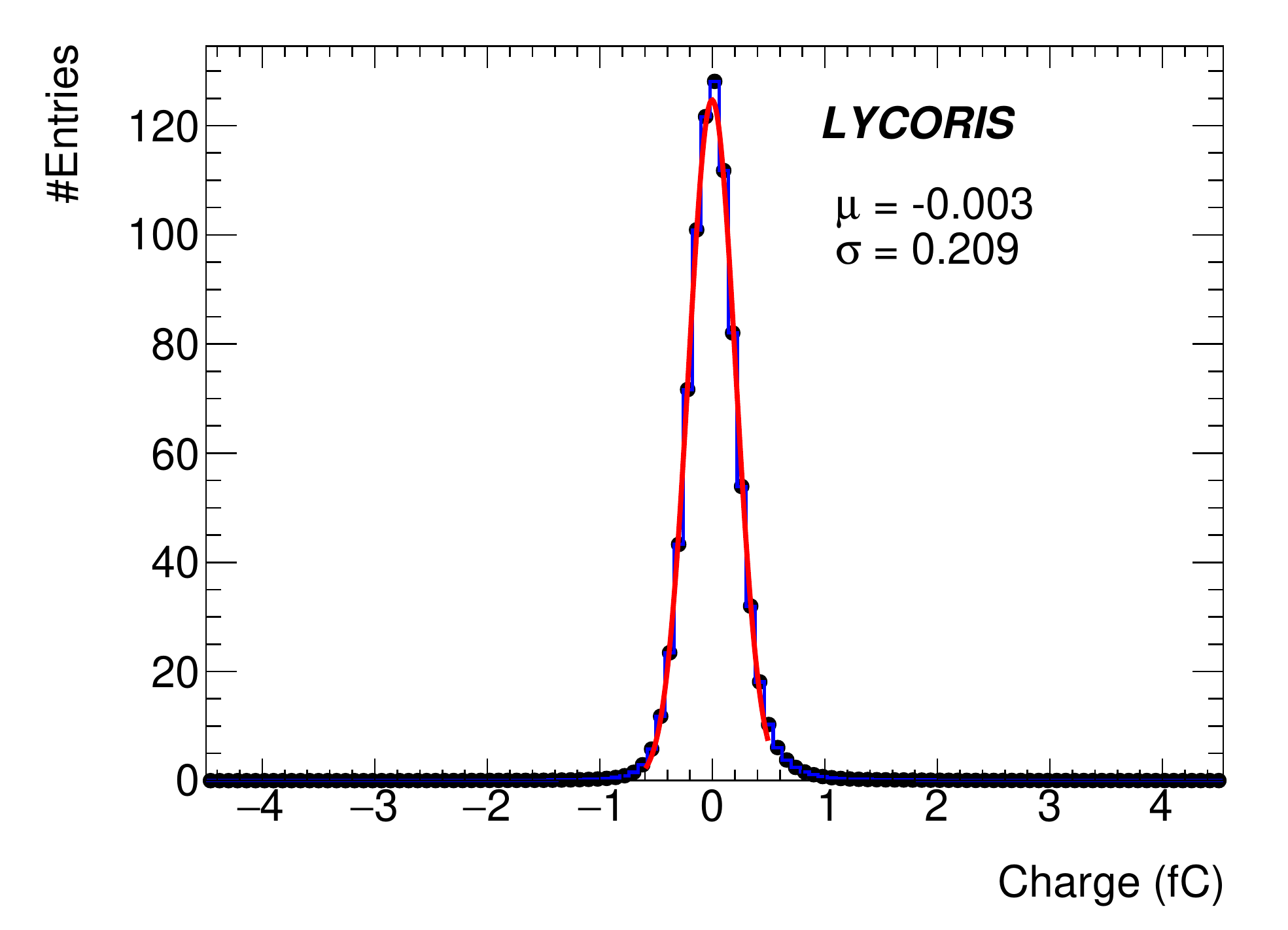}
  \includegraphics[height=5cm]{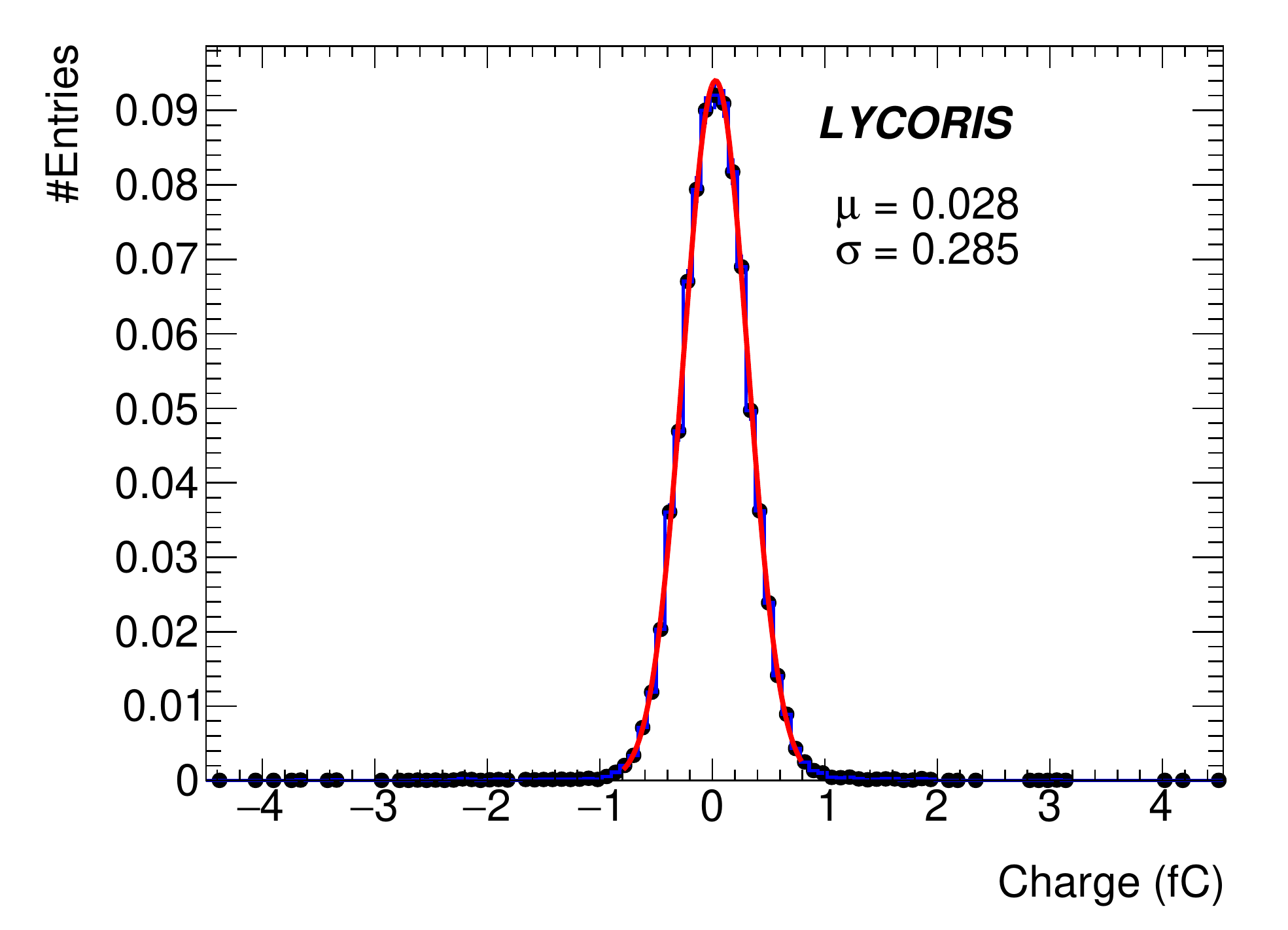} %
  \caption{\label{fig:performance:noise} Charge distribution after common-mode subtraction from one entire example \KPIX chip (left)
  and from one example \KPIX channel that is connected to a strip (right).} %
\end{center}
\end{figure}

After applying the common-mode subtraction, the charge distribution of the example chip shows an expected Gaussian shape with its
mean value pushed closer to $0$ by two orders of magnitude, see Figure~\ref{fig:performance:noise} (left).
Besides, its width is reduced by a factor of more than two because the common-mode noise subtraction further compensate a time
dependent pedestal drift which is found sufficiently uniform over time in each single \KPIX chip.
This pedestal drift is caused by the leakage of the charge stored in the memory buckets during the time between storage and digitization.
Therefore, the amount of the drift varies according to the amount of time that the data is stored in the memory bucket before digitization, i.e. when the data was stored during the acquisition period. As a consequence, this drift is uniformly distributed over time in single \KPIX chip, and thus contributing together with the common-mode noise to the asymmetric shape of the distribution in Figure~\ref{fig:performance:noise_ped} (left).

Figure~\ref{fig:performance:noise} (right) shows the Gaussian charge distribution read out from one example strip, i.e.\ one \KPIX channel
that connects to a strip. Its width of \SI{0.29}{\femto\coulomb} defines its noise level N and with the measured charge value Q the  significance can be
calculated as Q/N which is the input for the subsequent clustering step.

\subsubsection{Cluster reconstruction}
\begin{figure}[htbp]
  \begin{center}
  \includegraphics[width=0.49\textwidth]{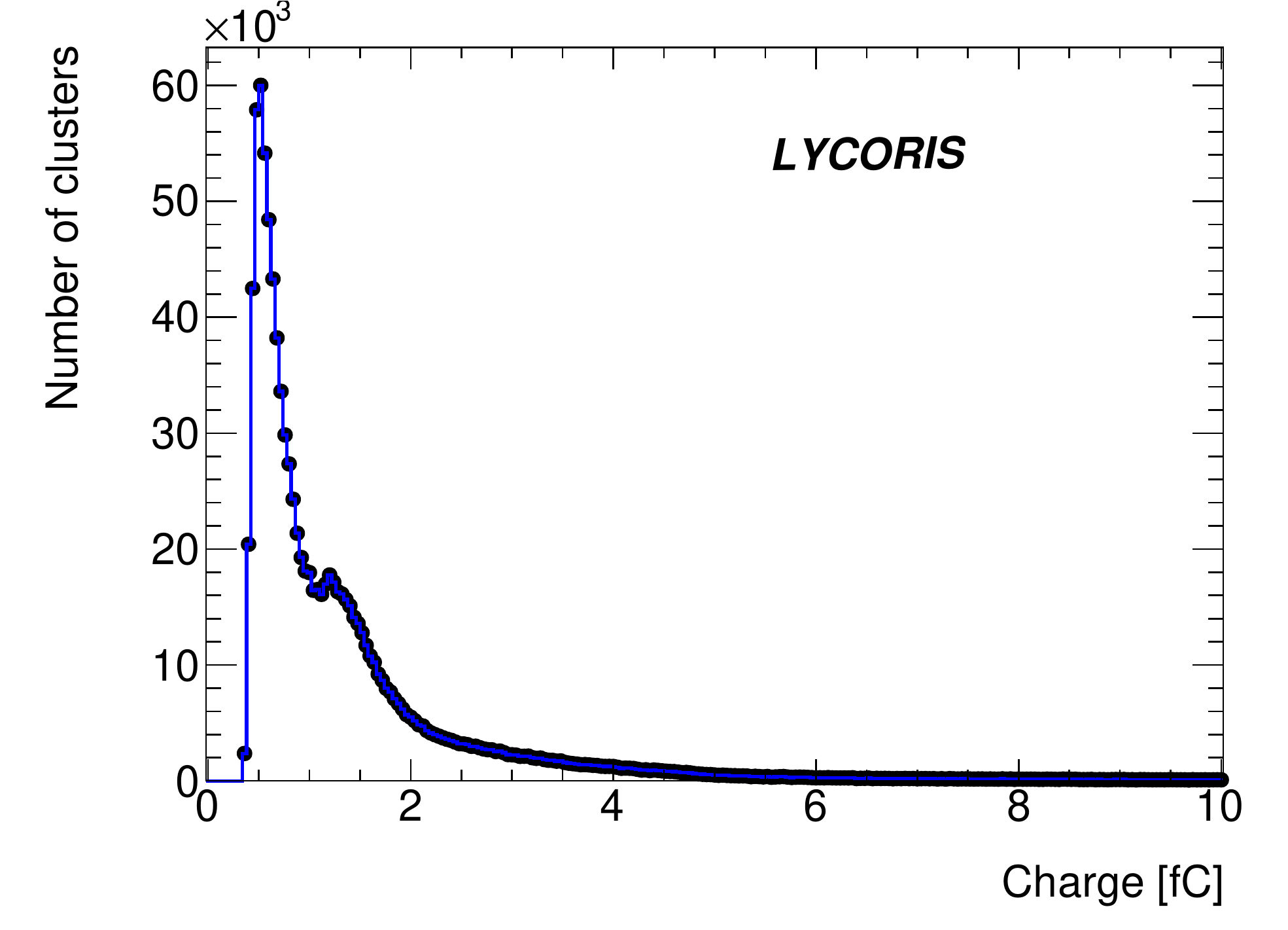}
  \includegraphics[width=0.49\textwidth]{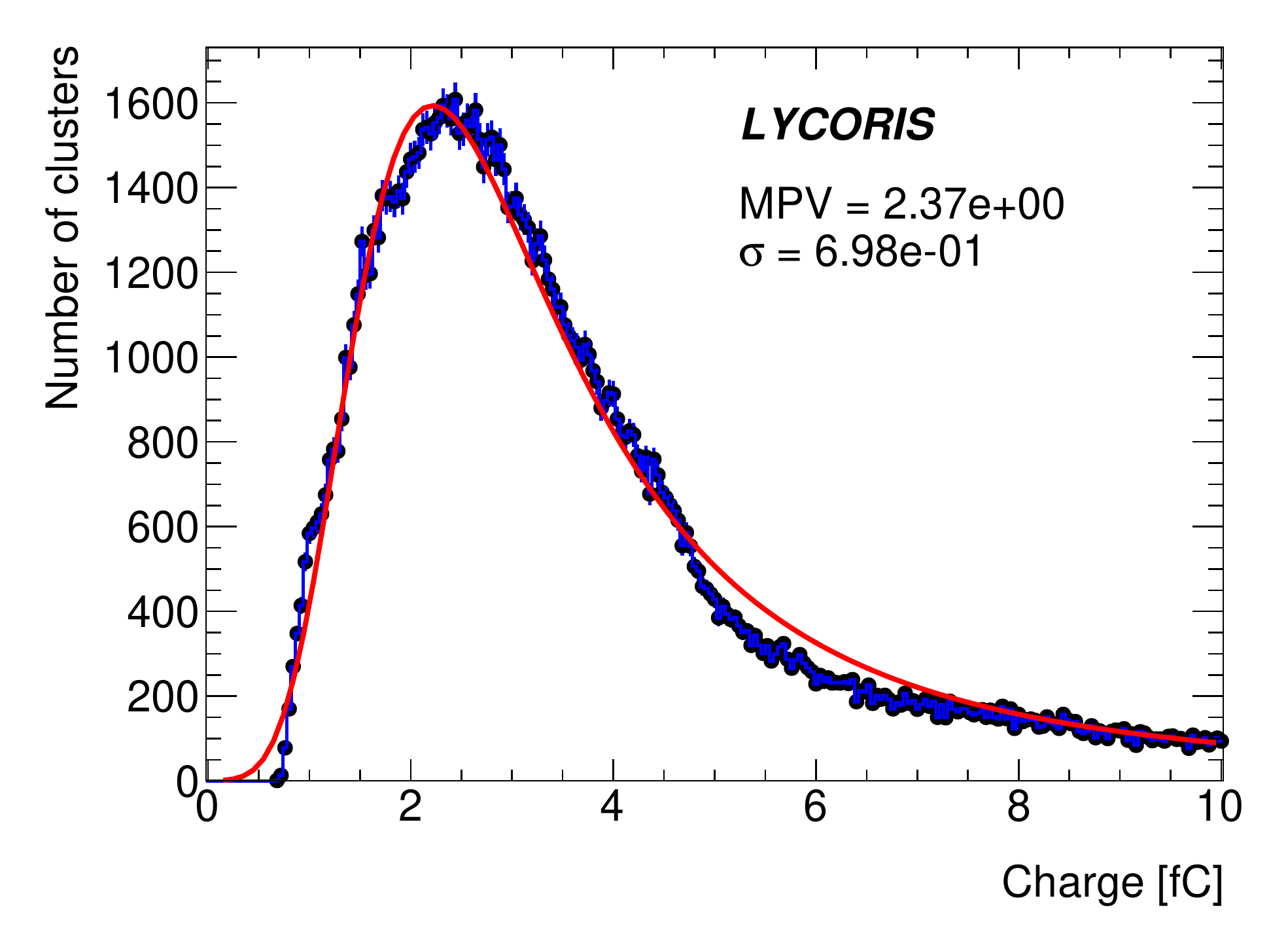}
  \caption{\label{fig:performance:cluster} Charge distributions of the reconstructed clusters from one example sensor before
  (left) and after (right) the selection of $S/N\ge4$. }
\end{center}
\end{figure}

The clusters are the reconstructed objects (see Section~\ref{sec:software:hitclustering}), which serve as the
input hits to the alignment and tracking algorithms for track reconstruction.
Figure~\ref{fig:performance:cluster} (left) shows the charge distribution of all the clusters found after clustering.
A large noise peak is presents at lower cluster charges, which can be suppressed by a cut on the cluster significance.
A signal amplitude of about \SI{3.1}{\femto\coulomb} is expected,
and the measured noise level is found to be $\sim$\SI{0.25}{\femto\coulomb} on average,
hence the signal-to-noise ratio ($S/N$) is expected to be $12:1$.
As only every second strip is read out and 20\% of the signal charge is expected to be lost to the backplane for hits on floating strips, the signal-to-noise ratio is reduced to $4.8:1$.
Therefore, we applied a criterion of $S/N\ge4$ on clusters, which serve as the input hits to the tracking algorithm.
The cluster charge distribution after this selection is shown in Figure~\ref{fig:performance:cluster} (right) demonstrating a good trade-off
between hit statistics and hit purity.

\subsection{\LYCORIS Performance} 

\subsubsection{Tracking with \EUDET telescope} 
In \LYCORIS performance studies, the \EUDET telescope \AZALEA was used to provide a reference track with which \LYCORIS hits can be associated to form a combined track.

Two track finding algorithms are used. They are similar to the ones used for \LYCORIS stand-alone track finding
(see Section~\ref{sec:software}) but specialized for the pixel telescope.
The two algorithms can be called in any order and all hits used to form a track candidate are not considered for the subsequent algorithm.
\begin{description}
  \item[Triplet finder:]
  In each arm of \AZALEA the first and the last layers are used to form doublets along the beam direction by applying a loose selection on slopes on x-z and y-z planes.
  To become a valid triplet, a doublet is interpolated to the middle layer to search for a matched third hit within a short distance.
  All valid triplets inherit the tracking parameters from the doublets and are later extrapolated to the center between the two \AZALEA arms for a match.
  When a triplet from one arm matches to another triplet in the other arm, requiring a small deviation in both x and y as well as in the track
  angle in x and y, then a valid triplet track has been found. Same as for the \striplet finder, a double uniqueness is required, meaning triplets
  from the same doublet but different third hits will be rejected here.

  \item[Road search:]
  The road search first forms track roads along the beam direction using the first and the last layers of the both arms.
  If there are at least two other uniquely matched hits found within a short distance to the track road, i.e.\ at least four
  hits can be associated with this track, then a valid track candidate has been reconstructed.
\end{description}

For the combined track finding, the road search is used to associate the \LYCORIS hits to the track road defined by the \AZALEA track candidate.
One \AZALEA track can be associated with up to six \LYCORIS hits that each is a unique match in a short distance.

\subsubsection{Track Hit Performance}
\begin{figure}[htbp]
  \begin{center}
  \includegraphics[height=5cm]{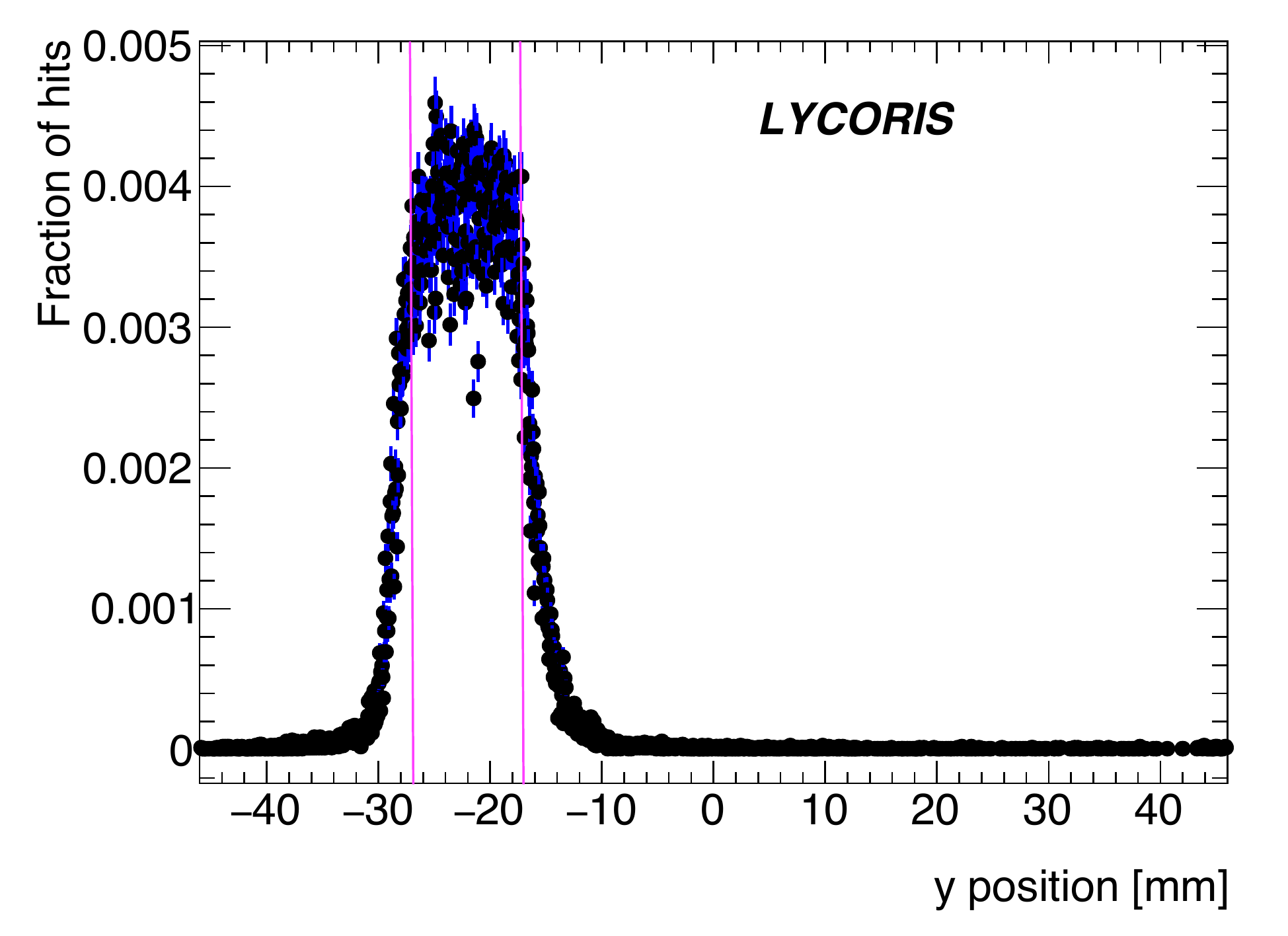}
  \includegraphics[height=5cm]{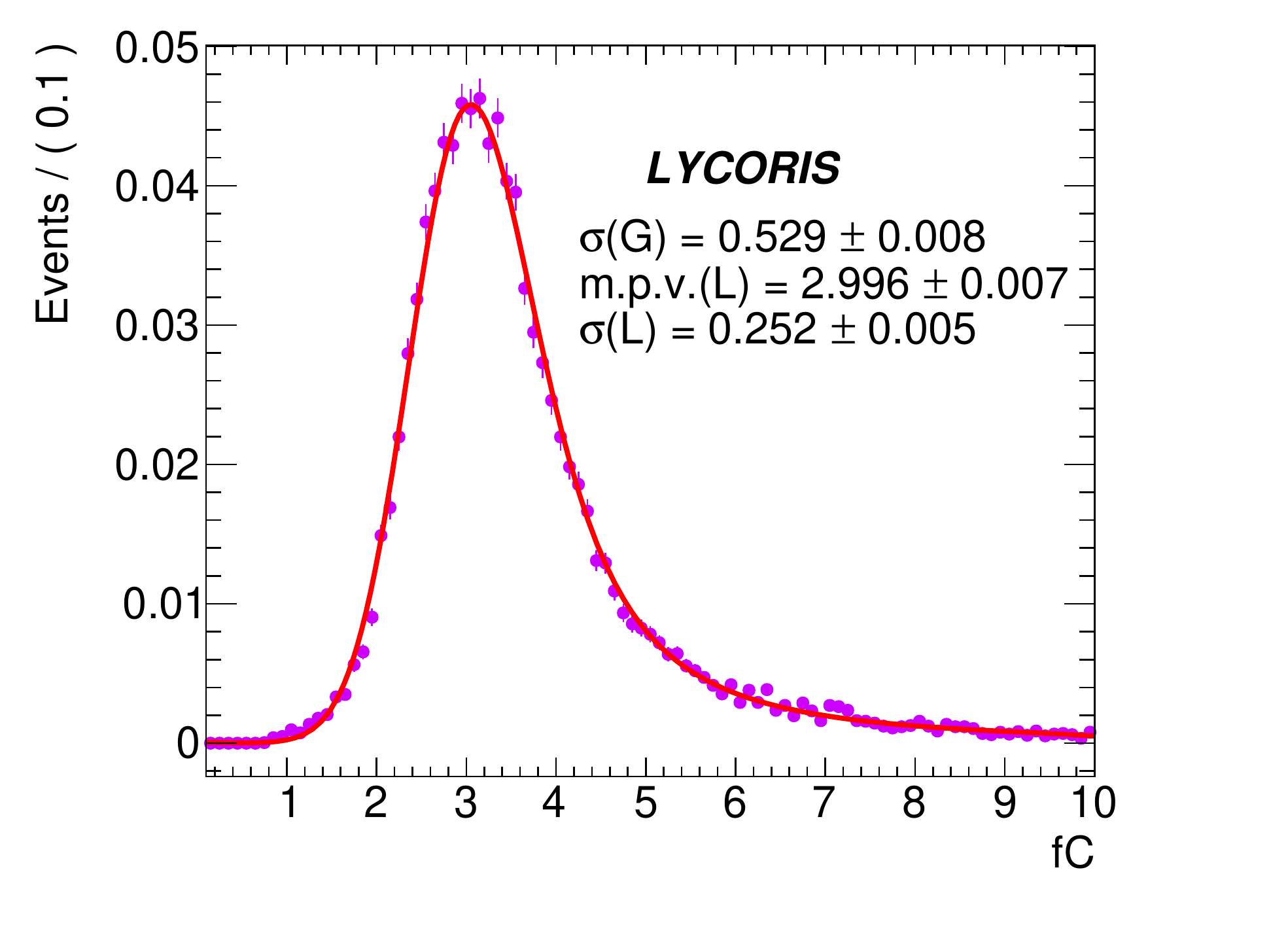} %
  \caption{\label{fig:performance:ypos-landau} The hit position in the local Y-axis of all six \LYCORIS layers according
  to data taken with a beam collimator of $x\times y=$\SI{20}{\milli\metre}$\times$\SI{10}{\milli\metre} (left), where the pink lines indicates the projection of the beam collimator size.
  The normalized charge distribution of all the hits on one example sensor plane associated to stand-alone \LYCORIS tracks, a Landau convoluted
  Gaussian fit is performed giving the Landau most probable value of \SI{3.0}{\femto\coulomb}.
}
\end{center}
\end{figure}

Track hits are the hits associated to a reconstructed track.
The track hit performance shown here is using the \LYCORIS stand-alone tracking algorithm described in Sec~\ref{sec:software}.
Figure~\ref{fig:performance:ypos-landau} (left) shows the position of hits in local $y$ coordinate, which shows a clear 1D profile of the beam spot at the \DIITBF.

\paragraph{Signal Amplitude}
Figure~\ref{fig:performance:ypos-landau} (right) shows the charge distribution of hits from one example sensor plane associated to stand-alone \LYCORIS tracks.
A Landau convoluted Gaussian fit is performed giving the Landau most probable value of \SI{2.9}{\femto\coulomb} and this is the measured signal amplitude for this sensor.
The measured signal amplitude is consistent with the basic expectation on a MIP in this \SI{320}{\micro\metre} thick sensor read by every other strip.

\begin{figure}[htbp]
  \begin{center}
  \includegraphics[width=0.45\textwidth]{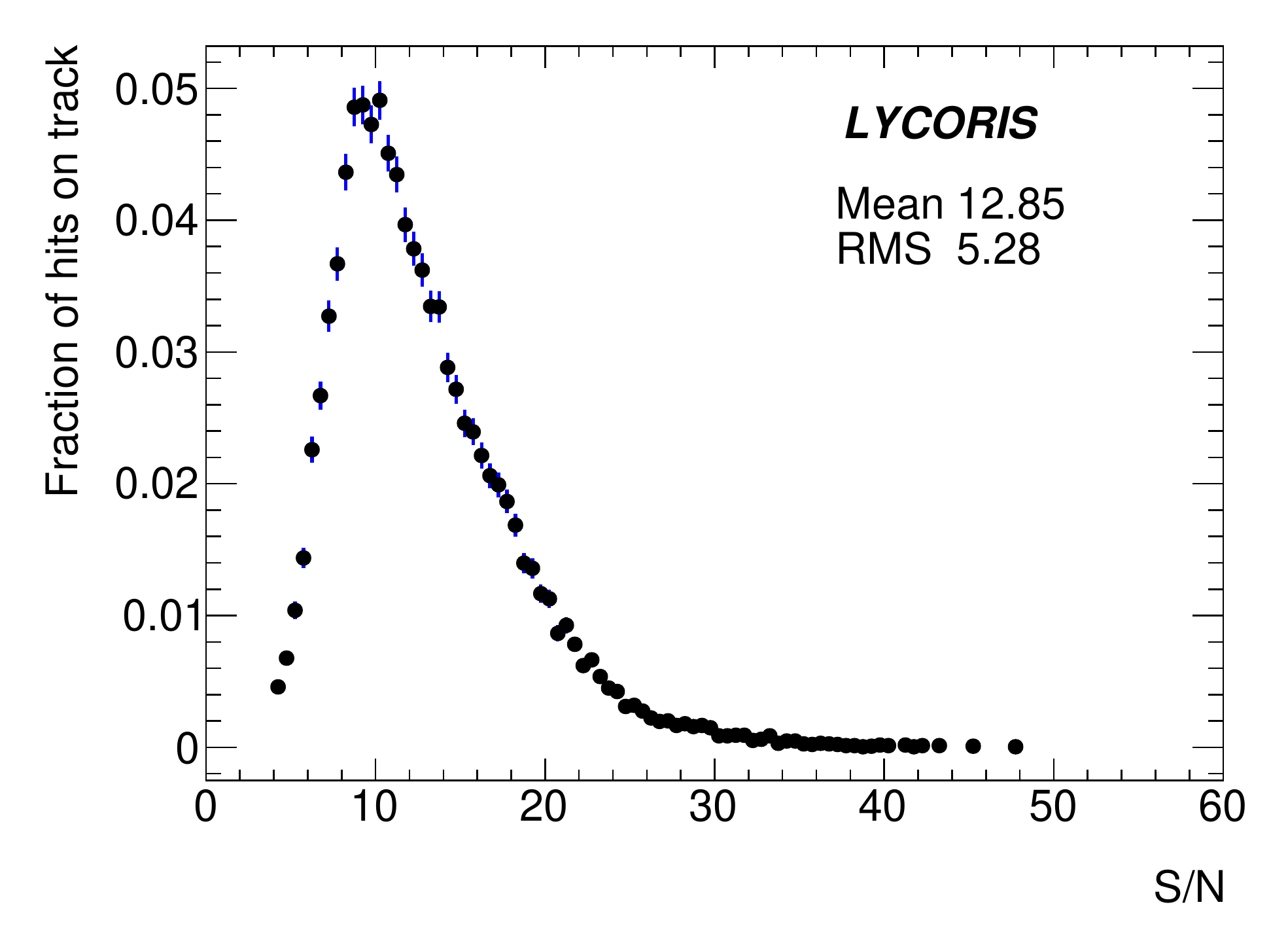} %
  \caption{\label{fig:performance:snr} The signal-to-noise ratio distribution of the hits on track from one sensor.}
\end{center}
\end{figure}

\paragraph{Signal-to-Noise Ratio}
The signal-to-noise ratio is calculated for each hit cluster that is associated to a track, and its distribution for
one of the \LYCORIS sensors is shown in Figure~\ref{fig:performance:snr}.
The cut-off to the left is due to the quality selection of the clusters, that are used as input to the tracking algorithm
while the majority of clusters are centered the mean value of $14.4$ giving a
good estimate of the signal-to-noise ratio.

\begin{figure}[htbp]
  \begin{center}
    \includegraphics[width=0.49\textwidth]{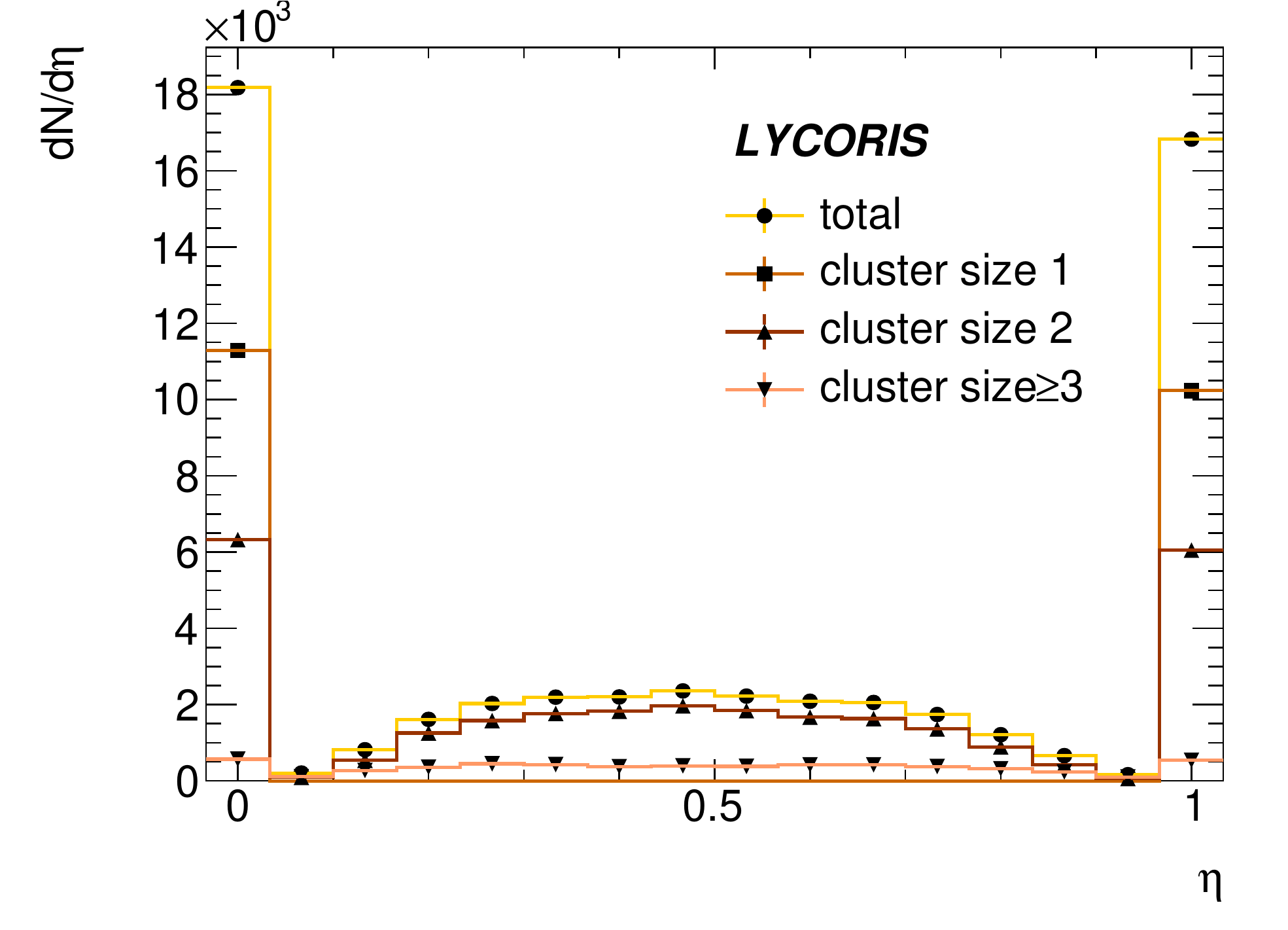} %
  \caption{\label{fig:performance:eta} $dN/d\eta$ of hits from all six \LYCORIS layers.}
\end{center}
\end{figure}

\paragraph{Charge Sharing}%
The charge sharing has been studied with two methods in this paper.
First, the spatial distribution of charge collected by the two neighbor readout strips is studied with the $\eta$ observable~\cite{etaSiDet:1983}:
\begin{equation}
\eta = \frac{Q_R}{Q_L+Q_R},
\end{equation}
where $Q_L$ and $Q_R$ are charge collected by the two neighbor strips L and R respectively.
A group of \LYCORIS hits with particles traversing between strips L and R are selected by demanding that its associated \AZALEA track projection is
between strips L and R. Figure~\ref{fig:performance:eta} shows an example $dN/d\eta$, where contributions from different hit cluster
sizes are overlaid.
The $\eta$ distribution for single strip hit (cluster size 1) shows almost no charge sharing between two readout strips. This is expected as the
diffusion is about a quarter of the readout pitch (half of the sense pitch).
Charge sharing between readout strip and the floating strip is shown by the $\eta$ distribution for hits composed by two strips (cluster size 2).
$\delta$ electron contributions is described by the curve for hits consisting of more than two strips.

\begin{figure}[htbp]
  \begin{center}
  \includegraphics[width=0.49\textwidth]{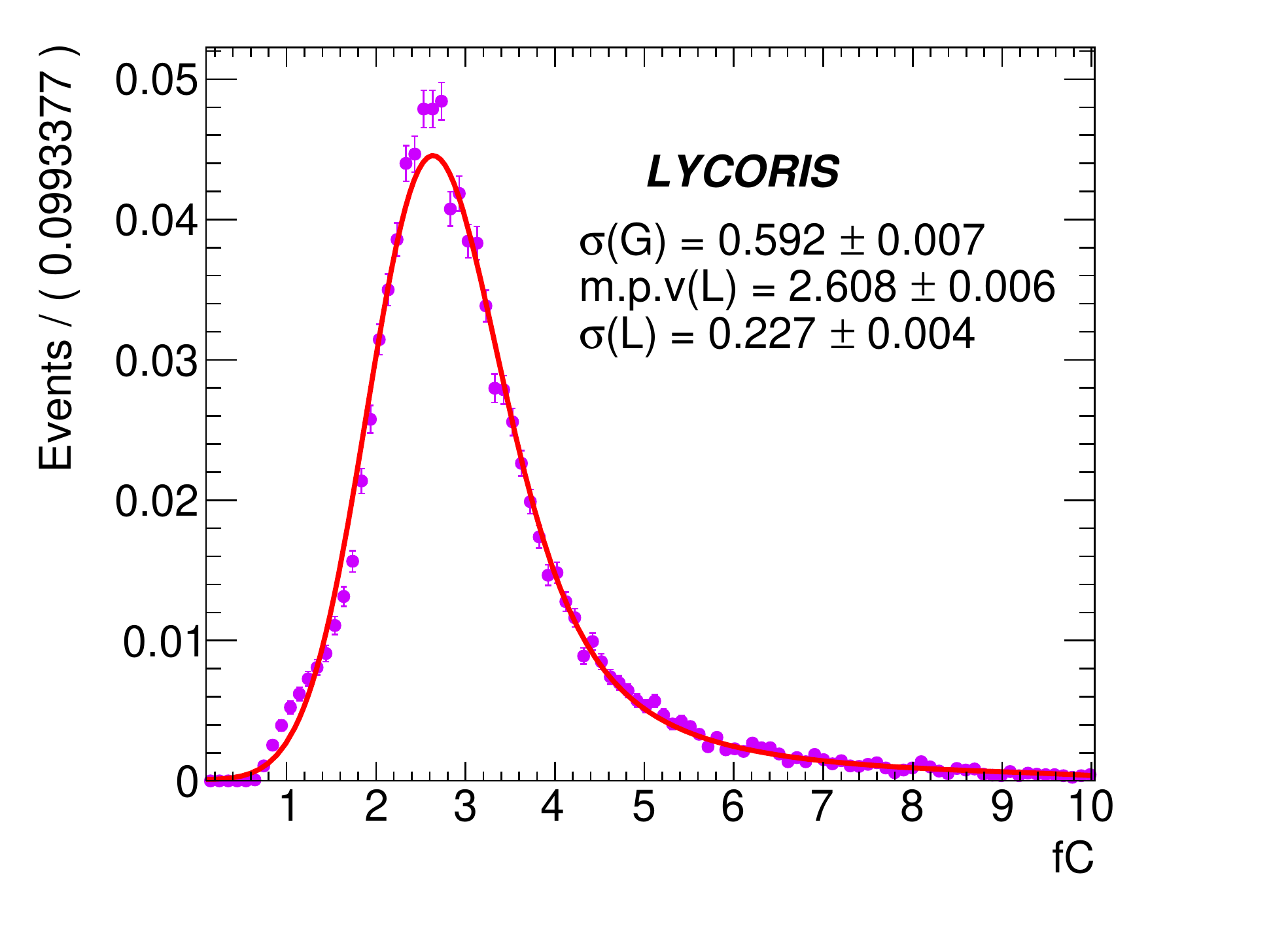}
  \includegraphics[width=0.49\textwidth]{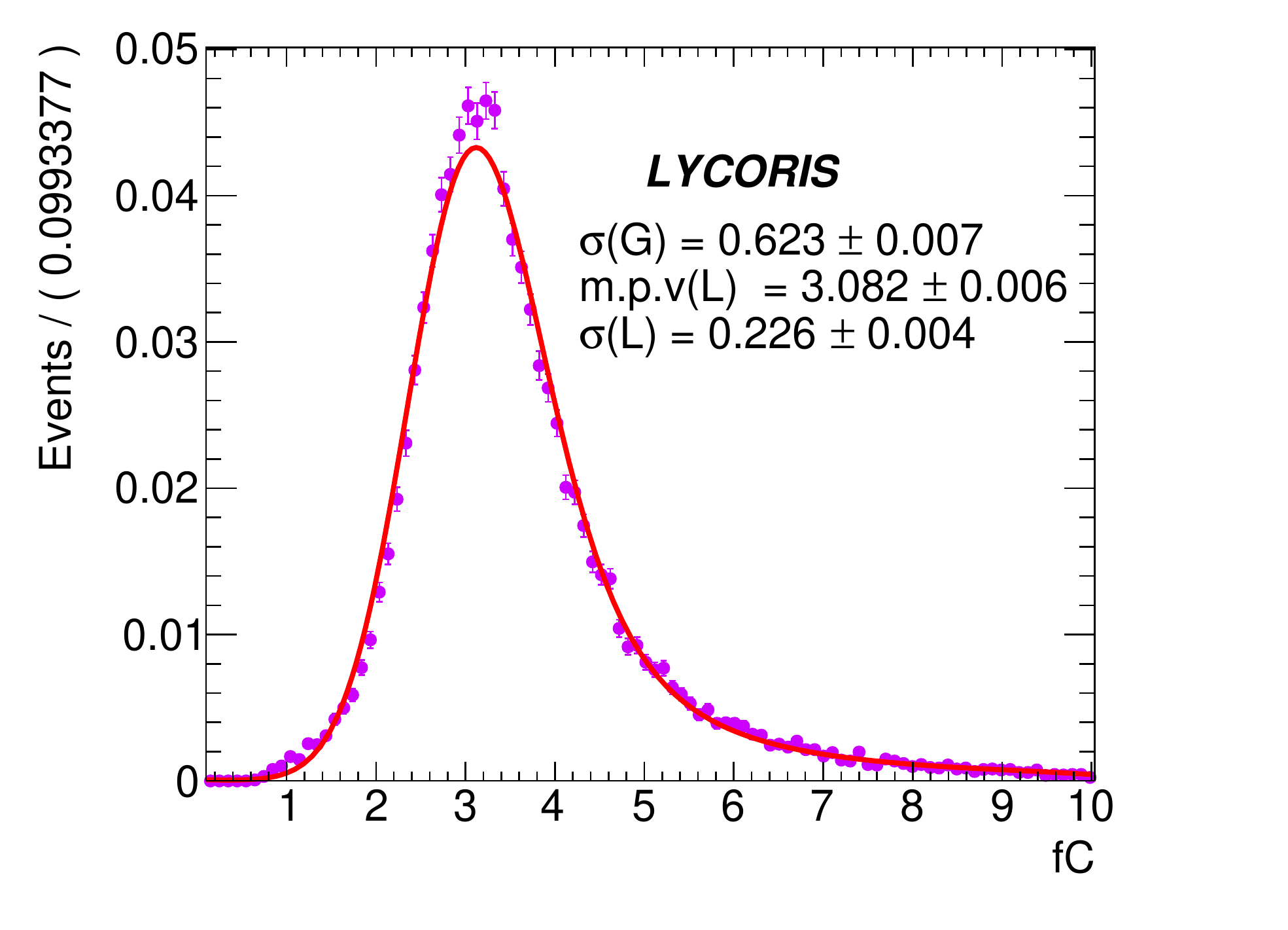}
\caption{\label{fig:performance:landau_sub} The charge distributions of \LYCORIS hits generated from floating strips (left) and
from readout strips (right) and the classification on each hit uses its associated \AZALEA track projection.}
\end{center}
\end{figure}

Secondly, the charge sharing is studied by comparing signal amplitudes of hits from floating strips to hits from readout strips.
The \LYCORIS hit is classified as hit generated from a readout strip if its associated \AZALEA track projection is within one pitch
(\SI{25}{\micro\metre}) window centered by a readout strip.
Figure~\ref{fig:performance:landau_sub} shows the charge distributions of hits generated from floating strips (left) and from readout strips (right).
The ratio of the amplitudes from the floating strip hits to the readout strip hits is about $0.85$.
This means about $15\%$ charge loss to the backplane for floating strips assuming no charge loss of the readout strip hits, and this
number is consistent with the designed value of $20\%$.

\begin{figure}[htbp]
  \begin{center}
  \includegraphics[width=0.49\textwidth]{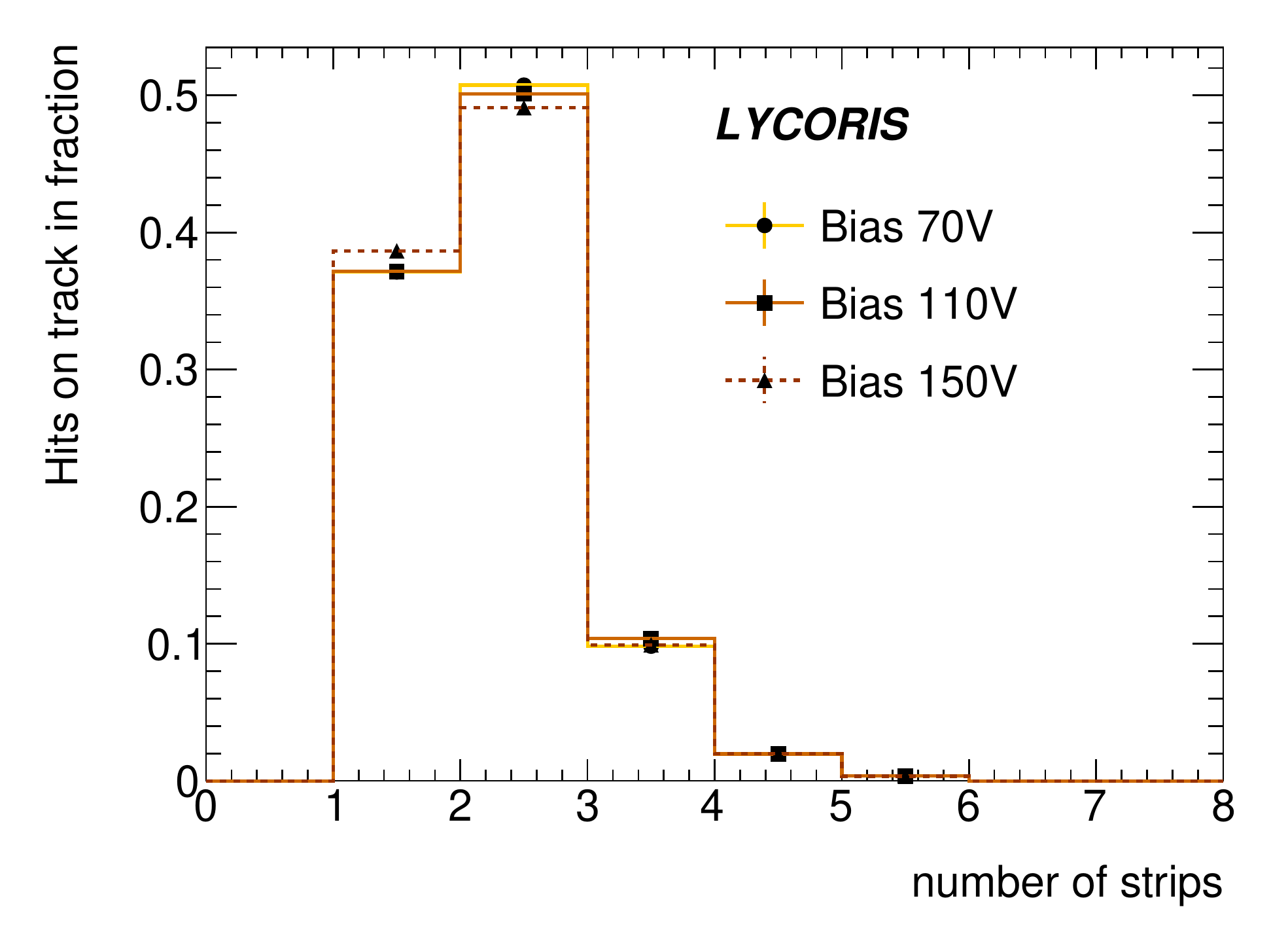} 
  \includegraphics[width=0.49\textwidth]{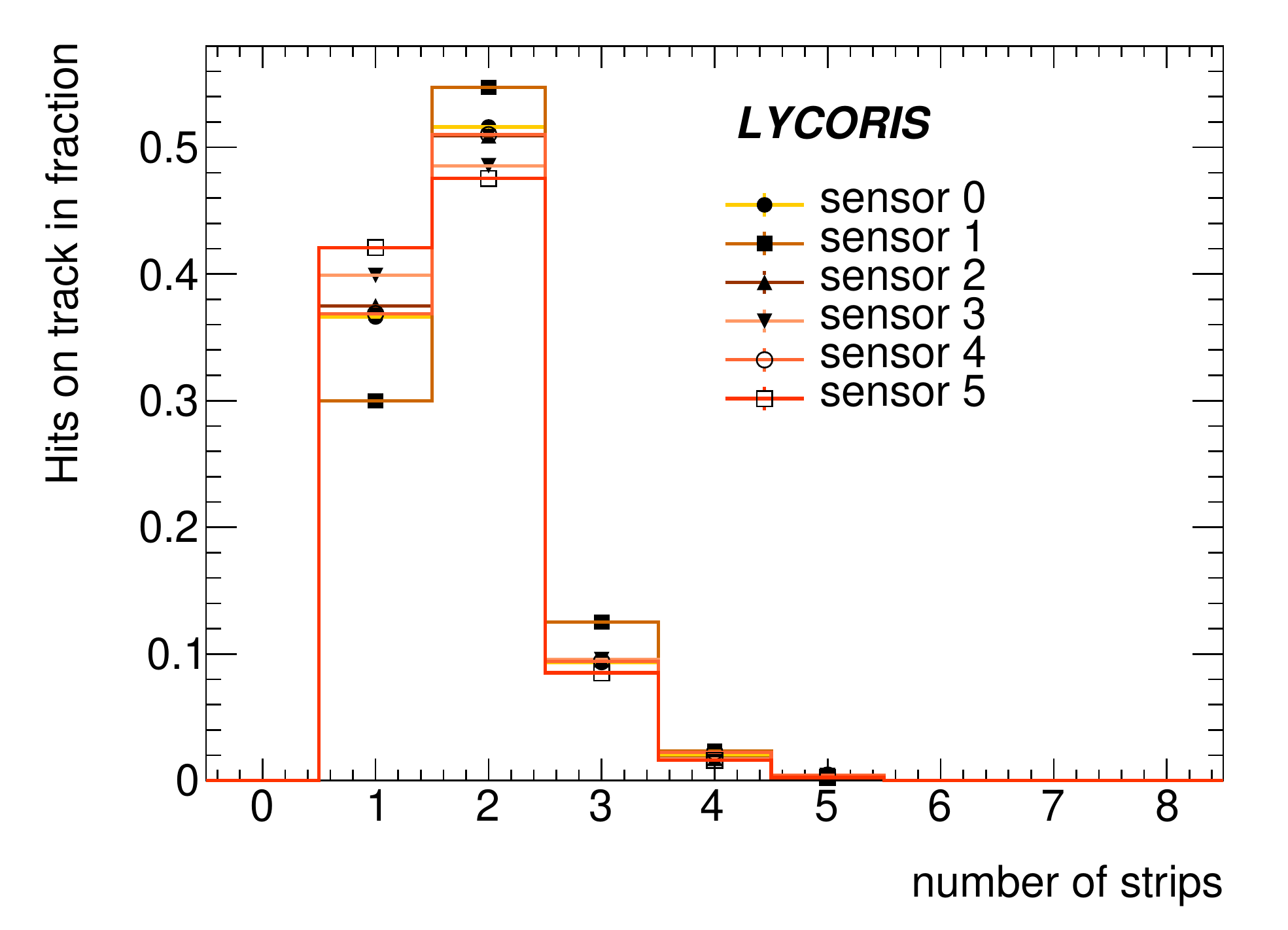} 
  \caption{\label{fig:performance:csize} The hit cluster size for \LYCORIS standalone tracks for one example sensor at three bias
  voltages (left) and the hit cluster size for all \LYCORIS planes at the default \SI{70}{\volt} bias voltage (right).}
\end{center}
\end{figure}

\paragraph{Cluster Size}%
Figure~\ref{fig:performance:csize} (left) shows the cluster size of hits on \LYCORIS stand-alone tracks from one example sensor at bias voltages of
\SI{70}{\volt}, \SI{110}{\volt}, and \SI{150}{\volt}.
A slight increase on the number of single-strip clusters and a slight drop on the number of two-strip clusters can be observed
with the increase of the bias voltage. It also demonstrates, that the default operation bias voltage \SI{70}{\volt} is sufficient.
Figure~~\ref{fig:performance:csize} (right) shows the hit cluster size for all six \LYCORIS sensor planes at the default bias voltage of \SI{70}{\volt}.

\paragraph{Hit Efficiency}

The hit efficiency of the sensor planes has been derived using two separate approaches, using stand-alone tracks from \LYCORIS and using tracks from
the \AZALEA telescope as probe-tracks.

\begin{figure}[htbp]
  \begin{center}
  \includegraphics[width=0.49\textwidth]{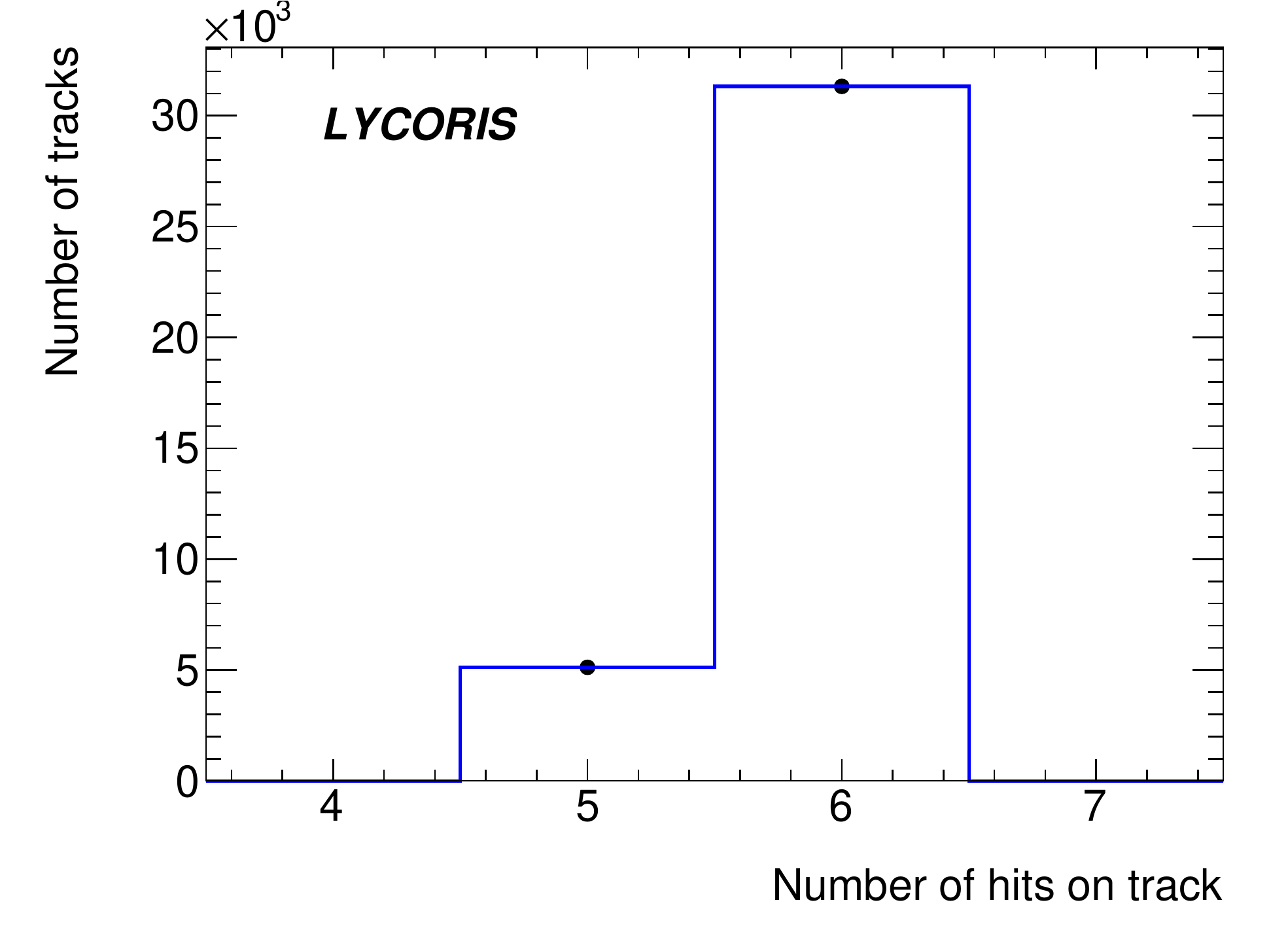} %
  \caption{\label{fig:performance:nhits} Hit multiplicity for \LYCORIS stand-alone reconstructed tracks.}
\end{center}
\end{figure}

Assuming a similar single hit efficiency $e$ for each of the six sensor planes, the probability to get a six hits on-track is $P(6) = e^6$
while for five hits on-track the efficiency is $P(5)=6\cdot e^5(1-e)$.
The ratio of six hits to five hits on-track is
\begin{equation}
f = \frac{P(6)}{P(6)+P(5)} = \frac{e}{6-5\cdot e},
\end{equation}
hence the hit efficiency is $e = 6f/(5f+1)$.
Figure~\ref{fig:performance:nhits} shows the hit multiplicity for \LYCORIS stand-alone reconstructed tracks using the default hit selection used in this paper,
from which the measured fraction is $0.86$ resulting in a single hit efficiency of approximately $97.3\%$.

The fake rate of the reconstructed tracks has been studied by applying the track reconstruction algorithm onto a pure noise dataset, i.e.\
data collected by random trigger without beam present.
From a $25000$ events sample, only $29$ tracks consisting of five hits are found with a very poor $\chi^2$ value, validating the excellent
noise rejection of the tracking algorithm developed for \LYCORIS.

An independent way to derive the single-plane tracking efficiency is to use tracks formed by \AZALEA and using them as probe tracks to derive the efficiency of each
\LYCORIS plane independently, removing any potential bias introduced by the stand-alone method.
The \AZALEA tracks are reconstructed using its standard reconstruction algorithms in the \EUTELESCOPE package and for the \LYCORIS hits to be matched,
the default hit selection is used. The single-plane efficiencies found for each plane are shown in Table~\ref{tab:performance:sigma_y}
and is very much compatible with the stand-alone method and also validating the previous assumption of very similar plane efficiencies.

\subsubsection{Tracking performance}

\paragraph{Spatial Resolution}
The spatial resolution is defined as the width of a Gaussian fit of the distribution of the residuals, i.e. the spatial distance from the \LYCORIS measurement to the projection of
the reference track provided by the \AZALEA telescope. A single hit measured by \LYCORIS has only one axis information, the other axis information can only be obtained through the
\striplet object. Therefore, the spatial resolution will be given separately.

\begin{figure}[htbp]
  \begin{center}
  \includegraphics[width=0.49\textwidth]{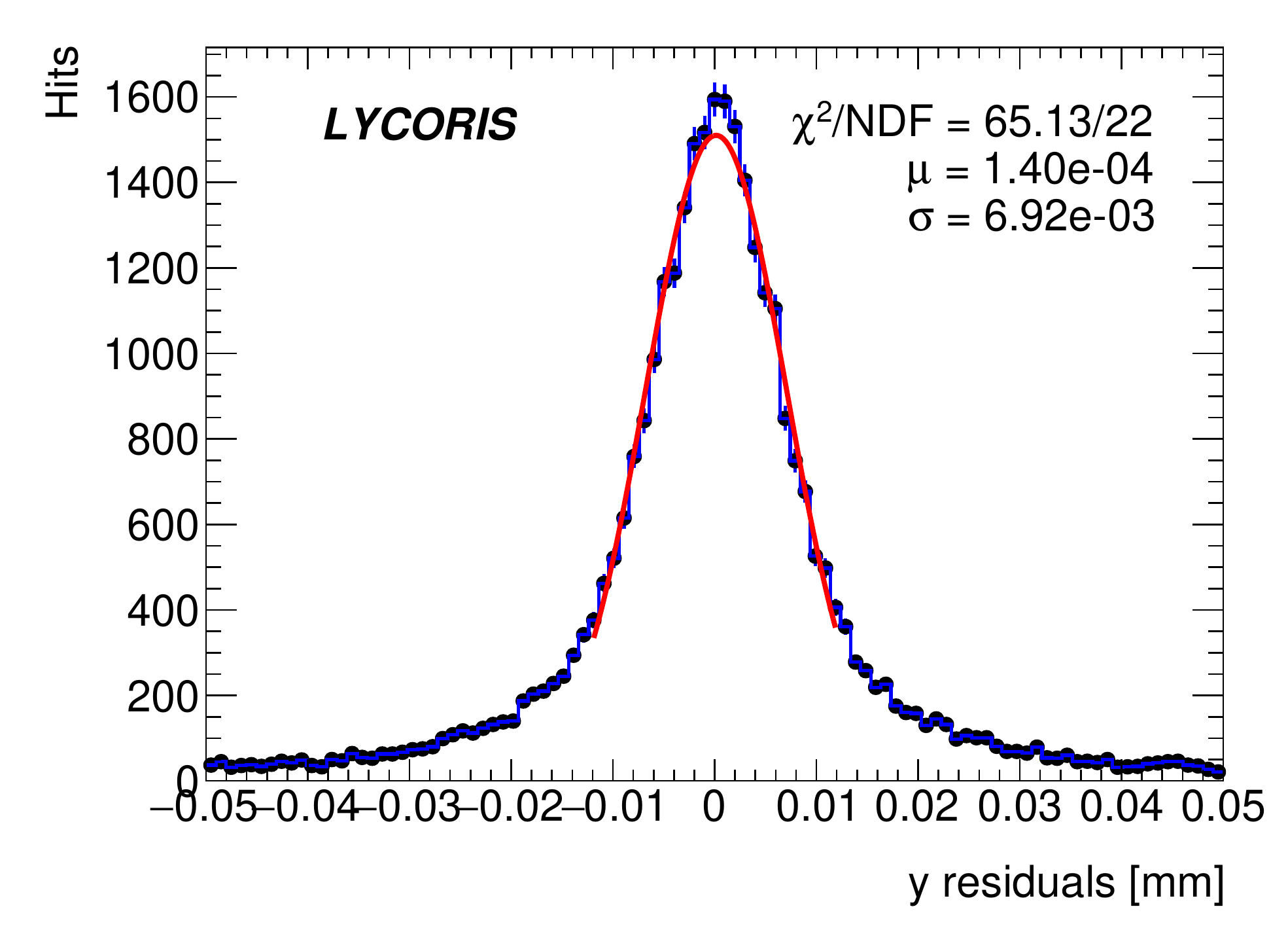}
  \includegraphics[width=0.49\textwidth]{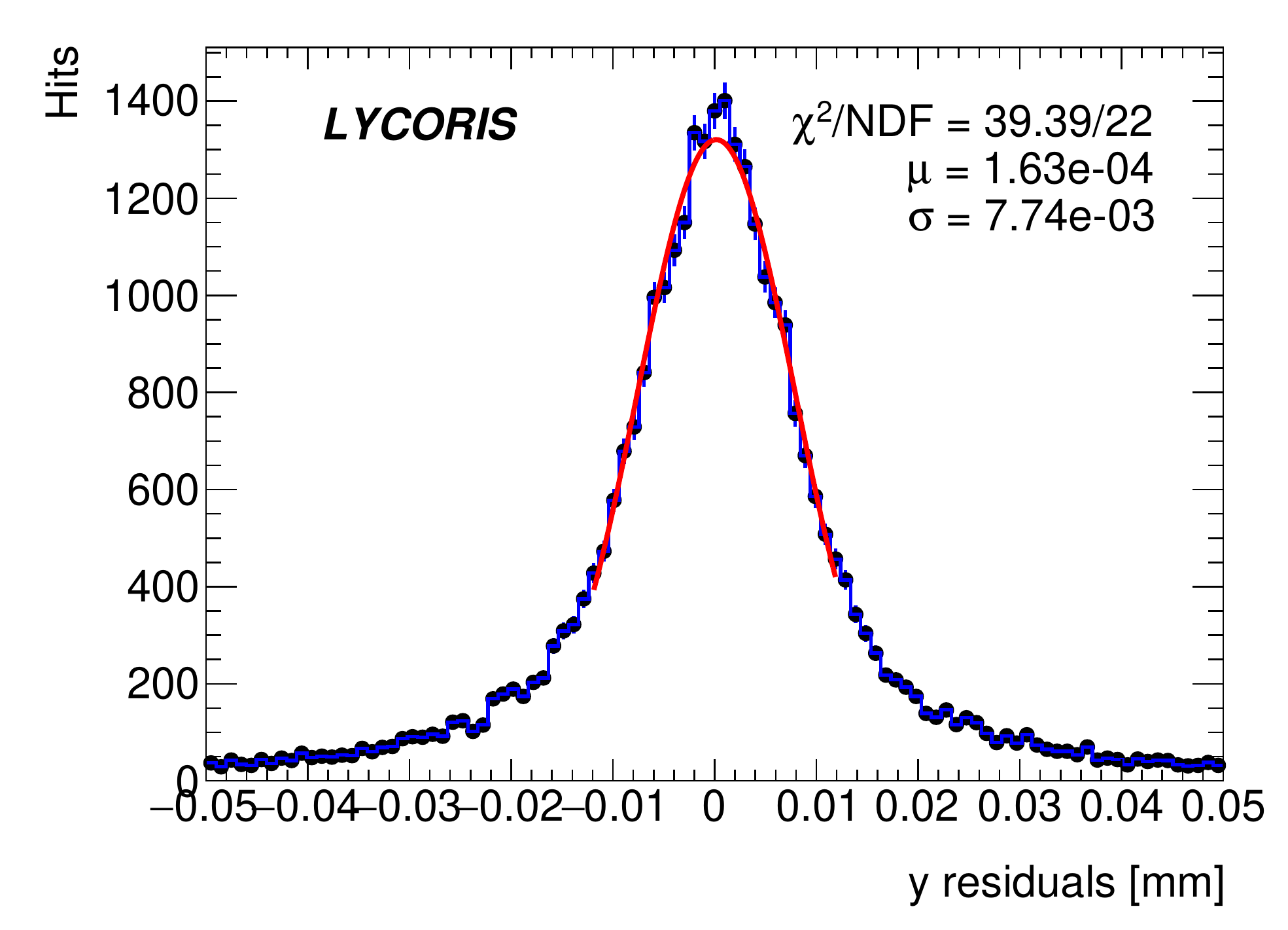} %
  \caption{\label{fig:performance:residual-dy} The biased (left) and the unbiased (right) residuals on the local y-axis (perpendicular to the strip direction) in \si{\milli\meter} for one example \LYCORIS plane.}
\end{center}
\end{figure}

The single point resolution for \LYCORIS is the resolution on the local y-axis (perpendicular to the strip direction) $\sigma_y$.
It is given by the y-axis residuals, measured by comparing every hit on a \LYCORIS plane with respect to the intersection of the reference \AZALEA track on this plane.
\AZALEA tracks can be reconstructed using \LYCORIS hits as input or not, respectively resulting in a biased or an unbiased measurement of the single point resolution.
Figure~\ref{fig:performance:residual-dy} shows the biased (left) and the unbiased (right) y-axis residuals for one example \LYCORIS plane, each of which fits to a Gaussian well.
The fitted results are both well centered at zero, showing a good alignment performance.
The width of the two Gaussian fits gives a biased resolution $\sigma_y^{biased}=$\SI{6.92}{\micro\metre} and an unbiased resolution $\sigma_y^{unbiased}=$\SI{7.74}{\micro\metre}.

The final single point resolution of each \LYCORIS sensor is estimated using the geometric mean~\cite{resolution_2005} of the biased and the unbiased resolutions:
\begin{equation}
\sigma_y = \sqrt{\sigma_y^{unbias} \cdot \sigma_y^{bias}}.
\end{equation}
Table~\ref{tab:performance:sigma_y} shows the unbiased, biased and the geometric mean of the single point resolution for all six sensor planes.
The final single point resolution ranges from \SI{6.92}{\micro\metre} to \SI{7.32}{\micro\metre}, significantly better than the required \SI{10}{\micro\metre}.

\begin{table}[htbp]
  \centering
  \begin{tabular}{c c c c c}
  Sensor ID &  Efficiency & $\sigma_y^{biased}$\SI{}{{\micro\metre}} & $\sigma_y^{unbiased}$\SI{}{{\micro\metre}} & $\sigma_y$\SI{}{{\micro\metre}} \\ \toprule
  upstream 0   & 95.0\%  &$6.92\pm0.06$  &$7.74\pm0.08$  &$7.32\pm0.07$ \\
  upstream 1   & 97.0\%  &$6.53\pm0.05$  &$7.33\pm0.07$  &$6.92\pm0.06$ \\
  upstream 2   & 96.2\%  &$6.81\pm0.06$  &$7.67\pm0.08$  &$7.22\pm0.07$ \\
  downstream 0 & 96.7\%  &$6.64\pm0.05$  &$7.41\pm0.07$  &$7.01\pm0.08$ \\
  downstream 1 & 96.6\%  &$6.69\pm0.05$  &$7.45\pm0.07$  &$7.06\pm0.06$ \\
  downstream 2 & 96.7\%  &$6.64\pm0.06$  &$7.59\pm0.08$  &$7.21\pm0.07$ \\
  \bottomrule
  \end{tabular}
  \caption{The single plane hit efficiency and the three (biased, unbiased and their geometric mean) y-axis resolutions for all six sensor planes.}
  \label{tab:performance:sigma_y}
\end{table}

\begin{figure}[htbp]
  \begin{center}
  \includegraphics[width=0.49\textwidth]{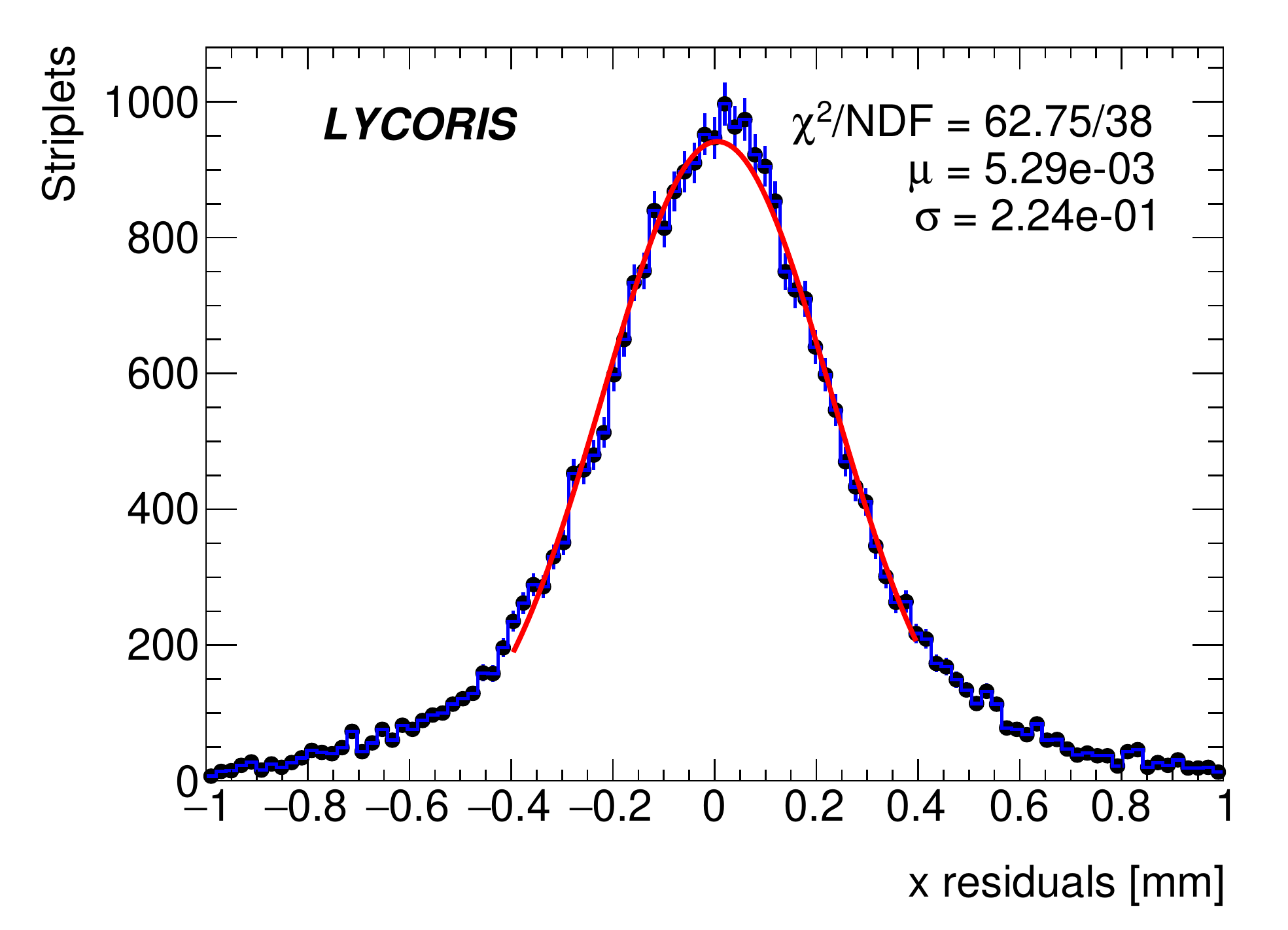} %
  \includegraphics[width=0.49\textwidth]{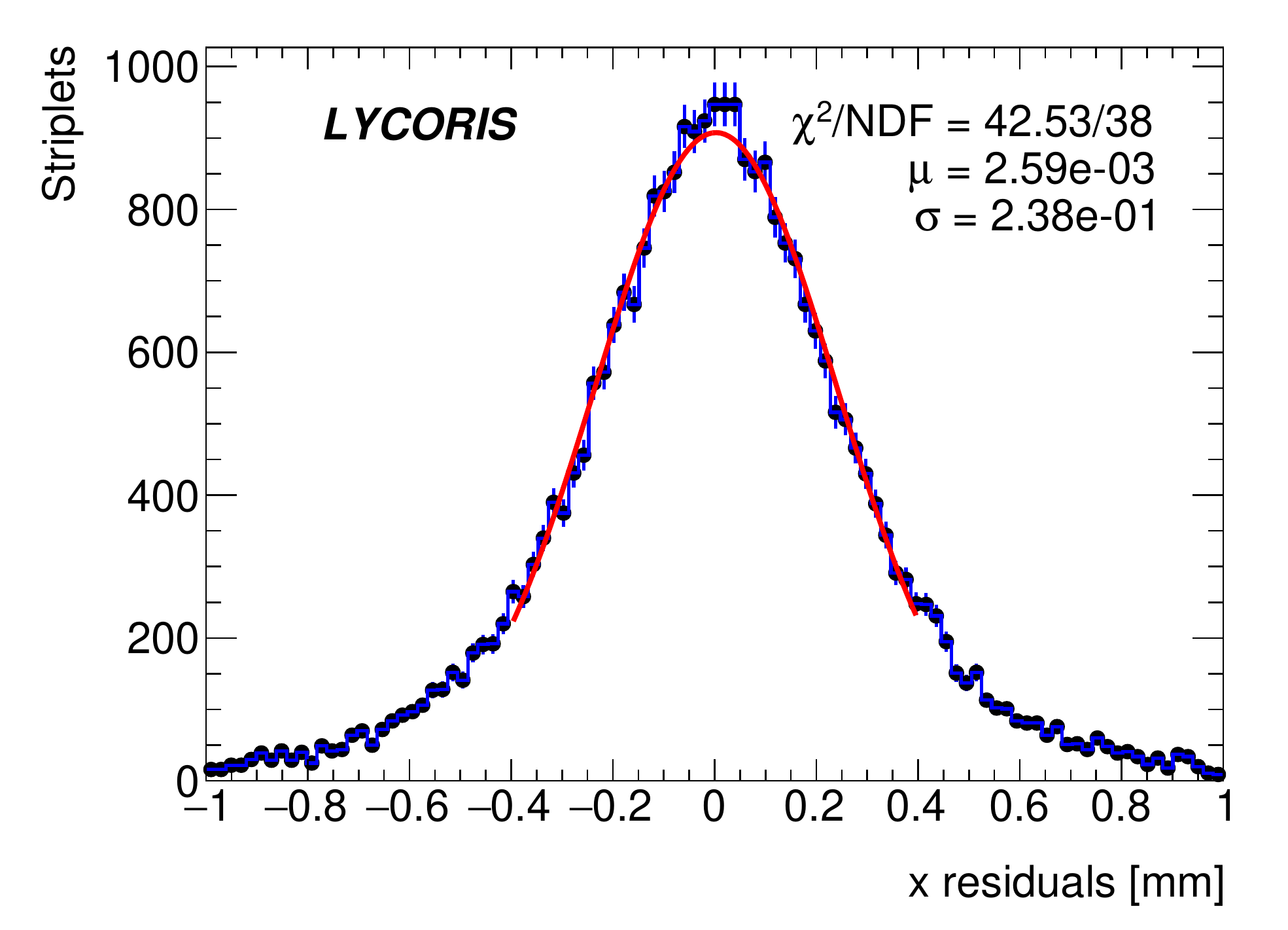}
  \caption{\label{fig:performance:residual-dx}Residuals on the local x-axis (along the strip direction) in \si{\milli\meter} of \striplet from up- (left) and downstream (right) cassettes.}
\end{center}
\end{figure}

A single \LYCORIS plane measurement is not capable to retrieve information on the local x-axis (along the strip direction),
while the \striplet object reconstructed from three planes in one cassette provides a 2D information with the \SI{2}{\degree} stereo angle arrangement.
Figure~\ref{fig:performance:residual-dx} shows the x-axis residuals of up- and downstream cassettes.
Both residual distributions are fitted to a Gaussian that is well centered at zero.
The width of the Gaussian fit gives an x-axis resolution $\sigma_x=$\SI{0.224}{\milli\metre} for the upstream cassette and $\sigma_x=$\SI{0.238}{\milli\metre} for the downstream cassette,
exceeding the design requirement of \SI{1}{\milli\metre} by a factor of around five.

\paragraph{Momentum Resolution} %
The momentum resolution is defined as the error propagated on the curvature measurement of \LYCORIS stand-alone tracks.
Data collected inside the \PCMAG with the two \LYCORIS cassettes well separated is the ideal configuration for momentum measurements.
As a result of a very low data rate inside the \PCMAG and a lack of test beam time, only $251$ valid tracks have been reconstructed
from the data collected by \LYCORIS inside the \PCMAG.
Therefore, a simulation is implemented to estimate the momentum resolution in this case, and the single point resolution is taken from the average measurement value of \SI{7}{\micro\metre} shown in this paper.
The simulation takes geometry input from the test beam setup of \LYCORIS installed inside the \PCMAG, where the distance is \SI{750}{\milli\metre}
from the first to the last \LYCORIS planes, and the distance between sensor planes in each cassette is taken from the technical drawings.
The simulated momentum resolution is \SI{3.60E-3}{\per\GeV} given the magnetic field is \SI{1}{\tesla}.
This aligns well with the desired momentum resolution of \SI{5.0E-3}{\per\GeV} expected from users for measurements inside \PCMAG.

\subsection{Summary}
The \LYCORIS strip telescope has been successfully tested at the \DIITBF. It has
demonstrated a single hit efficiency of approximately $97.3\%$. The  single-point resolution orthogonal to the strip ($\sigma_y$)
is on average \SI{7.07}{\micro\meter} and the resolutions parallel to the strip ($\sigma_x$) for the up- and downstream cassettes
are \SI{0.22}{\milli\meter} and  \SI{0.24}{\milli\meter} respectively.

\section{Possibilities for User DUT integration}\label{sec:dut}
The design of \LYCORIS was driven by the requirements of operating it inside the \PCMAG at the \DIITBF,
however, much care was taken to keep the design as flexible as possible. So \LYCORIS can also be operated at other test beam facilities world-wide, if there is a user need.
Provided there is a suitable mechanical support for the cassettes, the only requirement to the facility is a suitable signal to derive a \KPIXAcqstart, in order to start up \LYCORIS.

The \LYCORIS telescope can receive an external clock and timestamp its events using this clock as a time base.
Therefore, \LYCORIS can synchronize to a device that is able to issue or receive a common clock line and timestamp its events based on this clock.
Therefore, integrating a user DUT can be realized in the following two methods:
\begin{itemize}
  \item one is to synchronize the user DUT with \LYCORIS through an external trigger source, with which \LYCORIS is able to synchronize;
  \item the other is to synchronize the user DUT with \LYCORIS directly.
\end{itemize}
Both methods mentioned above can be used to operate \LYCORIS in a self or external trigger mode.

The \LYCORIS telescope has been designed from the beginning with a user DUT interface, that is fully compatible with the \AIDA TLU and \EUDAQII, which simplifies the
integration of a user DUT. By using this interface, the data taking sequencing is managed by \EUDAQII and the synchronization is realized with the \AIDA TLU.
Therefore, the user DUT needs to be integrated with the \AIDA TLU, which includes receiving the TLU triggers and synchronizing to the TLU events
by receiving the trigger ID or timestamping the triggers using the TLU common clock.

It is also possible to integrate a DUT differently instead of using the \AIDA TLU compatible interface.
This can be done by synchronizing with \LYCORIS by providing a common clock signal through the NIM or the BNC port on the \LYCORIS DAQ board.
Furthermore, a common external trigger can be fed through one of these ports as well.
This of course requires additional engineering efforts and extra tests to verify the synchronization status and thus will vary case by case.
Therefore, in this paper only basic guidelines are given for such an integration scheme:

\begin{itemize}
  \item \LYCORIS needs to receive a common clock signal with a clock counter start signal $T_0$, and the clock has to be a fraction or multiples of \SI{200}{\mega\hertz};
  \item If external triggers are going to be used, it is recommended but not necessary to veto all triggers outside the \KPIX active data taking window to simplify the synchronization.
\end{itemize}

\section{Summary and Outlook}\label{sec:conclusion}

The \LYCORIS telescope has been designed and built to meet the user requirements for a large-area coverage silicon-strip telescope
for the \DIITBF with a resolution of better than \SI{10}{\micro\meter}.

The key building blocks are the \LYCORIS modules using a so-called hybrid-less design with two \KPIX readout ASICs bump-bonded directly
to the silicon sensor, which has a strip pitch of \SI{25}{\micro\meter} with every second strip being read out.
The entire telescope consists of six planes located in two cassettes with three modules each.
The performance of \LYCORIS has been investigated in several test beam campaigns at the \DIITBF and the achieved resolution on the y-axis (orthogonal to the strip)
per plane is on average \SI{7.07}{\micro\meter}.
The results are consistent with the expected hit resolution of \SI{7.2}{\micro\meter} estimated using $d/\sqrt{12}$, with $d$ corresponding to \SI{25}{\micro\metre}.
This demonstrates the viability of the  approach selected for \LYCORIS of only reading out every second strip and still maintaining the expected hit resolution.
The resolutions on the x-axis (parallel to the strip) for up- and downstream cassettes are respectively \SI{0.22}{\milli\meter} and \SI{0.24}{\milli\meter},
exceeding the \SI{1}{\milli\metre} requirement by a factor of around five. The achieved resolutions are summarised in Table~\ref{tab:achieved}.

\begin{table}[htbp]
\begin{center}
\begin{tabular}{l l r r }
		                            &           & Required					& Achieved \\  \toprule
Resolution in bending plane     & $\sigma_y$& $<$ \SI{10}{\micro\metre} & $<$ \SI{7.2}{\micro\metre} \\
Resolution orthogonal to the bending plane  & $\sigma_x$&$<$ \SI{1}{\milli\metre} &$<$ \SI{0.24}{\milli\metre} \\
Area coverage                   & $A_{xy}$  &   10$\times$  \SI{10}{\centi\metre\squared} 	&   10$\times$\SI{10}{\centi\metre\squared} \\
Thickness of single station     & d         &  \SI{3.3}{\centi\metre} 			& $<$ \SI{3.5}{\centi\metre} \\ \bottomrule
\end{tabular}
\caption{\label{tab:achieved} The key requirements for \LYCORIS and the achieved vlaues, where the achieved area can be extended to 10$\times$\SI{20}{\centi\metre\squared} by installing three more sensors inside each station.}
\end{center}
\end{table}

\LYCORIS will now be rolled out to the users at the \DIITBF and further wishes for improvements and enhancements are to be expected.
One obvious way for an improvement is possible second-generation cassette design that uses four sensor layers, two axial layers and one stereo (back-to-back) double layer in the center, which would yield three unbiased measurements on the y-axis.

\section*{Acknowledgment}  
This project has received funding from the European Union's Horizon 2020 Research and Innovation programme under Grant Agreement no. 654168.
The authors would like to acknowledge
Deutsches Elektronen-Synchrotron DESY and Helmholtz Gesellschaft Funding for all the support.
Support at SLAC was funded by the US-Japan collaboration
and Research at the University of Oregon was supported by the U.S. Department of Energy Office of Science, Office of High Energy Physics under Award Number DE-SC0017996.
The measurements leading to these results have been performed at the test beam facility at DESY Hamburg (Germany),
a member of the Helmholtz Association.
The authors would like to thank the technical team at the \DESYII accelerator and the \DIITBF  for the smooth operation of the test beam
and the support during the test beam campaign.

\bibliography{lycbibfile}

\end{document}